\documentclass[aps,prd,twocolumn,noeprint,showpacs,nofootinbib,superscriptaddress]{revtex4-2}

\usepackage{caption}
\usepackage{dcolumn} 
\usepackage{bm} 
\usepackage{multirow}
\usepackage[table]{xcolor}
\usepackage{siunitx}
\usepackage{graphicx}
\usepackage{natbib}
\usepackage{epsfig}
\usepackage{float}
\usepackage{xcolor}
\usepackage{booktabs}
\usepackage{bibunits}
\usepackage{amsmath,amssymb,amsfonts}

\definecolor{lightgray}{gray}{0.95}
\definecolor{lightblue}{RGB}{230, 230, 250}

\begin{document}

\preprint{APS/123-QED}

\title{New Gravitational Wave Discoveries Enabled by Machine Learning}

\author{Alexandra E. Koloniari}
\affiliation{Department of Physics, Aristotle University of Thessaloniki, 54124 Thessaloniki, Greece}

\author{Evdokia C. Koursoumpa}
\affiliation{Department of Physics, Aristotle University of Thessaloniki, 54124 Thessaloniki, Greece}

\author{Paraskevi Nousi}
\affiliation{Swiss Data Science Center, ETH, Z\"urich, Switzerland}

\author{Paraskevas Lampropoulos}
\affiliation{Department of Physics, Aristotle University of Thessaloniki, 54124 Thessaloniki, Greece}

\author{Nikolaos Passalis}
\affiliation{Department of Informatics, Aristotle University of Thessaloniki, 54124 Thessaloniki, Greece}

\author{Anastasios Tefas}
\affiliation{Department of Informatics, Aristotle University of Thessaloniki, 54124 Thessaloniki, Greece}

\author{Nikolaos Stergioulas}
\affiliation{Department of Physics, Aristotle University of Thessaloniki, 54124 Thessaloniki, Greece}

\date{\today}

\begin{abstract}
The detection of gravitational waves has revolutionized our understanding of the universe, offering unprecedented insights into its dynamics. A major goal of gravitational wave data analysis is to speed up the detection and parameter estimation process using machine learning techniques, in light of an anticipated surge in detected events that would render traditional methods impractical.  Here, we present the first detections of new gravitational-wave candidate events in data from a network of interferometric detectors enabled by machine learning.
We discuss several new enhancements of our ResNet-based deep learning code, AresGW, that increased its sensitivity, including a new hierarchical classification of triggers, based on different noise and frequency filters. The enhancements resulted in a significant reduction in the false alarm rate, allowing
AresGW to surpass traditional pipelines in the number of detected events in its effective training range (single source masses between 7 and 50 solar masses and source chirp masses between 10 and 40 solar masses), when the new detections are included. We calculate the astrophysical significance of events detected with AresGW using a logarithmic ranking statistic and injections into O3 data. Furthermore, we present spectrograms, parameter estimation, and reconstruction in the time domain for our new candidate events and discuss the distribution of their properties.  In addition, the AresGW code exhibited very good performance when tested across various two-detector setups and on observational data from the O1 and O2 observing periods. Our findings underscore the remarkable potential of AresGW as a fast and sensitive detection algorithm for gravitational-wave astronomy, paving the way for a larger number of future discoveries.

\end{abstract}

\maketitle

\section{Introduction}

The cosmos whispers its secrets through gravitational waves, subtle ripples in the fabric of spacetime that carry profound insights into the universe's most enigmatic phenomena. In the quest to decipher these cosmic murmurs, the convergence of machine learning (ML) algorithms and gravitational wave astronomy has ushered in a new era of discovery, promising unprecedented sensitivity and efficiency in detecting these elusive signals.

During the initial three observing runs (O1-O3) conducted initially by the LIGO-Virgo Collaboration \cite{LIGO_ref, Virgo_ref}, with the later addition of Kagra \cite{KAGRA_ref}, around 90 gravitational wave (GW) events were confidently identified and published in the GWTC catalogs \cite{GWTC1, GWTC2, LIGO2022GWTC21, GWTC3}. These events primarily consisted of binary black hole (BBH) mergers, alongside a minority of binary neutron star (BNS) and neutron star-black hole (NSBH) systems. Additional events were published in the OGC catalogs \cite{OGC-1,OGC-2,OGC-3,OGC-4} and the IAS catalogs \cite{IAS_O3a,IAS_O3b,IAS_higher_harmonics} and updated significance of events was discussed with the pycbc\_KDE pipeline in \cite{2024arXiv240310439K}. As we find ourselves in the middle of the fourth observing run (O4) and anticipate the dawn of upgraded or next-generation detectors, such as LIGO-India \cite{LIGO_India_2022}, Voyager \cite{LIGO:2020xsf}, and Virgo nEXT, Cosmic Explorer \cite{2019BAAS...51g..35R}, Einstein Telescope \cite{2020JCAP...03..050M}, and NEMO \cite{Ackley:2020atn}, the need for more efficient gravitational wave detection algorithms becomes crucial \cite{2021arXiv211106987C}. Traditional matched-filtering methods face computational challenges for near-real-time processing, compounded by the complexities of accurately detecting systems with non-aligned spins. Unmodeled search techniques, while promising, exhibit variable sensitivity to different gravitational wave sources, necessitating further exploration, especially in light of theories beyond general relativity.

In recent years, there has been an increase in the use of machine learning approaches for the analysis of gravitational wave data (see \cite{cuoco2020review,app13179886,2023arXiv231115585Z} for reviews). The implementation of convolutional neural networks (CNN), auto-encoders, and other machine-learning methods has been investigated as an attractive solution to the problem of detecting gravitational waves (GWs), see e.g. \cite{PhysRevLett.120.141103,PhysRevD.97.044039,2019PhRvD.100f3015G,CORIZZO2020113378,PhysRevD.102.063015,2020PhRvD.101j4003W,2020PhLB..80335330K,2020arXiv200914611S,2021PhRvD.103f3034L,DODIA,2021NatAs...5.1062H,2021MNRAS.500.5408M,2021PhLB..81236029W,2021CQGra..38o5010A,2021PhRvD.104f4051J,10.3389/frai.2022.828672,2022arXiv220208671C,PhysRevD.105.043003,2022arXiv220606004B,schafer2022training,PhysRevD.106.042002,2022arXiv220704749A,2022arXiv220612673V,2022PhRvD.106h4059A, 2022MNRAS.516.3847G, 2023PhRvD.107h2003A,2023MNRAS.519.3843A,2023PhRvL.130q1402L,2023PhRvL.130q1403D,2023CQGra..40m5008B,2023arXiv230615728T,2023PhRvD.108d3024M,2023MLS&T...4c5024B,2022arXiv220111126M,2023PhLB..84037850Q,2023arXiv230200295L,2023arXiv230519003J,2023CQGra..40s5018F,2023arXiv230716668G,2023PhRvD.107d4032Y,2023arXiv230808429B,2023MLS&T...4d5028J,2024ChJPh..88..301T,2024arXiv240318661M,2024PhRvD.109d3011S,2024arXiv240207492Z}. In \cite{2024CQGra..41l5003T}, a new candidate event was found with a  machine-learning algorithm in a single-detector search.

In the past, it has been difficult to assess the effectiveness of such efforts in a realistic setting. For this reason, the inaugural Machine-Learning Gravitational-Wave Mock Data Challenge (MLGWSC-1) was concluded \cite{challenge1}, establishing an objective framework to assess the sensitivity and efficiency of ML algorithms on modeled injections in both Gaussian and real O3a detector noise, compared to traditional algorithms. In \cite{AresGW_model}, AresGW, the leading ML algorithm for BBH template waveform injections in real O3a noise, was detailed, demonstrating that with further enhancements, it surpassed, for the first time, the results achieved with standard configurations of traditional algorithms in a specific context. This success was achieved for a component mass range of $7-50 M_\odot$ (covering $70\%$ of the reported events in the cumulative GWTC catalog \cite{GWTC3}) and a relatively low false-alarm rate (FAR) as low as one per month.

Here, we present several new improvements in the AresGW algorithm and assess its performance in comparison to the existing GWTC, OGC, IAS and pycbc\_KDE pipelines on O3 data. This includes the definition of a logarithmic ranking statistic, inspired by \cite{2024CQGra..41l5003T} and the evaluation of the astrophysical probability, $p_{\rm astro}$ of candidate events.
Furthermore, we identify {\it eight new} GW candidates in the O3 data, with $p_{\rm astro}>0.5$ and a cumulative astrophysical probability of 5.94. This is the {\it first identification of new coincident events in GW searches with a network of detectors by a machine-learning-based algorithm}. 

A key aspect of AresGW's algorithm that allowed us to uncover new GW candidate events in the O3 period, is that fact that it is trained to evaluate simultaneously the time-series data of the two LIGO detectors, producing directly a single ranking statistic. In this way, the network can learn patterns, relationships, or features from the combined data of the two detectors,  extracting valuable information that might not have been apparent when each detector's data is first ranked independently of the other, as is the case in traditional searches.

The paper is organized as follows: 
In Section II we present the methods used in our analysis. In Section III we discuss the background model and statistics. In Section IV we discuss the foreground model and statistics and in Section V we present the calculation of the astrophysical probability. In Section VI we demonstrate the confirmation of known gravitational wave events with AresGW and in Section VII we present the new candidate events found with AresGW. The paper concludes with a discussion in Section VIII.

\subsection{Effective Training Range}\label{sec:chirp_mass}

\begin{figure}[t]
  \centering
  \includegraphics[width=0.99\linewidth]{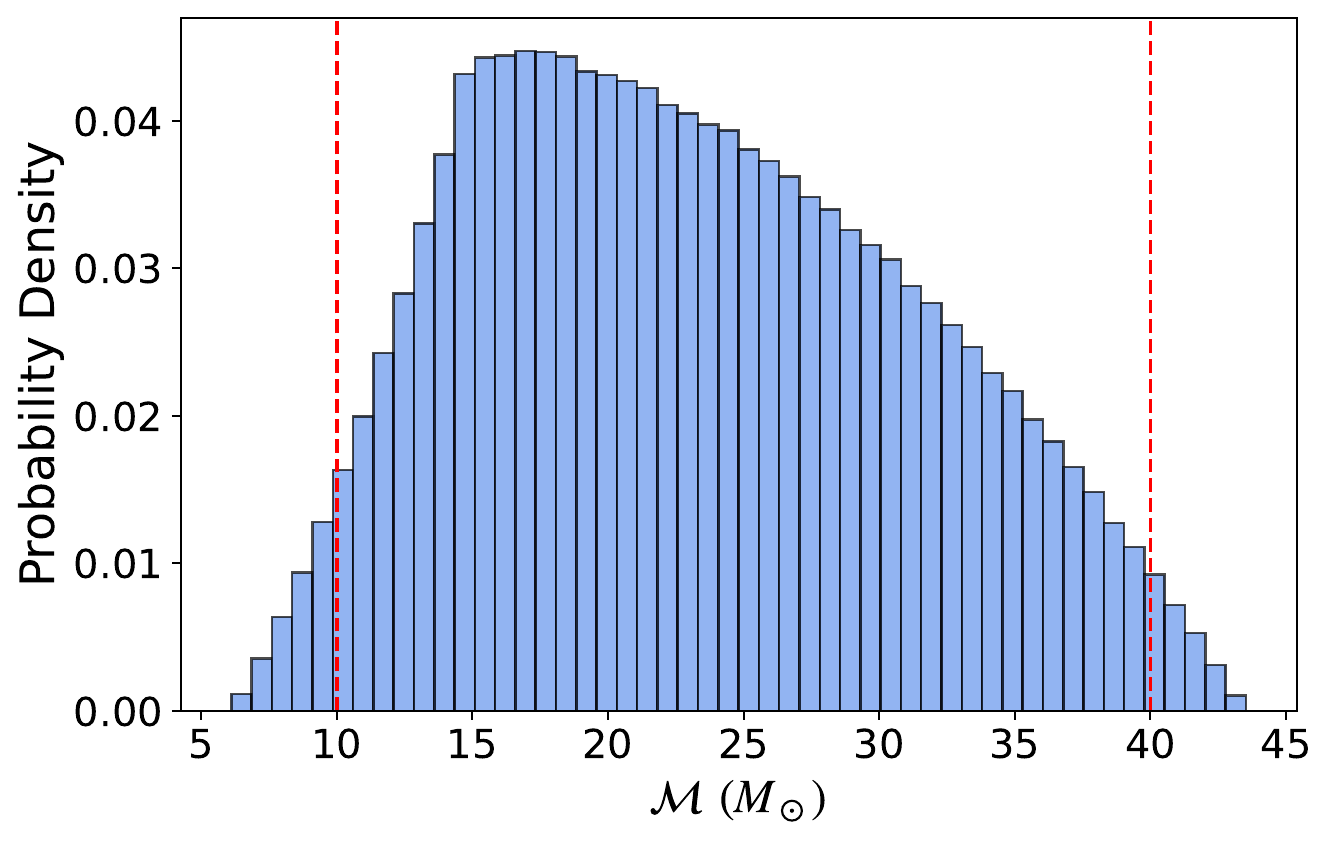}
    \caption{Due to the relationship between the individual masses $m_1$ and $m_2$ and the chirp mass (${\cal M}$), the resulting distribution of the chirp mass appears skewed, even though $m_1$ and $m_2$ both follow a uniform distribution. Considering the final distribution of ${\cal M}$, the probability for a modeled waveform used in training to have ${\cal M} < 10$ is approximately 0.027, while the probability for ${\cal M} > 40$ is approximately 0.017. The dashed red lines on the plot represent the ${\cal M}$ values of 10 and 40. We consider only the ${\cal M}$ values between those two lines to be within the training dataset range since the network hasn't been exposed to enough samples from outside that range to be effectively trained on them.}
    \label{fig:Mchirp_hist}
\end{figure}

\section{Methods}

In the following subsections, we will introduce the different methods we used in our analysis. Starting from the training data, we will emphasize the range of chirp mass values $\cal M$ and other parameters in which our code AresGW was trained to detect BBH merger events. We will then describe the various enhancements that we implemented with respect to the initial version of AresGW in \cite{AresGW_model}, the removal of known glitches, the parameter estimation procedure and, finally, the various consistency checks of new candidate events.

\subsection{Training Data} \label{sec:Training_Data}

The training dataset consists of real O3a noise from the two LIGO detectors, with a duration equivalent to 35 days, generated with the code used in \cite{challenge1}, with a random seed of 42 and with a start time on day 45 of the initial dataset with an equivalent duration of 81 days. This represents a significant increase in the duration of the training set, which was 12 days in our previous version of AresGW \cite{AresGW_model,AResWG}. Since a partial overlap of the noise used in the training with the O3 data may have a negative impact on the accuracy of the code, we mitigate this effect, by randomly shuffling 40\% of the noise segments between the two detectors in each epoch, leading to only a small percentage of noise segments in the training samples that overlap with the test data.  

The O3a noise was downloaded from the Gravitational Wave Open Science Center (GWOSC) \cite{KAGRA:2023pio}, keeping the
"data'' quality flag active and requiring the internal "CBC\_CAT1\_VETO," "CBC\_CAT2\_VETO," "CBC\_HW\_INJ", and "BURST\_HW\_INJ" flags inactive.  Only segments with a minimum duration of 2 hours where both LIGO detectors were observing were included, while 10-second intervals around detections listed in GWTC-2 were excluded (see \cite{challenge1} for further details).

We injected model signals into the noise that were generated employing masses uniformly drawn in a range spanning from 7 $M_\odot$ to 50 $M_\odot$, spins characterized by an isotropic distribution with magnitudes ranging from 0 to 0.99, and random orientation.  Additionally, the signals were  uniformly distributed across coalescence phase, polarization, inclination, declination, and right ascension, 
see \cite{challenge1,AresGW_model} for further details. The chosen waveform model, IMRPhenomXPHM \cite{IMRPhenomXPHM_ref}, incorporates higher-order modes and includes precession effects.

The chirp mass distribution in the dataset used for training, as shown in Fig. \ref{fig:Mchirp_hist}, has a mean value of 23.2 $M_\odot$ and a standard deviation of 8.08 $M_\odot$. This encompasses a large fraction of the inferred distribution from published GW observations, see discussion in Section \ref{sec:Population}. Note that although individual mass components were chosen uniformly in the range 7 $M_\odot$ to 50 $M_\odot$, the corresponding distribution of $\cal M$ is nonuniform, with $\sim 95\%$ of the signals included in the range of $10 M_\odot <{\cal M}< 40 M_\odot$. We consider models outside of this range as outliers.

The justification for this is as follows: Neural networks, when trained on a dataset with a predominant distribution, tend to focus more on learning patterns within the majority of the data. In the context of a binary classification problem, where the network aims to distinguish between gravitational waves and pure noise, this means the model may become less sensitive to signals that deviate significantly from the common chirp mass values.
Standard loss functions, such as mean squared error or cross-entropy, are designed to minimize errors across the entire dataset. However, when dealing with imbalanced distributions or rare events at the tails, these loss functions might not adequately penalize the misclassification of rare instances. As a result, the neural network may struggle to learn and generalize well to signals at the edges of the chirp mass distribution.

Consequently, we posit that the {\it effective training range} lies within ${\cal M} \in (10 M_\odot, 40 M_\odot)$, acknowledging the potential limitations posed by the network's reduced exposure to samples beyond these boundaries.

\begin{figure}[t]
  \centering
  \includegraphics[width=0.95\linewidth]{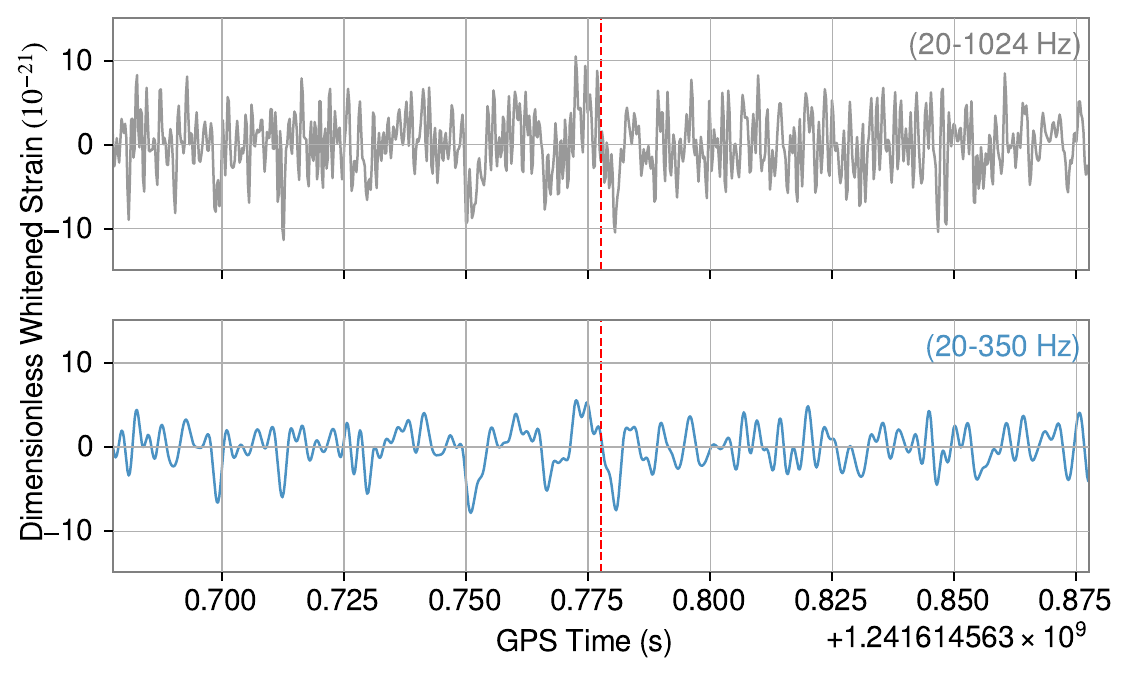}
  \caption{An example of a time-domain segment from the Livingston detector, which includes one of the new events identified by AresGW (GW190511\_135545). {\it Top panel:} The data is bandpassed with a 20-1024 Hz filter and appears noisy in the time domain. {\it Bottom panel:} With a 20-350 Hz filter the time series is less noisy, allowing for a better training of AresGW's neural network. The red dashed line  represents the merger time.}

  \label{fig:350-1000Hz}
\end{figure}

\subsection{Enhanced AresGW Code} 

Our enhanced AresGW code is based on the one-dimensional ResNet structure of its previous versions \cite{AresGW_model,AResWG} with several critical modifications that allowed us to achieve significantly higher accuracy, that is, to better separate noise signals from signals with waveforms hidden within the noise. 
For instance, within the whitening module of the network, we have added a low-pass filtering mechanism alongside the existing high-pass filtering, which occurs during the inverse spectrum truncation process to compute the power spectral density. This modification is discussed in greater detail in Section~\ref{subsec:High cut-off}.

The output of AreGW is a ranking statistic ${\mathcal R}$ ranging from 0 to 1. For astrophysically relevant triggers, the values of ${\mathcal R}$ are typically extremely close to 1 and we therefore follow \cite{2024CQGra..41l5003T} and define a logarithmic ranking statistic as
\begin{equation}
{\mathcal R_s} = -\log_{10} \left(1 - {\cal R} \right
).
\label{Eq:Rs1}
\end{equation}
In this way, an exact value of ${\mathcal R}=1$ would correspond to ${\mathcal R_s}=+\infty$. However, due to the finite precision of arithmetic operations, there exist a maximum meaningful value of ${\mathcal R_s}$, as we elaborate below.

\subsubsection{Double precision}

The network's parameters, and every computation within it, was in single-precision floating-point format (FP32), as is typically the case for deep neural network frameworks. However, a kind of numerical underflow was observed in the output layer of the network, i.e., the final softmax function which converts the network's predictions to a probablity distribution over the two classes (noise vs. noise plus signal). Setting the entire network to double-precision floating point format (FP64) fixed the issue but at the cost of significantly increased running time (over twice the time needed with FP32). To avoid this overhead, the use of FP64 on only the final softmax layer was investigated and was found to be successful in avoiding numerical problems as well as in adding virtually zero overhead to the overall detection framework. Thus, the output of the network is converted from FP32 to FP64 before the computation of the softmax function, whose output is also in FP64 format. Note that the above change to double precision output was only implemented during deployment of the code on the O3 data. During training, only single precision was used.

Due to the algorithm's precision extending up to 16 decimal places we make a slight modification of the definition in Eq. (\ref{Eq:Rs1}) as follows:
\begin{equation}
{\mathcal R_s} = -\log_{10} \left(1 - {\cal R} + 10^{-16} \right
).
\label{Eq:Rs}
\end{equation}
In Eq. (\ref{Eq:Rs}), the small value of $10^{-16}$ was added to avoid NaN errors when $\mathcal{R}$ was equal to 1.0 with double precision (which was the case for most of the loudest triggers). With this choice, the maximum meaningful value of ${\mathcal R_s}$ is 16. We emphasize that this maximum value is dependent on the above choice and represents an accumulation point of triggers that may have had even higher ${\mathcal R_s}$, if we had use an even higher number of digits as precision.

\subsubsection{Low-pass filter in both training and data analysis} \label{subsec:High cut-off}

When evaluating the efficacy of our prior model \cite{AresGW_model}, we observed instances where signals with moderate to low signal-to-noise ratios (SNR) were not effectively detected by the network. Upon closer inspection, we determined that the lack of detection stemmed from the presence of higher frequency noise masking the signals' waveforms.
To mitigate this issue, we adopted a low-pass filter set at 350 Hz to filter out extraneous noise, thereby enhancing the "visibility" of the signals for the network. To optimize the network's performance, we used the low-pass filter on both training and O3 datasets. This decision was made to ensure improved signal detection accuracy. As illustrated in Fig. \ref{fig:350-1000Hz}, we compare the effects of applying different cut-off frequencies on one of our new detections, GW190511\_135545, on the Livingston detector data frame. The upper panel represents the signal with a 1024 Hz cut-off, while the lower panel depicts the signal with a 350 Hz cut-off. In the case of the 350 Hz cut-off frequency, the signal remains discernible, whereas with the 1024Hz cut-off frequency, it becomes contaminated with larger noise variations. It should be noted that this signal exhibited an SNR of 7.34 at the Livingston detector.

In this context, it is important to note that while the choice of a 350 Hz threshold for the majority of signals in our training dataset is generally effective, it may not be optimal for all cases. Specifically, the maximum frequency of signals with smaller ${\cal M}$, such as GW190925\_232845, may exceed this threshold. This also holds true for signals originating from two black holes with smaller component masses, exemplified by GW191204\_171526 or GW191216\_213338. However, such signals lie outside the scope of our current training dataset, as elaborated in Section \ref{sec:chirp_mass}. For future searches in other mass ranges, it is likely that this threshold may need adjustment.

Despite these considerations, it appears that the neural network adeptly recognises signals resembling those mentioned, even when the low-pass filter truncates part of the signal in the frequency domain. For instance, the network successfully identifies GW191204\_171526 with the highest achievable ranking statistic.

\begin{figure}[t]
  \centering
  \includegraphics[width=0.95\linewidth]{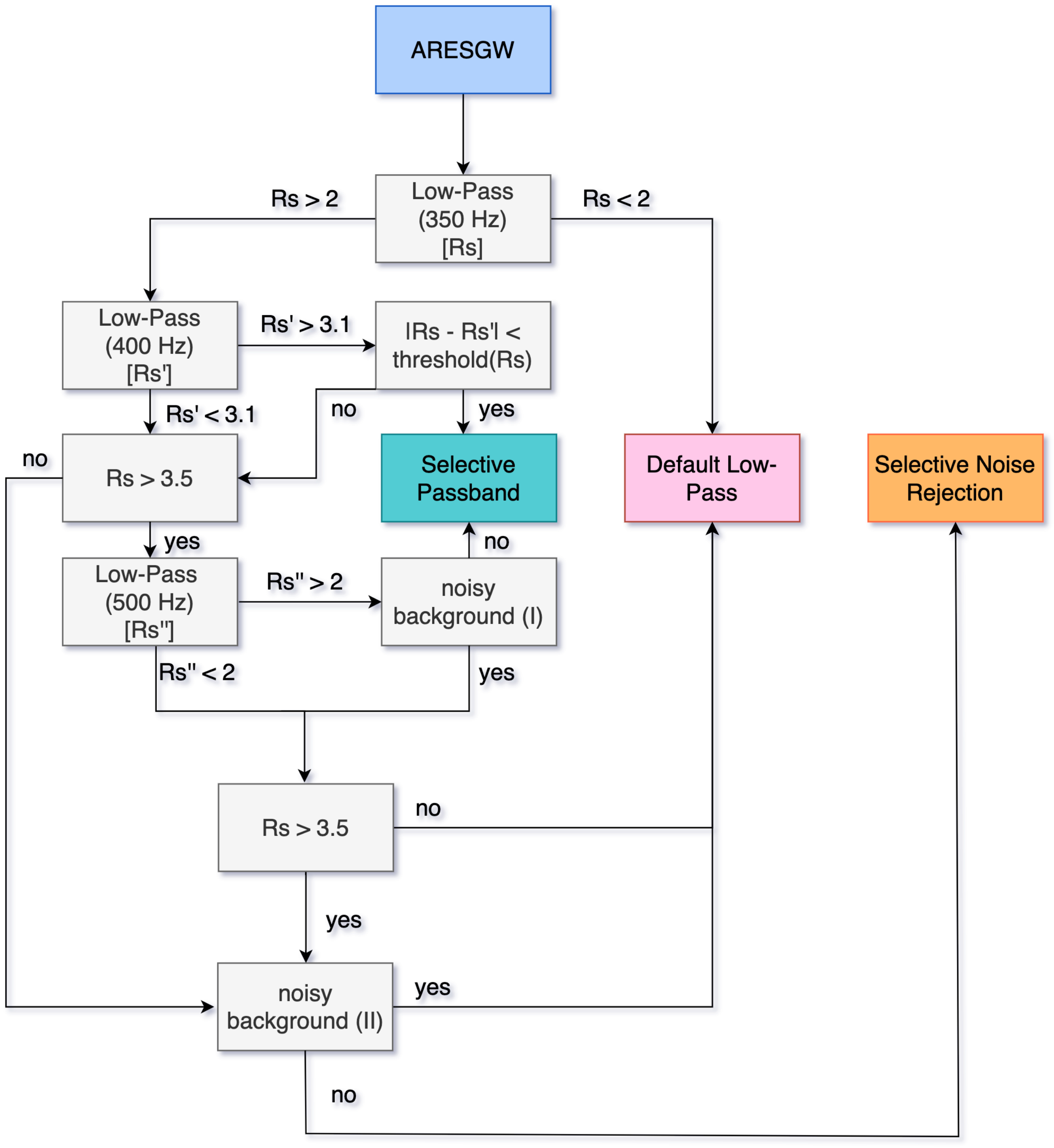}
  \caption{Flowchart of our hierarchical classification of triggers into three different classes (see text for a detailed description)} 
  \label{fig:flowchart}
\end{figure}

\subsubsection{Hierarchical Classification of Triggers} \label{sec:New_filters}

After we obtain triggers with the default low-pass filter of 350Hz, we focus on the subset of triggers with ${\mathcal R_s}>2$ and classify them according to a number of conditions, following a hiearchical scheme. This leads to three classes of triggers: a) {\it Default Low-Pass triggers}, b) {\it Selective Noise Rejection triggers}, and c) {\it Selective Passband triggers}. A trigger needs to satisfy certain conditions to be accepted in the Selective Noise Rejection class and even more strict conditions to be accepted in the Selective Passband class. As we will discuss in Sec. \ref{sec:background} we analyze the background statistics separately for each of these three classes and find that those belonging to the Selective Passband class have significantly lower FAR than the other two classes. 

Fig. \ref{fig:flowchart} describes in detail our hierarchical trigger classification scheme for triggers with ${\mathcal R_s}<16$. In the first step, triggers with ${\mathcal R_s}<2$ are classified as Default Low-Pass, whereas those with ${\mathcal R_s}>2$ are examined again, using a cut-off frequency of 400 Hz. If, with this new cutoff frequency, their ranking statistic becomes ${\mathcal R_s'}>3.1$ and the difference $|{\mathcal R_s} - {\mathcal R_s'}|$ is less than a required threshold (which depends on ${\mathcal R_s}$), then these triggers are classified as Selective Passband. If $|{\mathcal R_s} - {\mathcal R_s'}|$ is larger than the threshold or ${\mathcal R_s'}<3.1$, then if ${\mathcal R_s}>3.5$, the trigger is examined using a third cut-off frequeny of 500 Hz. If its ranking statistic is then ${\mathcal R_s''}>2$, we examine the noise triggers in its vicinity. If no noise triggers with high ranking statistic are found near the candidate trigger (with either of the cut-off frequencies), then it is classified as belonging to the Selective Passband class. Otherwise, or if ${\mathcal R_s''}<2$, then triggers that have ${\mathcal R_s}<3.5$ are classified as Default Low-pass. Triggers with ${\mathcal R_s}>3.5$ or ${\mathcal R_s}<3.5$ but with ${\mathcal R_s'}<3.1$, are reexamined for nearby noise triggers with high ranking statistics, but with a somewhat less strict condition. If they pass this test, they are classified as Selective Noise Rejection, otherwise as Default Low-Pass.  

\begin{figure*}[t]
  \centering
  \includegraphics[width=0.95\linewidth]{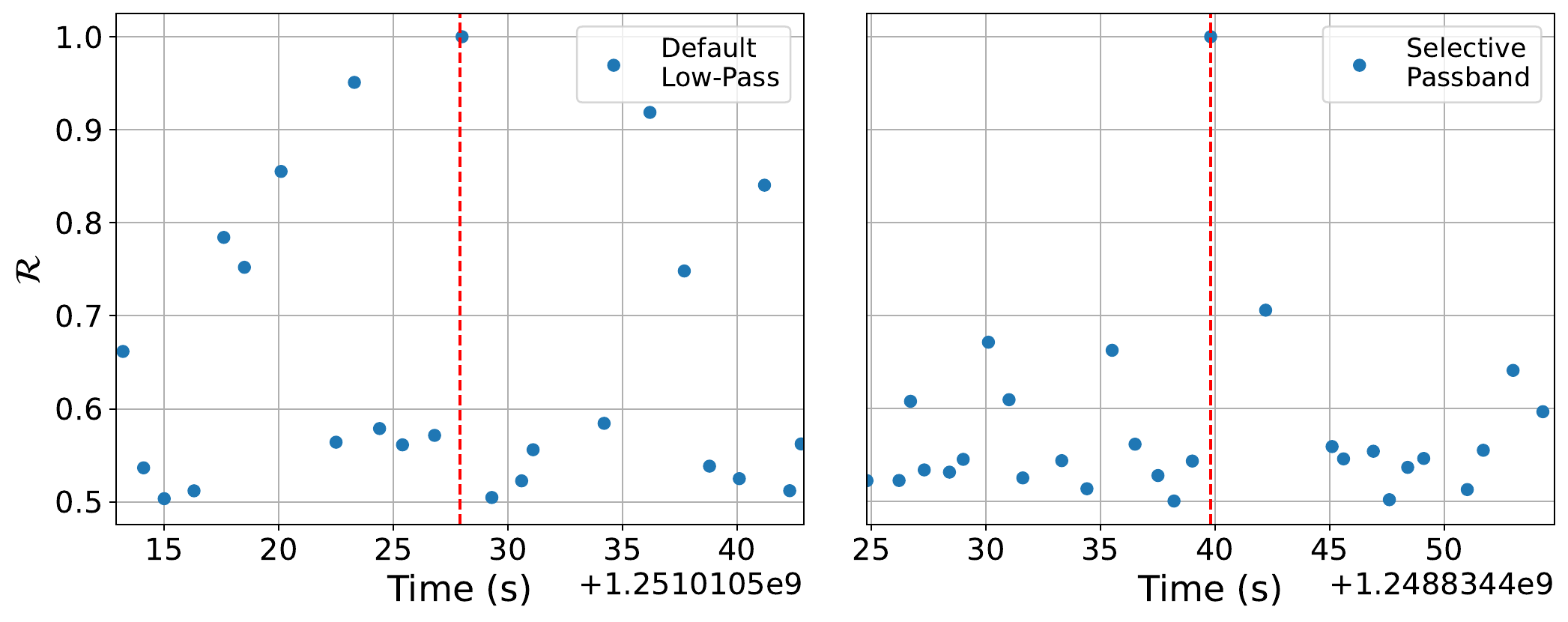}
  \caption{{\it Left panel}: Ranking statistic ${\cal R} >0.5$ of triggers in a time range of $\pm 15$s around the GWTC-3 event  GW190828\_065509. Due to an excessive number of noise triggers with high ranking statistic, the trrigger of the candidate event is classified as Default Low-pass. {\it Right panel}: Same as left panel, but for the GWTC-3 event GW190828\_063405. This trigger is classified as Selective Passband, due to the absence of nearby noise triggers with high ranking statistic and the fact that it still has high ${\cal R}$ values for different low cutoffs.} 
  \label{fig:default_example}
\end{figure*}

Note that, the majority of candidate triggers are reported by AresGW with the maximum possible ranking statistic of ${\mathcal R_s}=16$ with the default cut-off frequency of 350 Hz. For these cases, we follow a simplified classification scheme, instead of the general scheme shown in Fig. \ref{fig:flowchart}. Specifically, we classify such triggers as Selective Passband, if in a range of 30s around the trigger there are fewer than 30 triggers with ${\mathcal R_s}>0.3$ (i.e. ${\mathcal R}>0.5$, which is the threshold value for a trigger to be classified as positive instead of noise). If that is not the case, such triggers are tested for the Selective Noise Rejection class and if they do not qualify, they are classified as Default Low-Pass triggers.

In the left panel of Fig. \ref{fig:default_example} we show an example of the triggers produced by AresGW within $\pm 15$ s of a candidate trigger. When using the default cut-off frequency of 350 Hz, the candidate trigger is reported with ${\mathcal R}=0.9999976$, which corresponds to ${\mathcal R_s}=5.62$. When the cut-off frequency is raised to 400 Hz, the ranking statistic reduces to ${\mathcal R_s'}=4.05$. The difference of $|5.62-4.05|$ is larger than the threshold value we set, which in this case is approximately 0.99. When examined with a 500 Hz cut-off frequency, the ranking statistic becomes ${\mathcal R_s''}=2.79$. Next, the candidate trigger is rejected as belonging to either the Selective Passband or the Selective Noise Rejection classes, because of other noise triggers with high ranking statistic in its vicinity and if classified as Default Low-Pass.  The right panel of Fig. \ref{fig:default_example}, shows a different example, with a low-noise background, that was classified as Selective Passband.

It is important to mention that the various threshold values at the different stages of our hierarchical classification scheme were established through empirical testing and experimentation. 
For the Selective Noise Rejection triggers, we adjust the duration around the trigger based on the value of ${\mathcal R_s}$, with smaller values of ${\mathcal R_s}$ requiring a longer time period to be examined. 
Signals with smaller masses tend to yield lower ${\mathcal R_s}$ due to their morphological characteristics, such as smaller amplitudes and longer durations. It's important to note that the network evaluates one-second intervals at a time. However, the dataset's signal masses correspond to a maximum duration of 20 seconds, approximately reflecting the range of triggers checked on either side of the main trigger point for smaller ${\mathcal R_s}$ values. 
Similar considerations apply to the Selective Passband triggers, where higher sensitivity for events with low $\cal M$ is achieved by raising the low-pass cut-off frequency to 400 Hz or 500 Hz, effectively taking into account a larger part of the signal.

\subsection{ Ranking statistic optimization}
\label{sec:meanrs}

 In this subsection, we introduce an innovative approach based on ${\mathcal R_s}$, which leads to an improved calculation of the astrophysical probability $p_{\rm astro}$ (refer to Sec. \ref{sec:far}). AresGW processes 1-second data segments, utilizing a 4-second window for the power spectral density (PSD) estimation. The analysis window advances in 0.1-second increments, with clustering occurring every 0.3 seconds as described in \cite{AresGW_model}. Consequently, an event can be detected by the network at any point within this 1-second data window.
	
This variability in an event's position within the data frame is crucial because it influences the ranking statistic generated by the network. Several factors contribute to this effect. For instance, a significant portion of an event could be truncated, or variations in the start time of the 1-second window can alter the PSD, thus changing the network's input. As a result, the event may yield different ranking statistics depending on the start time of the data frame. This principle applies equally to background noise triggers, affecting their ${\mathcal R_s}$ values as well. 

\begin{figure}[t]
  \centering
  \includegraphics[width=0.99\linewidth]{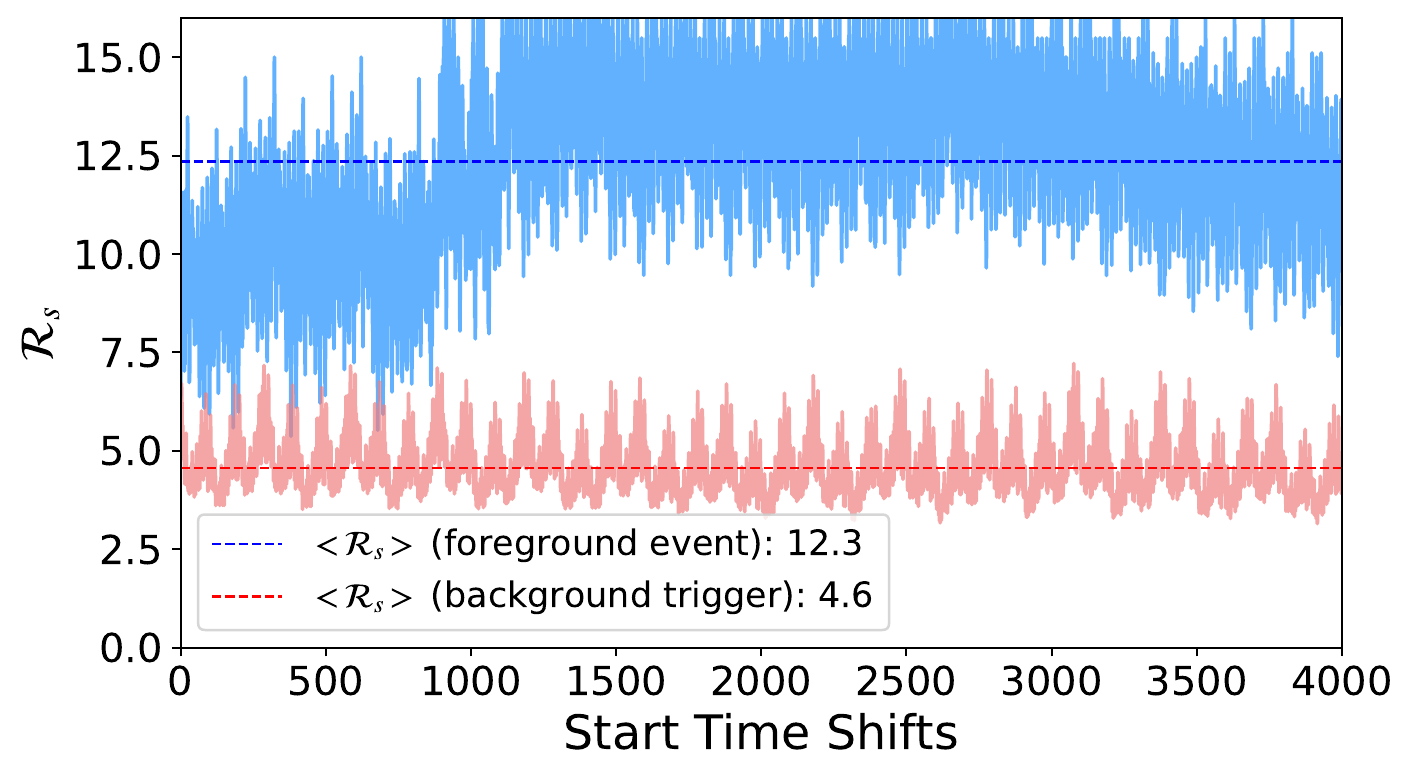}
  \caption{Ranking statistic $\mathcal{R}_s$ (light blue) obtained for different start times close the event GW190413\_052954. In addition, we show $\mathcal{R}_s$ for a representative noise trigger (see text for details). The blue and red dashed lines represent the ensemble-averaged values $\langle
{\mathcal R_s}\rangle$,  respectively.} 
  \label{fig:Event-Background-timeseries}
\end{figure}

To account for the variance in ${\mathcal R_s}$, we define the {\it ensemble-averaged} 
$\langle
{\mathcal R_s}\rangle$ in the following way. The start time shift for each trigger was applied beginning 5 seconds before the coalescence time for events and 5 seconds before the trigger time for background noise triggers.  The shift step for the start time of the dataset was 0.001 seconds. The outcome of this process is illustrated in Fig. \ref{fig:Event-Background-timeseries}.  It is evident that the 
${\mathcal R_s}$
values for the event GW190413\_05295 (depicted in light blue) are substantially higher compared to those of the background trigger (depicted in pink). Additionally, one can observe that the 
${\mathcal R_s}$
 values for the event fall within the range 
[5.37,16.0], highlighting the variability in the network's output for the same event. This fact underscores the unfairness of considering only the 
 ${\mathcal R_s}$
  value from a single run, whereas using the ensemble-averaged $\langle
{\mathcal R_s}\rangle$ effectively mitigates this kind of randomness.

\begin{figure}[t]
  \centering  \includegraphics[width=0.95\linewidth]{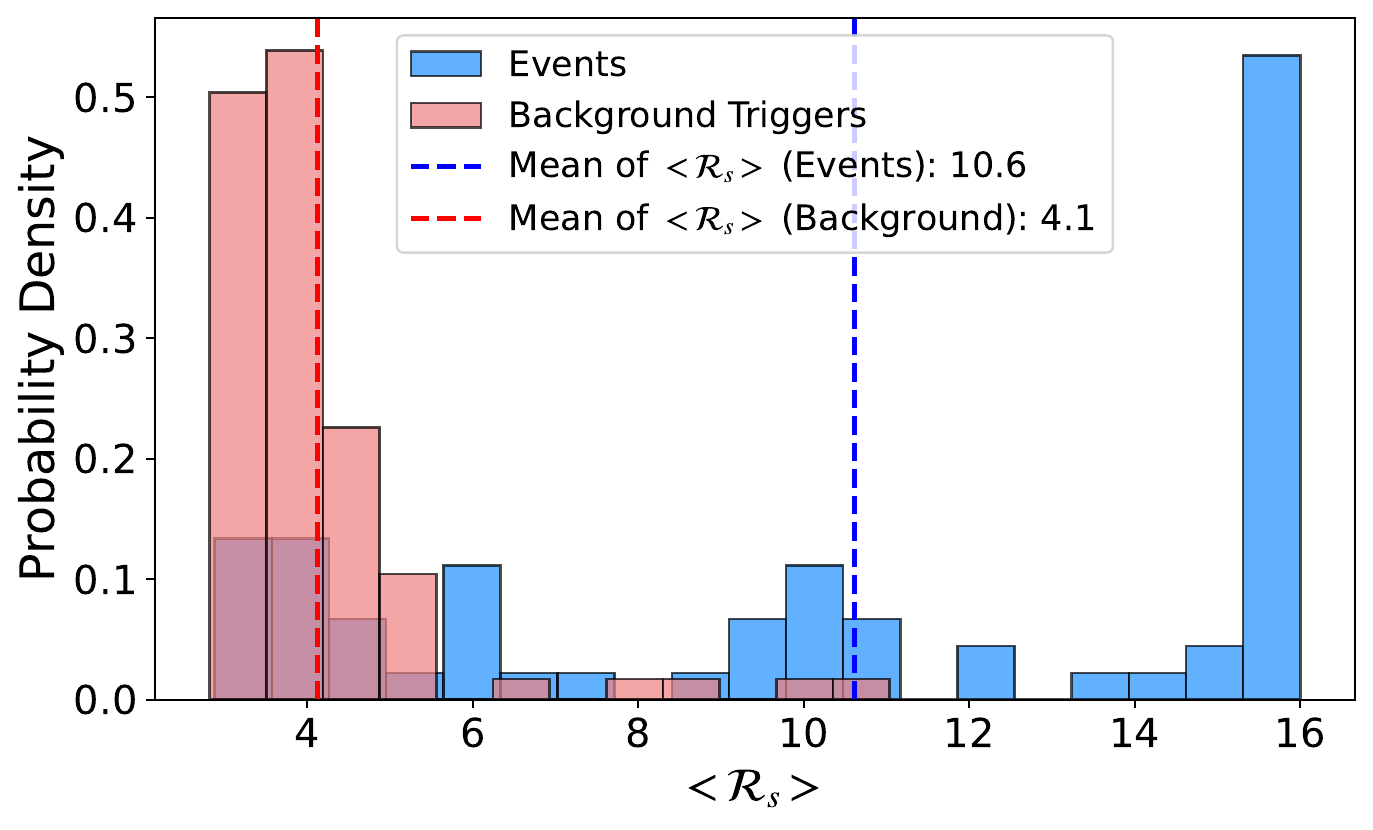}
  \caption{Histograms of the ensemble-averaged ranking statistic $\langle{\mathcal R_s}\rangle$ for the foreground events (light blue) and background triggers (pink). The blue and red dashed vertical lines represent the mean value for each distribution, respectively.} 
  \label{fig:mean_histogram}
\end{figure}

Why is this information important? Our analysis demonstrates that, statistically, the mean of all the possible ranking statistic values for astrophysical events is significantly higher than that for noise triggers. This distinction is displayed in Fig. \ref{fig:mean_histogram}, where the light blue histogram shows the distribution of the $\langle
{\mathcal R_s}\rangle$ values for all events\footnote{Here, we include all previously published events that AresGW detects with  $\langle
{\mathcal R_s}\rangle >3$ and our new candidate events in Table \ref{table:AresGW_New_Detections} with $p_{\rm astro}\geq 0.5$.}, while the pink histogram represents the same for all background noise triggers. The mean of the $\langle
{\mathcal R_s}\rangle$ values for all events is 10.6, compared to 4.1 for all background noise triggers.

The rationale for using the ensemble-averaged $\langle
{\mathcal R_s}\rangle$ values is based on the direct relationship between the ranking statistic and the network's confidence in identifying a trigger as an event. By constructing $\langle
{\mathcal R_s}\rangle$, we effectively aggregate the network's predictions across different start times, yielding a more robust and comprehensive assessment of the signal by considering it in all possible positions, with a chosen timestep, within the data frame. This approach appears to be a better criterion than the initial ${\mathcal R_s}$, likely because it leads to a more accurate representation of the trigger's significance. 
In fact, this method appears to be particularly critical for threshold events, as demonstrated by the data: with the initial ${\mathcal R_s}$, AresGW would have detected 38 events within its effective training range. In contrast, using the ensemble-aveaged $\langle
{\mathcal R_s}\rangle$ values, it detects 42 events. The additional four events represent new near-threshold detections.

Notice that the quasi-periodicity seen in Fig. \ref{fig:Event-Background-timeseries} in the variations of ${\mathcal R_s}$ for both candidate events and noise triggers, is probably associated with the morphology of a real signal or a glitch in the time-frequency space. This could be exploited for the classification of triggers (even in an online search) and is the subject of further investigation. This observation is relevant not only for AresGW, but also for other machine-learning codes that generate a ranking statistic in discrete time steps.

In the remainder of the text, we explicitly distinguish between $R_s$ and $\langle{\mathcal R_s}\rangle$.

\subsection{Glitch removal with Gravity Spy} \label{sec:Glitch_removal}

Gravity Spy \cite{Gravity_Spy,2021CQGra..38s5016S,2023CQGra..40f5004G,2024EPJP..139..100Z} employs a convolutional neural network (CNN), a specialized deep learning algorithm designed for image recognition tasks, to classify glitches in LIGO detector data based on their time-frequency morphology. We utilized the dataset containing glitches detected by Gravity Spy in the O3a and O3b data, which can be found in \cite{Gravity_Spy_Dataset}, to minimize their influence on the background FAR calculated using AresGW. In practice, we excluded all triggers of the Omicron \cite{2020SoftX..1200620R} dataset release for the O3a+O3b periods, since 97\% of these triggers have been classified by Gravity Spy as belonging to one of the known glitch classes with ranking statistic $> 0.5$. We utilized this approach both in creating a 10-year nearly glitch-free background, essential for determining the FAR, and in rejecting triggers detected by AresGW that turned out to be known glitches.

For trigger rejection, we focused on three main parameters: the start time and duration of the glitch, and the trigger time of AresGW. 
We examined whether a trigger was within a time interval equal to the duration of a known glitch $\pm 1$ s, in which case it was removed from the data. The buffer of $\pm 1$ s is set by the 1 s duration of the time segments examined by AresGW.

\subsection{Parameter estimation} \label{sec:PE}

For the purpose of parameter estimation of new candidate gravitational wave signals, we used the Bayesian inference library for gravitational-wave astronomy, {\tt Bilby} \cite{bilby}, with uniform distributions for priors. In particular, the priors for the detector-frame component masses $m_1,m_2$ correspond to a range of $[4 M_{\odot}, 150 M_{\odot}]$, and since we are sampling for chirp masses $\cal M$ and mass ratios $q=m_2/m_1\leq 1$, we set the component masses as constraints on the latter. In some cases, where the posterior distribution tails did not seem to drop to zero, we experimented with expanding the prior component mass range and/or the luminosity distance. The relation between redshift and luminosity distance is based on Planck15 cosmology \cite{planck}.

Since multidimensionality is one of the main concerns when it comes to Bayesian inference applications in gravitational-wave astronomy, we chose {\tt dynesty} \cite{dynesty} as a sampler. This method utilizes (Dynamic) Nested Sampling, combining the advantages of Markov Chain Monte Carlo (MCMC) algorithms, which concentrate on posterior estimation, with the ability of nested sampling to estimate evidences and sample from complex, multi-modal distributions. This is achieved by adaptively allocating samples according to the posterior structure.

Each trigger was initially inspected using the IMRPhenomPv2 \cite{2014PhRvL.113o1101H} waveform approximant, but our final results are produced with the IMRPhenomXPHM \cite{IMRPhenomXPHM_ref} model, which was also the waveform model for the injections on which AresGW was trained. Time and distance marginalization were also applied to each run (phase marginalization is formally invalid for precessing approximants). Finally, prior to conducting post-inference analysis on the candidate triggers, we verified the robustness of the obtained posterior distributions, by examining their resemblance to the Gaussian normal distribution (as assumed by Bilby's Gaussian likelihood). 

\begin{figure}[t]
  \centering
  \includegraphics[width=1\linewidth]{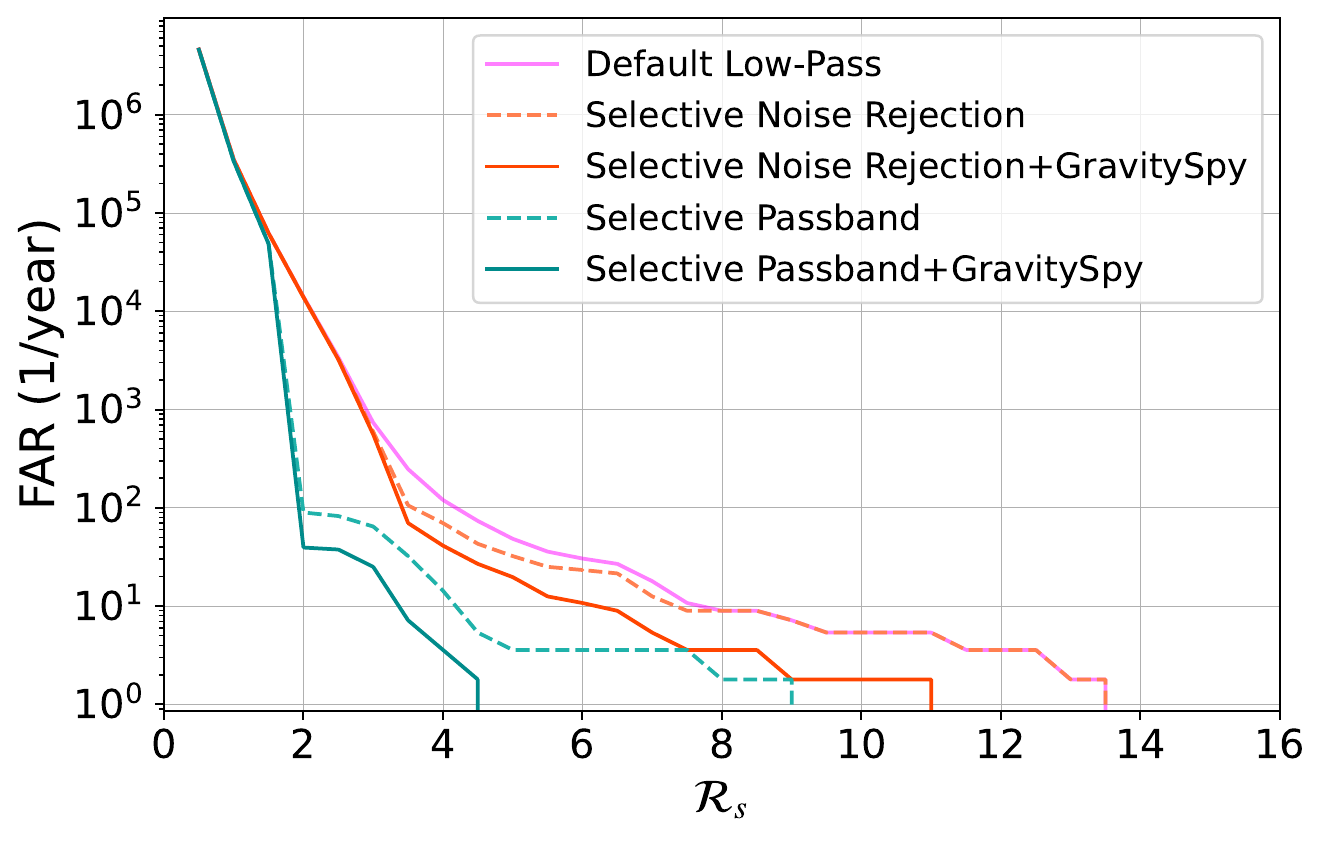}
    \caption{False alarm rate (without time shifts) vs. ranking statistic for the 6.7-month O3 dataset. A reduction of FAR when applying our classification scheme and removal of known glitches, is observed. Notice that the classification into Selective Noise Rejection and Selective Passband classes was only applied for $R_s>3.5$ and $R_s>2$, resplectively (for smaller $R_s$, the lines are simply joined to the results of the Default Low-Pass class). }
    \label{fig:False Alarm Rate vs Ranking Statistic}
\end{figure}

\subsection{Consistency tests of candidate events}

Each promising trigger we analyzed was subjected to several consistency tests, mainly utilizing routines from {\tt PyCBC} \cite{usman2016pycbc}. One such test requires that the light-travel time between the two LIGO detectors is less than 10 ms, with an additional 5 ms window included to account for uncertainty in the estimated coalescence time at each detector, as discussed in \cite{time_delay}. A second important test is the waveform consistency test, which is based on the statistical veto $\chi^{2}$, and tests for non-Gaussianity of the residual, when a signal template is subtracted from the data. Specifically, we use the version presented in \cite{allen}, with the number of bins chosen as
\begin{equation}
n = [0.4(f_{\text{peak}} / \text{Hz})^{2/3}].
\label{eq:nbins}
\end{equation}
In Eq. (\ref{eq:nbins}), $f_{\text{peak}}$ is the frequency observed at the point of maximum power output of the observed signal, corresponding to the time of merger \cite{empirical}. These tests allow us to effectively reject a number of triggers that could be due to previously unidentified glitches. 

Having performed the $\chi^2$ test, we also used the \textit{ reweighted} SNR \cite{usman2016pycbc} as our main detection statistic. First, we calculate the SNR using matched filtering with the median reconstructed waveform, as obtained with parameter estimation using Bilby. This is done to identify any non-noise-like characteristics that may persist. If such features are found, it indicates that the model waveform $h$ does not accurately match the non-Gaussian feature present in the data. Consequently, the SNR, $\rho$, is adjusted downwards. This can be expressed as \cite{usman2016pycbc}
\begin{equation}
\hat{\rho} = \rho \times 
\begin{cases} 
1 & \text{if } \chi^2_{r} \leq \nu, \\
\left[\frac{1}{2} + \frac{1}{2}(\chi^2/\nu)^3\right]^{-1/6} & \text{if } \chi^2_{r} > \nu,
\end{cases}
\end{equation}
where $\nu=2n-2$ are the degrees of freedom. For real signals that have been accurately modeled, this is close to the true SNR, while for glitches it is much smaller, as a higher $\chi^{2}$ down-weights it. For the network of the LIGO Livingston and LIGO Hanford detectors
\begin{equation}
    \hat \rho = \sqrt{\hat \rho_{\rm L}^2 + \hat \rho_{\rm H}^2},
\end{equation}
is the reweighted network SNR, which we are going to evaluate for both known GW events and new candidate GW events.

\begin{figure}[t!]
  \centering
  \includegraphics[width=1.0\linewidth]{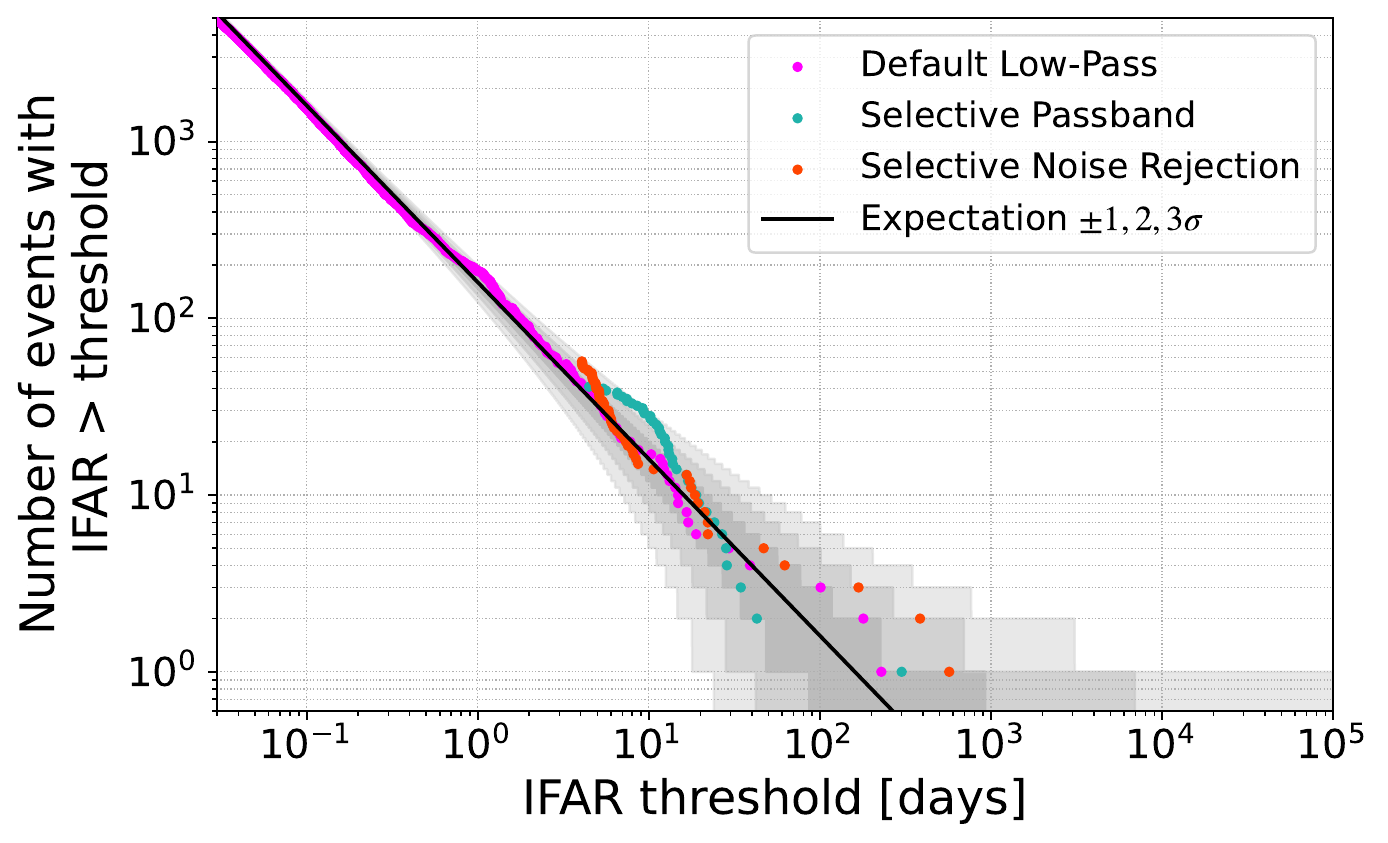}
  \caption{Cumulative distribution of the inverse false alarm rate (without time shifts) of noise triggers using the O3 data. All three classes of our hierarchical classification yield distributions within 3$\sigma$ of the expectation.} 
  \label{fig:IFAR}
\end{figure}

\section{Background Model and Statistics}
\label{sec:background}

\subsection{False alarm rate from O3 data }

After running AresGW on the entire 6.7-month data-quality dataset from the O3 observational period we then remove all known events previously published in the 
GWTC catalogs \cite{GWTC1, GWTC2, LIGO2022GWTC21, GWTC3},  the OGC catalogs \cite{OGC-1,OGC-2,OGC-3,OGC-4} and the IAS catalogs \cite{IAS_O3a,IAS_O3b,IAS_higher_harmonics}, as well as the eight new events we identify here with $p_{\rm astro}>0.5$. From the remaining triggers (and after rescaling to a duration of one year) we obtain an estimate for the FAR distribution (without time shifts) at a ranking statistic larger than ${\cal R}_s$, as shown in Fig. \ref{fig:False Alarm Rate vs Ranking Statistic} (the initial ranking statistic was use here). Applying the Selective Noise Rejection filter, we find a small reduction in FAR. A larger reduction is obtained, when the Selective Passband filter is applied. Furthermore, the removal of known glitches using Gravity Spy is very effective in reducing the false alarm rate. Thus, with the Selective Passband filter, combined with glitch removal, a reduction between one and two orders of magnitude in FAR is obtained.

 Fig. \ref{fig:IFAR} shows the inverse FAR (IFAR) cumulative distribution \cite{2021CQGra..38i5004A} (without time shifts), when the previously published and new GW candidate events have been removed from the O3 data.
For all three classes, the distribution of noise triggers is within 3$\sigma$ of the expected distribution.

\subsection{False alarm rate from time shifts} \label{sec:far}

\begin{figure}[t]
  \centering
  \includegraphics[width=0.95\linewidth]{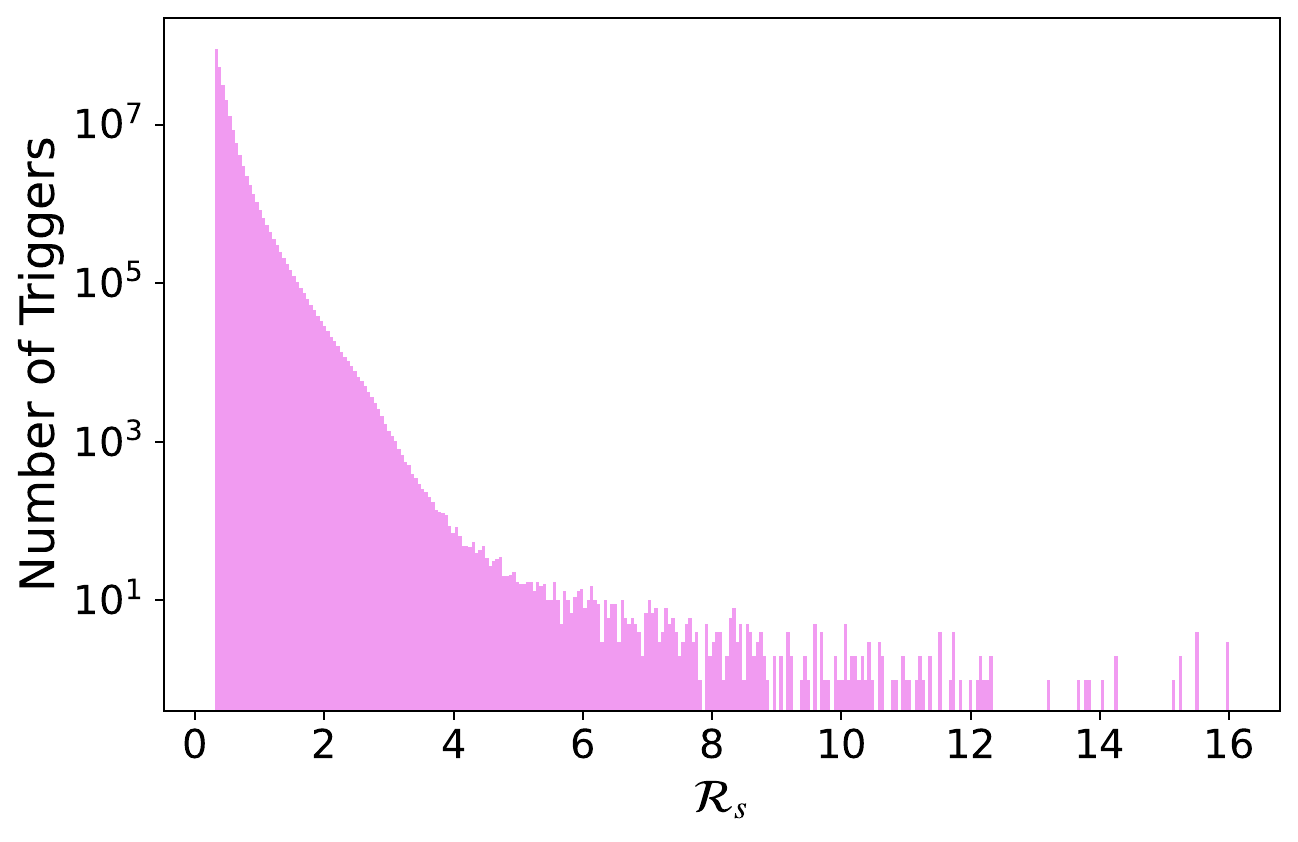}
  \caption{Histogram of the Ranking Statistic for background noise over a 10-year period, based on time shifts of data-quality noise from O3a (see Sec. \ref{sec:Training_Data}) without the glitches identified by Gravity Spy.}
  \label{fig:Background_hist}
\end{figure}

To better estimate the false alarm rate, we conducted an investigation on a time-shifted 10-year background datased, from which known GW events, as well as glitches, identified in the Gravity Spy datasets, were removed (see Section \ref{sec:Glitch_removal}). We also removed two additional obvious glitches of known class that were detected by AresGW, but were not included in the Gravity Spy database for O3. However, it is worth mentioning here, that despite removing those glitches, the dataset used for creating the background still contained some glitches. This fact, combined with the exclusive use of noise from O3a (even for O3b events), likely contributes to a somewhat higher (conservative) estimated ${\mathrm{F A R}}$ when using this method. The cumulative false alarm rate for all triggers with ranking statistic larger than a certain value  ${\mathcal R_s}$ will be denote as ${\mathrm{F A R}}(>{\mathcal R_s})$. This will be different for each of the three trigger classes introduced in Section \ref{sec:New_filters}. 
Fig. \ref{fig:Background_hist} shows a histogram of triggers generated with AresGW evaluating the 10-year background dataset, as a function of the ranking statistic ${\mathcal R_s}$ and using a bin size of 0.053 (the initial ranking statistic was use here). All triggers shown in Fig. \ref{fig:Background_hist} were treated as Default Low-Pass and were {\it not} classified in one of the other two classes, as shown in Fig. \ref{fig:flowchart}.

Next, we applied our hierarchical classification scheme presented in 
Section \ref{sec:New_filters} to all triggers above a certain ranking statistic threshold. More specifically, for the trigger classes Default Low-Pass and Selective Noise Rejection we used a threshold of ${\mathcal R_s} = 3.5$, whereas for the Selective Passband class we used a lower threshold of ${\mathcal R_s} = 2$ due to its significantly lower FAR. 

\begin{figure}[t]
  \centering
  \includegraphics[width=1\linewidth]{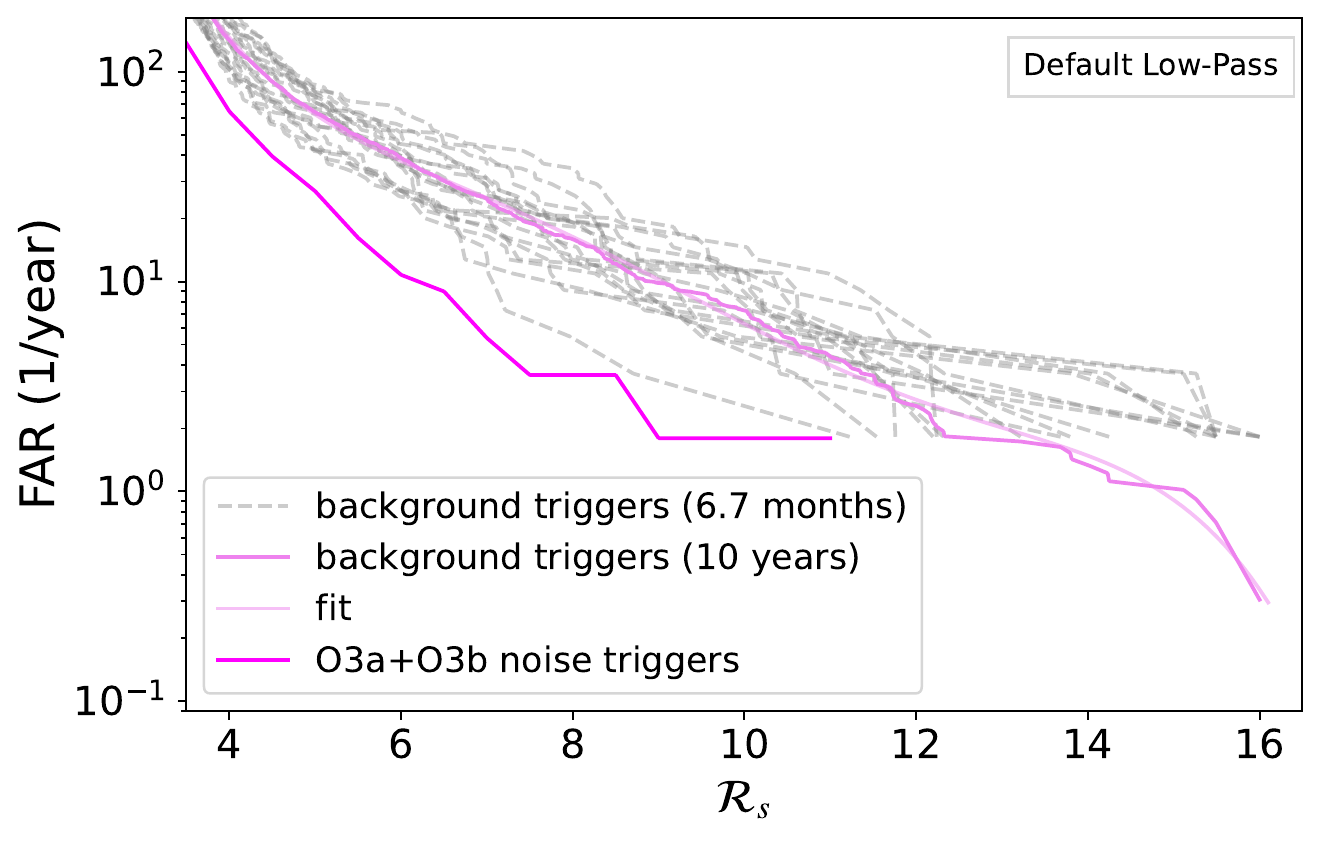}
  \caption{False alarm rate vs. ranking statistic. Each dashed gray line represents a unique background noise period lasting 6.7 months. In contrast, the solid light-pink line represents the analytical relation of the False Alarm Rate (FAR) concerning the ranking statistic. The combination of these eighteen dashed grey instances results in the complete 10-year background, illustrated by the darker pink line, while the fuchsia line represents the outcome of AresGW applied to the authentic O3 data, with a duration corresponding to approximately 6.7 months for the two LIGO detectors.}
  \label{fig:initial_cumulative plot}
\end{figure}

Fig. \ref{fig:initial_cumulative plot} shows the cumulative false alarm rate for triggers classified as Default Low-Pass.
Each dashed gray line represents 18 different realizations of a time-shifted background noise with a duration of 6.7 months. Adding all these different noise realizations to a single 10-year background, we obtain the cumulative FAR shown as the link line. With a lighter color, we also show the following analytic fit 
\begin{equation}
\begin{split}
\log_{10}(\text{FAR}
) = & -1.16 \times 10^{-4} {\mathcal R_s}^5 + 5.63 \times 10^{-3} {\mathcal R_s}^4 \\
& - 1.05 \times 10^{-1} {\mathcal R_s}^3 
+ 9.51  \times 10^{-1} {\mathcal R_s}^2  \\
&- 4.33 {\mathcal R_s} 
+ 9.69 .
\end{split}
\label{eq:Far_Initial}
\end{equation}
For ${\mathcal R_s=3.5}$ the FAR is 280/year, whereas for ${\mathcal R_s=16}$ the FAR drops to 0.34/year.

The darker pink line in Fig. \ref{fig:initial_cumulative plot} represents the outcome of AresGW applied to the actual O3 data when both LIGO detectors were observing (and with data-quality flag), with a real duration corresponding to approximately 6.7 months. Notice that the FAR for the actual O3 data is lower than any of the different 6.7 month noise realizations produced with time shifts of the 81-day O3a noise only. This could be explained, considering that the actual O3 noise encompasses both the O3a and O3b periods. In the latter period, the noise had significantly lower levels than in the former, due to some enhancements that were implemented to the LIGO detectors between O3a and O3b.  
A more suitable approach would involve splitting the analysis into two distinct periods: one for O3a and another for O3b, but due to time restrictions, we chose to use our O3a-based 10-year time-shifted background for producing a conservative FAR estimate also for triggers in the O3b period.

\begin{figure}[t]
  \centering
  \includegraphics[width=1\linewidth]{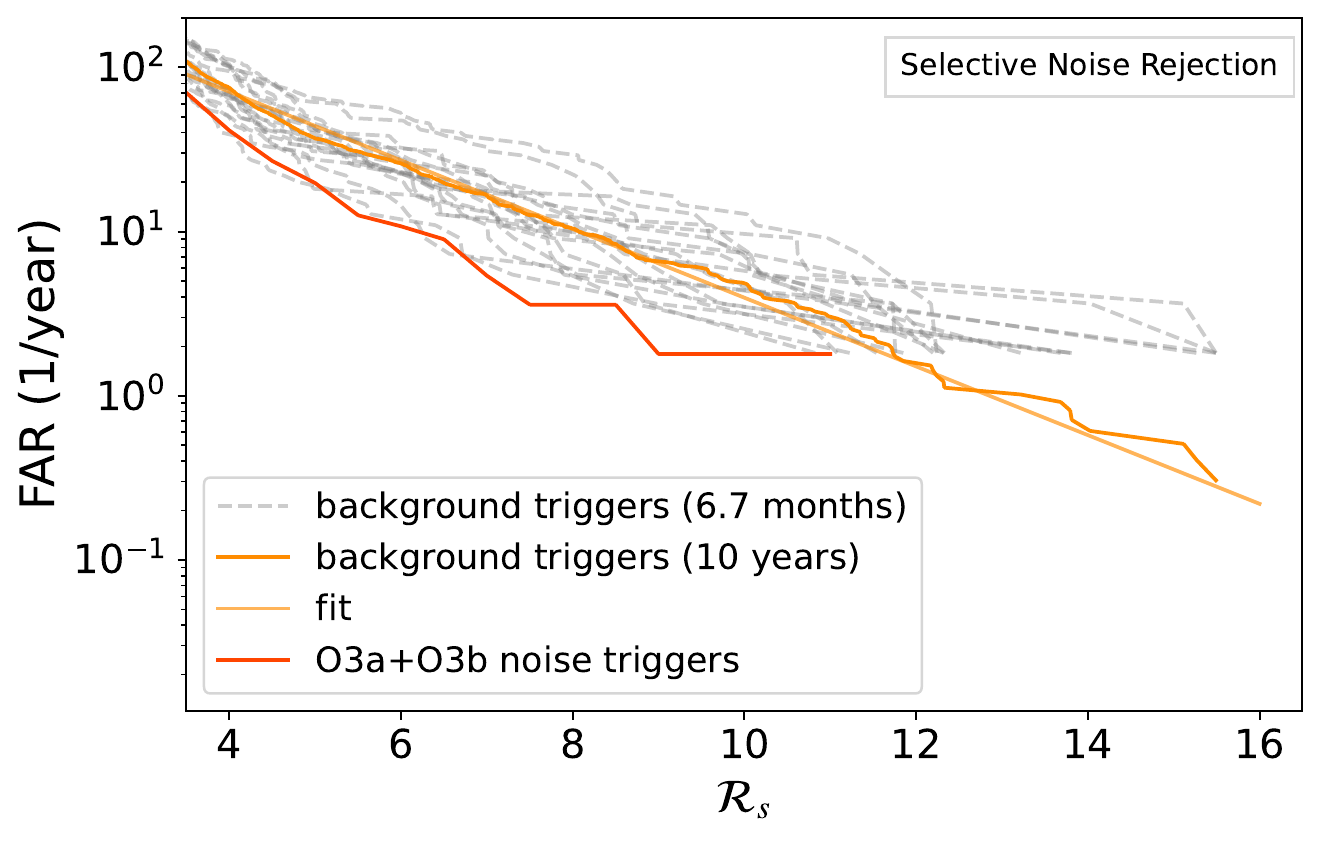}
  \caption{ Same as Fig. \ref{fig:initial_cumulative plot}, but for the Selective Noise Rejection class of triggers.}
  \label{fig:low_background_cumulative plot}
\end{figure}

Similarly, Fig. \ref{fig:low_background_cumulative plot} shows  the cumulative false alarm rate for triggers classified as
Selective Noise Rejection, including its analytic fit
\begin{equation}
\begin{split}
    \log_{10}(\text{FAR}) =  &-0.21 {\mathcal R_s}  
    +  2.69.
\end{split}
\label{eq:Far_Selective Noise Rejection}
\end{equation}
For this class, the FAR is 90/year for ${\mathcal R_s=3.5}$ and 0.22/year for ${\mathcal R_s=16}$.

\begin{figure}[t]
  \centering
  \includegraphics[width=1\linewidth]{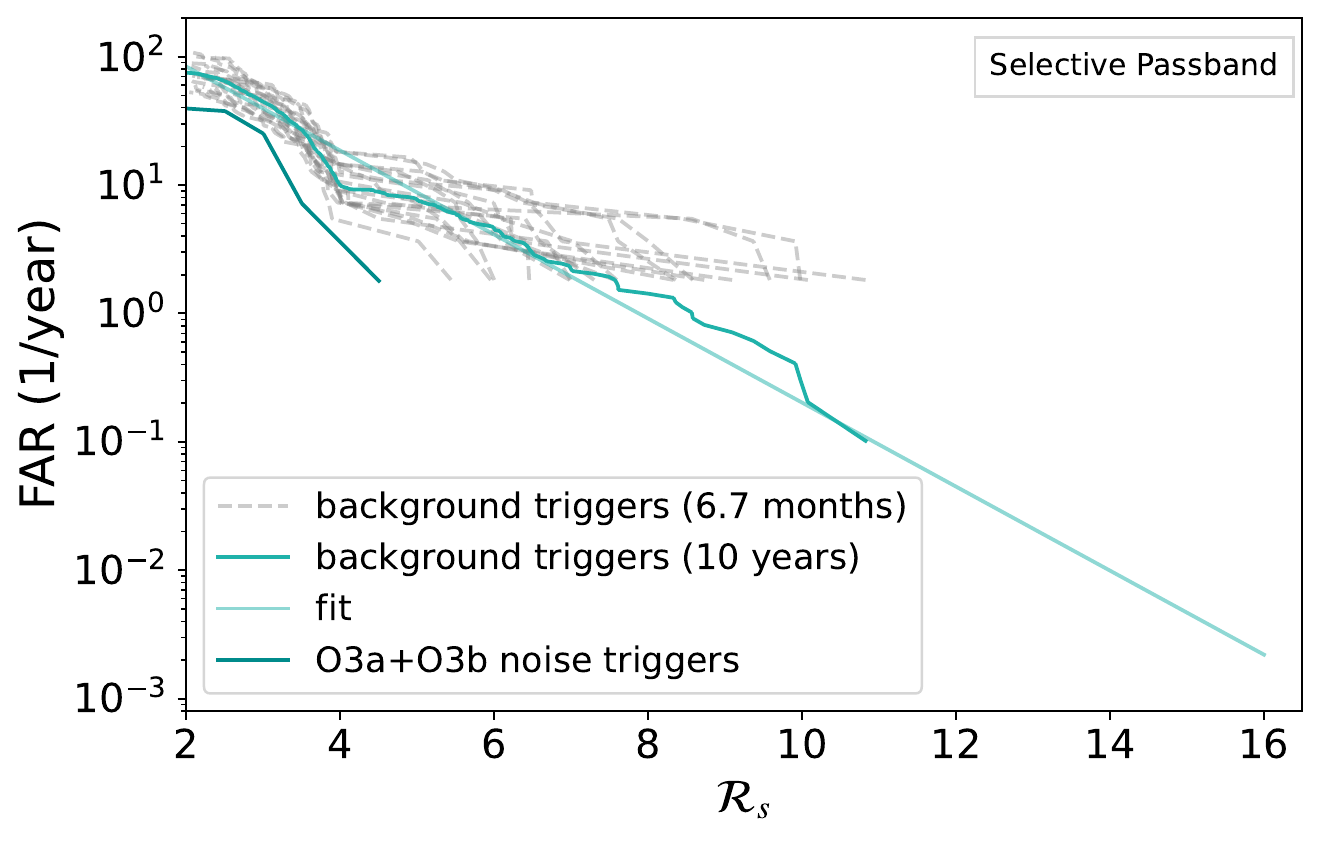}
  \caption{Same as Fig. \ref{fig:initial_cumulative plot}, but for the Selective Passband  class. }
  \label{fig:Selective Passband _cumulative plot}
\end{figure}

Finally, the cumulative false alarm rate for triggers classified as
Selective Passband is shown in \ref{fig:Selective Passband _cumulative plot}, including its analytic fit
\begin{equation}
\begin{split}
    \log_{10}(\text{FAR}) =  &-0.33 {\mathcal R_s} 
    +  2.58.
\end{split}
\label{eq:Far_Selective Passband }
\end{equation}
At ${\mathcal R_s} = 3.5$, the FAR is 27/year, which motivated us to examine this class at even smaller ranking statistics values, reaching ${\mathcal R_s} = 2.0$, where the FAR was 84/year. On the other hand, at ${\mathcal R_s} = 16$ the FAR was just 0.0022/year, which means that this is the most sensitive class of triggers.

Investigating the gray lines in Figs. \ref{fig:initial_cumulative plot}, \ref{fig:low_background_cumulative plot} and \ref{fig:Selective Passband _cumulative plot}, it is evident that the variance of false alarms as a function of ${\mathcal R_s}$ for different noise realizations widens notably for the 6.7-month backgrounds as one examines higher values of ${\mathcal R_s}$. On the other hand, the cumulative FAR for the 10-year period shows a more consistent, almost straight-line dependence on ${\mathcal R_s}$, implying a smaller variance. This phenomenon stems from the increased likelihood of additional noise interference, which can distort the outcomes. For example, consider a scenario where, over a span of 100 years, 10 false alarms are recorded at ${\mathcal R_s}$ = 10.0. These occurrences may cluster within the initial ten years, or they might distribute evenly, one every ten years. In the former case, if the analysis were limited to those initial 10 years, false alarms at ${\mathcal R_s}$ = 10.0 would appear as 1 per year, while the actual rate would be 0.1 per year. Therefore, longer-duration backgrounds yield more reliable results. 

Notice that for the calculation of the false alarm rate in this subsection, the initial ranking statistic was used.

\subsection{Background model}

Consider a data set of $N_b$ background events (either signals or background noise) that are ordered in terms of a ranking statistic $x$ as $x_1<x_2<\cdots<x_N$, where $x_1 = x_{\rm min}$ and $x_N = x_{\rm max}$.  If we assume that the  background events are samples from an inhomogeneous Poisson process and that $\theta$ are parameters that affect the shape of the distribution, then the differential rate for the background events can be defined as a function  $b(x, \theta)$ as
\begin{equation}
\frac{d N_b}{d x}=b(x, \theta).
\end{equation}
The cumulative rate of background events below a ranking statistic $x$ is 
\begin{equation}
B(x, \theta) \equiv \int_{x_{\rm min}}^x d x' b(x', \theta).
\label{eq:B}
\end{equation}
If $R_b$ is the number of background  events
\begin{equation}
R_b = B(x_{\rm max}, \theta),
\end{equation}
we can normalize 
\begin{equation}
{\hat B}(x,\theta) = B(x,\theta)/R_b.
\end{equation}
For background triggers, the distribution can be modeled as 
\begin{equation}
\hat{B}(x)=\left(\frac{1+\operatorname{erf}\left(\frac{x}{\sqrt{2}}\right)}{2}\right)^N,
\end{equation}
where $N$ is a parameter, or
\begin{equation}
\hat{B}(x)=\frac{\left(1+\operatorname{erf}\left(\frac{x}{\sqrt{2}}\right)\right)^N-\left(1+\operatorname{erf}\left(\frac{x_{\min }}{\sqrt{2}}\right)\right)^N}{2^N-\left(1+\operatorname{erf}\left(\frac{x_{\min }}{\sqrt{2}}\right)\right)^N},
\label{eq:backgroundmodel}
\end{equation}
when the data are available above a threshold, $x > x_{\rm min}$ \cite{PhysRevD.91.023005}. This expression was derived when the ranking statistic $x$ is assumed to be the matched-filtering SNR. Since AresGW uses a different ranking statistic, we adopt the above expression as our background model, assuming $x={\cal R}_s / c$ , where we  introduce the scaling parameter $c$. In addition, we found that assuming $R_b -1$ or $R_b -2$, instead of $R_b$ yields bettter fits, as it offsets the larger variance at high ${\cal R}_s$.

\begin{figure}[t]
  \centering
  \includegraphics[width=1.0\linewidth]{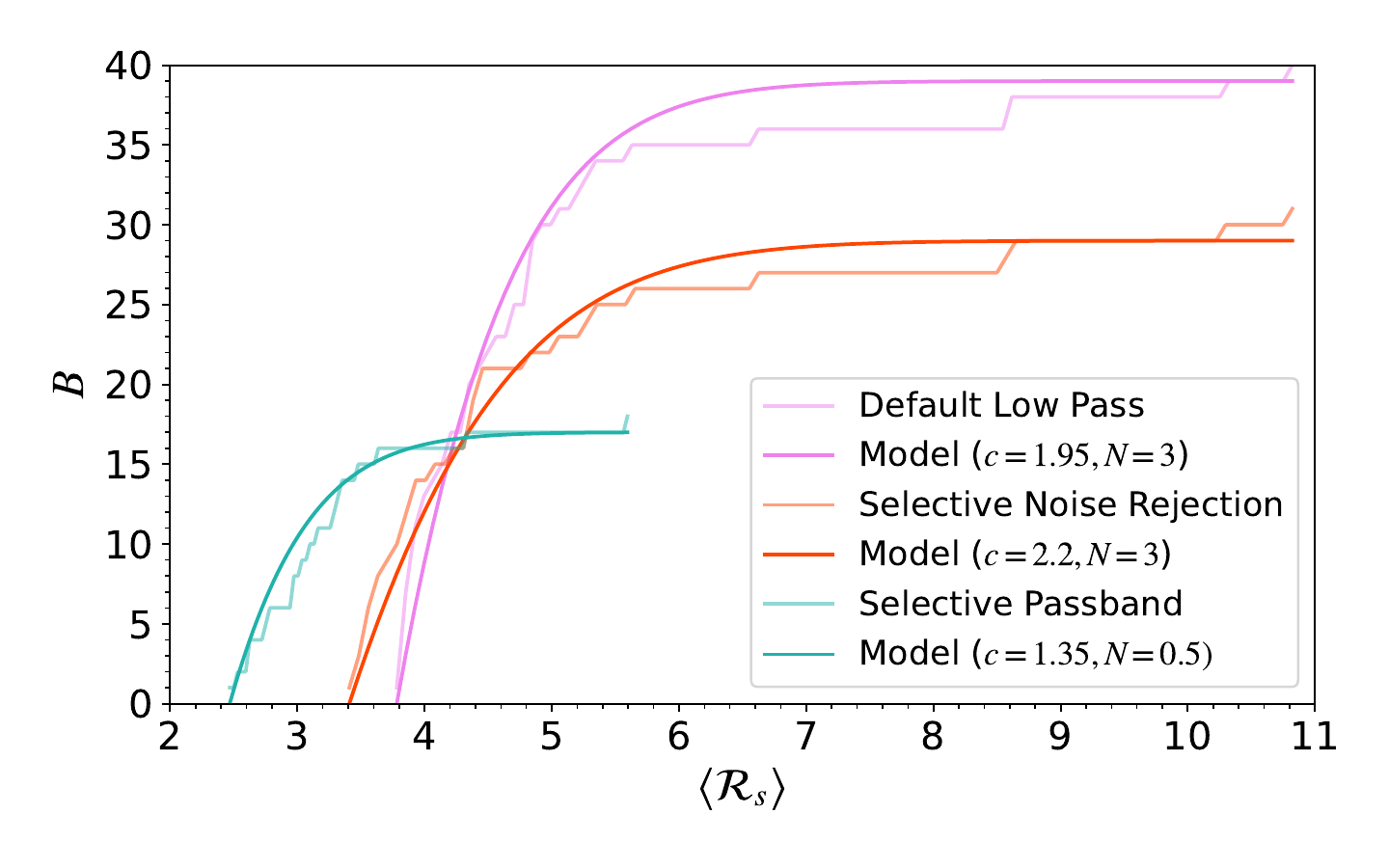}
  \caption{ Background distribution of the O3 data, for the different classes of our classification scheme. In all three cases, there is good agreement with our background model of Eq. (\ref{eq:backgroundmodel}). } 
  \label{fig:CDF_collective_noise}
\end{figure}

In Fig. \ref{fig:CDF_collective_noise} we show the background distribution of the O3 data, for the different classes of our classification scheme. In all three cases, our background model of Eq. (\ref{eq:backgroundmodel}) fits the data well. We will used this analytics model for the background in the calculation of the astrophysical probability in Sec. \ref{seq:pastro}.

    \section{Foreground Model and Statistics}

\subsection{Cumulative IFAR of all triggers}

\begin{figure}[h]
  \centering
  \includegraphics[width=0.95\linewidth]{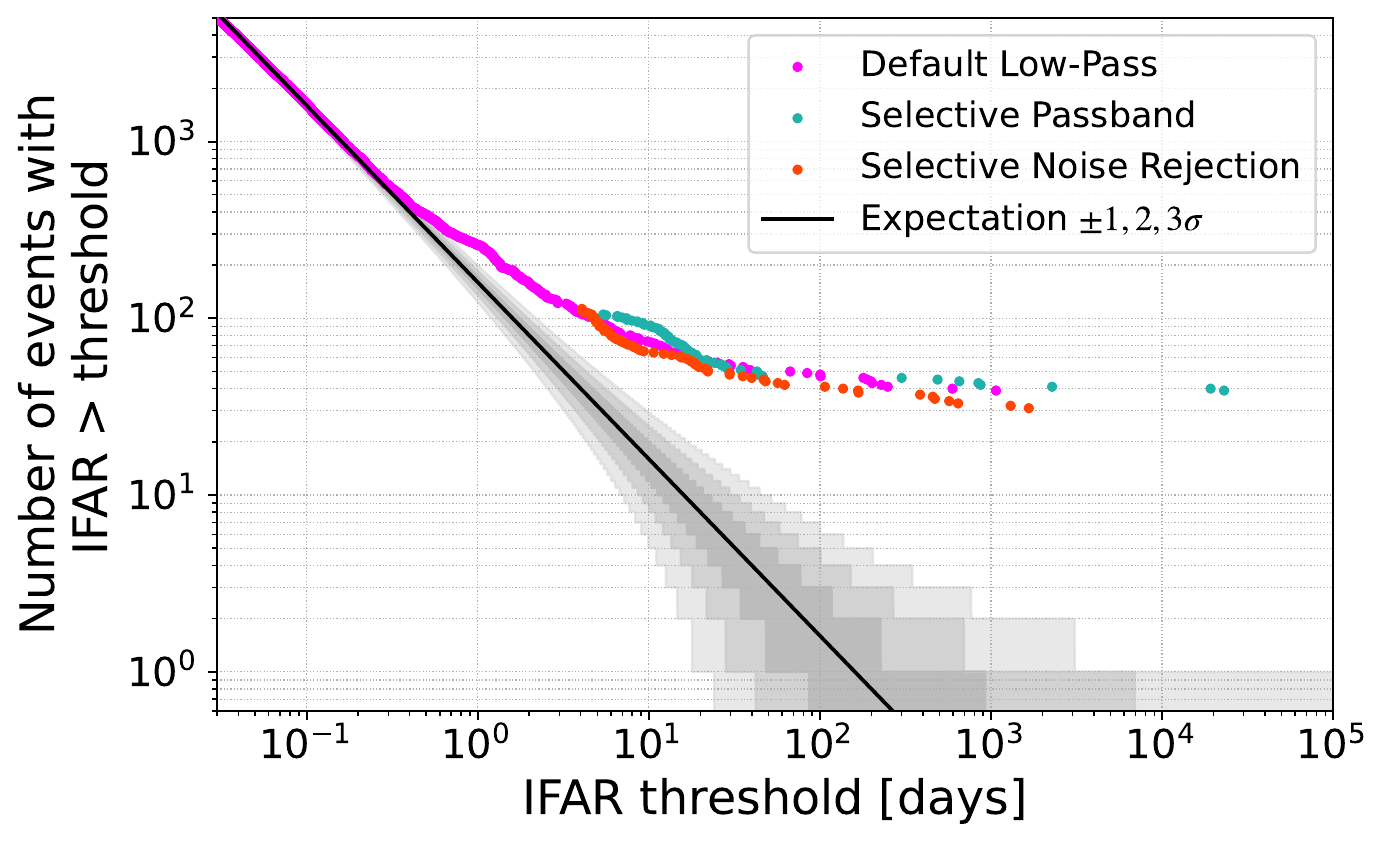}
 \caption{Cumulative distribution of the inverse false alarm rate of all triggers (including known events) using the O3 data.}  
  \label{fig:IFAR_events}
\end{figure}

Similarly to Fig. \ref{fig:IFAR}, the cumulative IFAR distribution for all triggers (without removing known events) is shown in Fig. \ref{fig:IFAR_events}. For all three classes, the distribution is greater than 3$\sigma$ of the expected noise distribution for ${\rm IFAR}=1/{\rm d}$.

\begin{figure*}[t]
  \centering
  \includegraphics[width=0.47\linewidth]{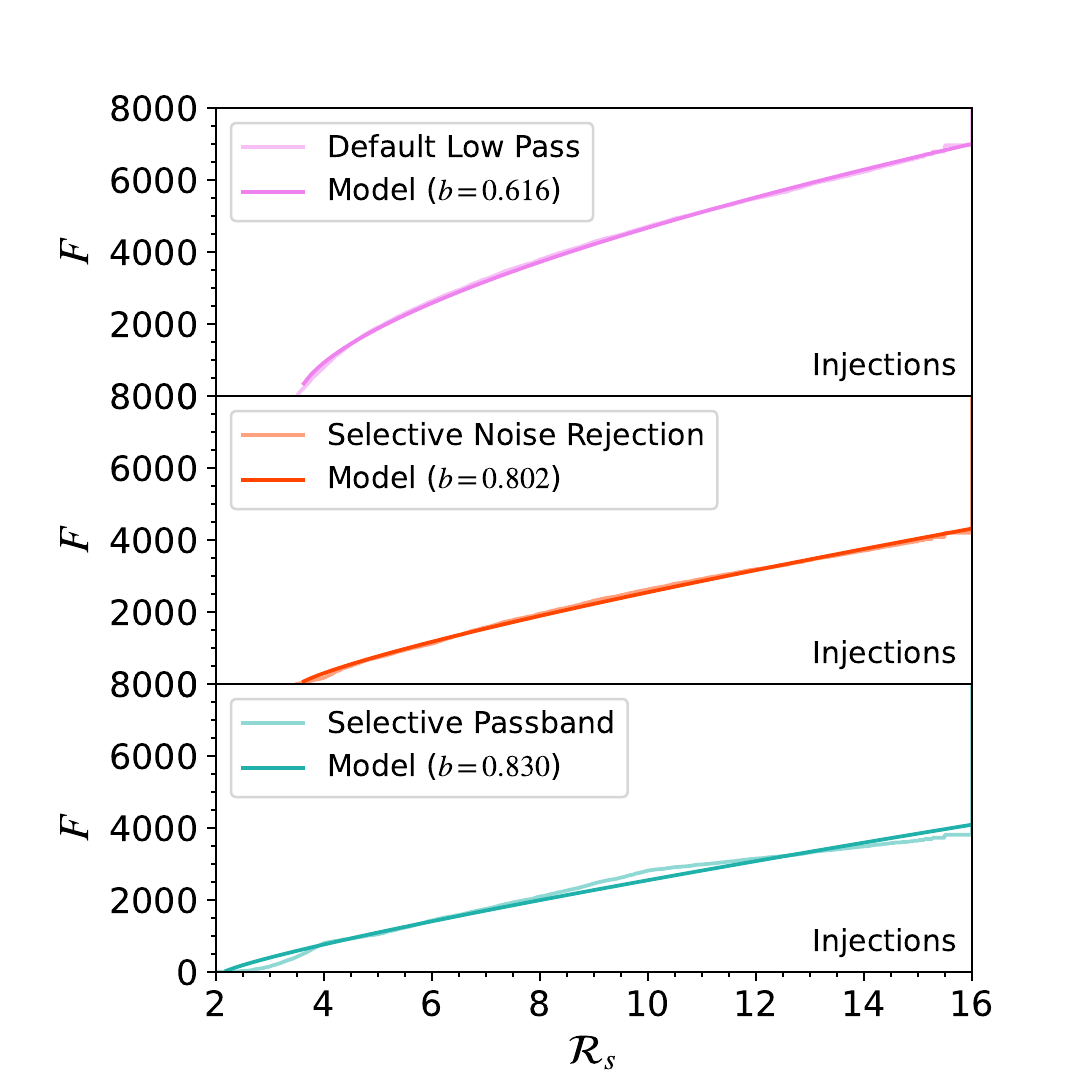}
  \includegraphics[width=0.485\linewidth]{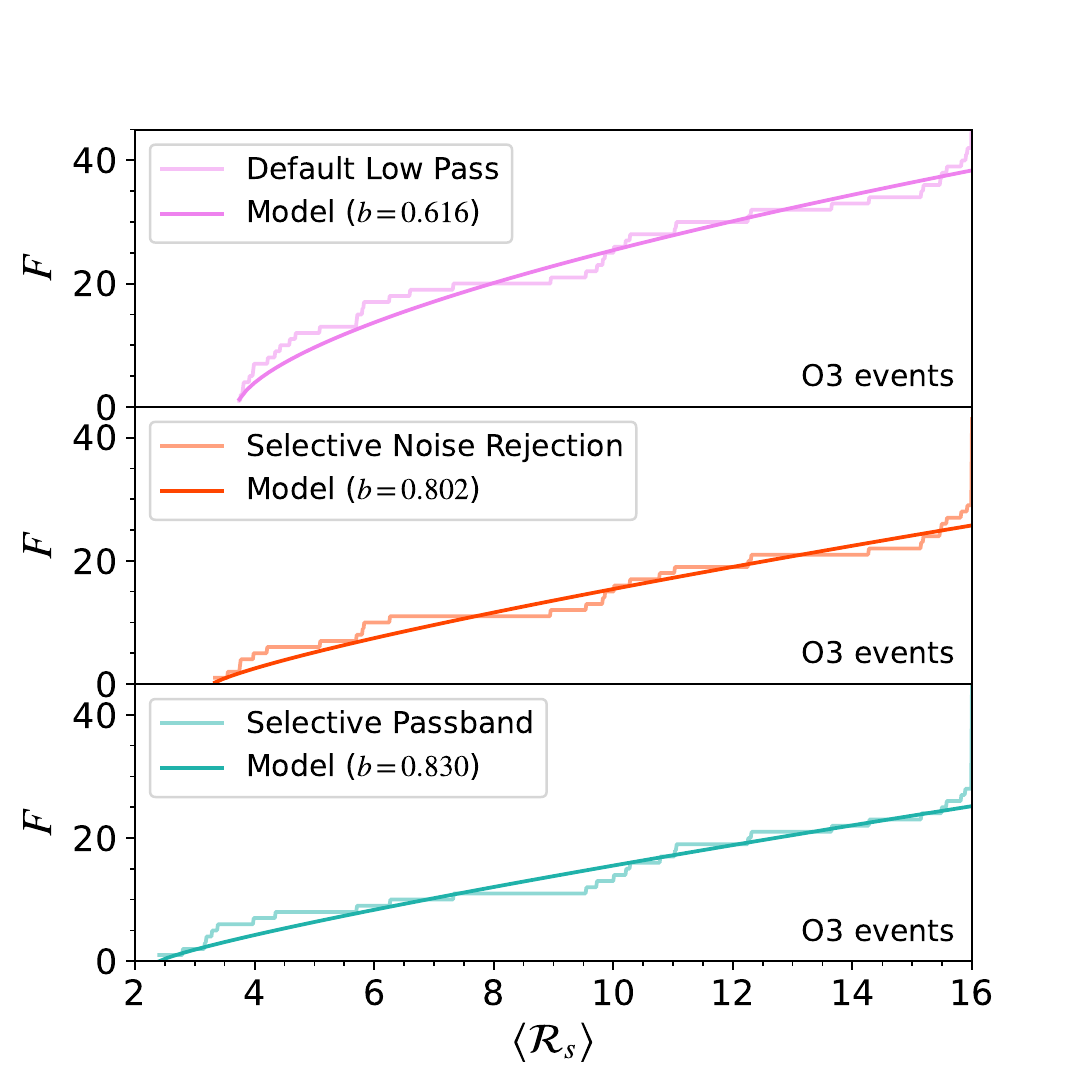}
  \caption{{\it Left panel:} Cumulative rate $F$ of foreground events for injections in O3a noise. There is excellent agreement with the  foreground model Eq. (\ref{eq:fmodel}). {\it Right panel:} Cumulative rate $F$ of known foreground O3 events. There is very good agreement with the  foreground model Eq. (\ref{eq:fmodel}), when the same exponent $b$ is used as obtained fom the injections.} 
  \label{fig:CDF_collective}
\end{figure*}

\subsection{Foreground Model from Injections}

If $N_f$ are foreground events ordered in terms of a ranking statistic $x$ as $x_1<x_2<\cdots<x_N$, where $x_1 = x_{\rm min}$ and $x_N = x_{\rm max}$, then the differential rate for the foreground events can be defined as a function  $f(x, \theta)$ as \cite{PhysRevD.91.023005}
\begin{equation}
\frac{d N_f}{d x}=f(x, \theta).
\end{equation}
The cumulative rate of foreground events below a ranking statistic $x$ is 
\begin{equation}
F(x, \theta) \equiv \int_{x_{\rm min}}^x d x' f(x', \theta).
\label{eq:F}
\end{equation}
If $R_f$ is the total number of foreground  events
\begin{equation}
R_f = F(x_{\rm max}, \theta),
\end{equation}
we can normalize 
\begin{equation}
{\hat F}(x,\theta) = F(x,\theta)/R_f.
\end{equation} 
Since our ranking statistic does not coincide with the SNR used in other foreground models (see \cite{PhysRevD.91.023005}), we determined the foreground model ${\hat F}(x,\theta)$ using the LVK O3 injections \cite{LIGO2021GWTC3} in the 80-day O3a noise created for the MLGWSC-1 challenge  \cite{challenge1}. 

The left panel of Fig. \ref{fig:CDF_collective} shows the un-normalized data $F({\cal R}_s)$ for the three classes of our classification scheme, as obtained by analyzing the injections with AresGW. We find that the data is in excellent agreement with the model
\begin{equation}
F(x) = a (x-x_{\rm min})^b,
\label{eq:fmodel}
\end{equation}
where $x={\cal R}_s$ (for the injections we used the initial ranking statistic of each trigger). The exponent $b$ is obtained as $b=0.616$ for the Default Low-Pass class,  $b=0.802$ for the Selective Noise Rejection class,  and $b=0.830$ for the Selective Passband class.

The right panel of Fig. \ref{fig:CDF_collective} shows the corresponding un-normalized data $F(\langle{\cal R}_s\rangle)$ for the three classes of our classification scheme, as obtained by analyzing the known O3 events with AresGW (notice that a significant fraction of known events accumulate at the highest ranking statistic $\langle{\cal R}_s\rangle=16$ and are not included in the model). For the O3 events, we used the mean ranking statistic, as discussed in Sec. \ref{sec:meanrs}. In each panel, we also show the agreement of the foreground model of Eq. (\ref{eq:fmodel}) with the O3 data, when using the same value for the exponent $b$, as obtained with injections in the left panel of Fig. \ref{fig:CDF_collective}. The agreement is very satisfactory, and hence we adopt Eq. (\ref{eq:fmodel}) as our foreground model for the O3 events, using their ensemble-averaged $\langle{\mathcal R_s}\rangle$ and with the exponent $b$ determined by injections.

\section{Astrophysical Probability}
\label{seq:pastro}

Inverting Eqs. (\ref{eq:B}) and (\ref{eq:F}), we obtain the differential rates
\begin{equation}
    b(\langle{\cal R}_s\rangle) = \frac{dB}{d\langle{\cal R}_s\rangle},
\end{equation}
and
\begin{equation}
    f(\langle{\cal R}_s\rangle) = \frac{dF}{d\langle{\cal R}_s\rangle}.
\end{equation}
The astrophysical probability of a candidate event is then \cite{PhysRevD.91.023005,2021CQGra..38i5004A,LIGOScientific:2016dsl}
\begin{equation}
p_{\rm astro} = \frac{f(\langle{\cal R}_s\rangle)}{b(\langle{\cal R}_s\rangle) + f(\langle{\cal R}_s\rangle)}.
\end{equation}
Since AresGW has been trained in a specific mass range, we treat this range as a single band. In the future, we aim at combining different versions of AresGW, each trained in a different segment of partially overlapping bands and use a multi-band expression for defining the astrophysical probability, as in \cite{2021CQGra..38i5004A}. 

\begin{figure}[H]
  \centering
  \includegraphics[width=1.0\linewidth]{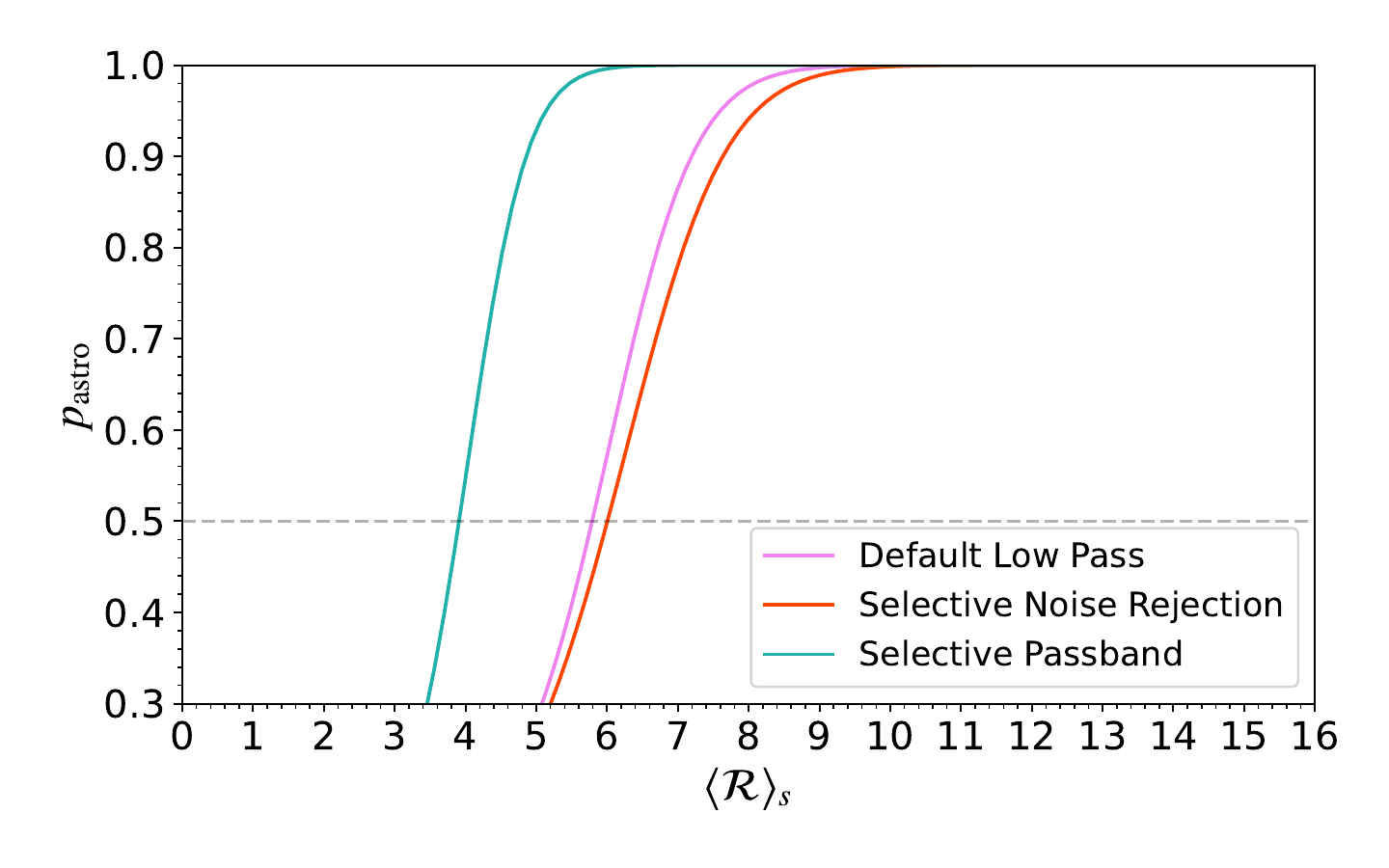}
  \caption{The astrophysical probability $p_{\rm astro}$ as a function of the (ensemble-avaraged) ranking statistic $\langle{\mathcal R_s}\rangle$ for the three classes of our classification scheme.} 
  \label{fig:pastro_collective}
\end{figure}

In Fig. \ref{fig:pastro_collective} we show $p_{\rm astro}$ as a function of the (ensemble-averaged) ranking statistic $\langle{\mathcal R_s}\rangle$ for the three classes of our classification scheme. Candidate events have $p_{\rm astro}>0.5$ for $\langle{\cal R}_s\rangle=5.79$ (Default Low-Pass), $\langle{\cal R}_s\rangle=6.0$ (Selective Noise Rejection), and $\langle{\cal R}_s\rangle=3.91$ (Selective Passband).

\section{Confirmation of known GW Events in O3 Data} \label{sec:Detection_of_known_GW_Events}

In this section, we will examine the performance of AresGW on gravitational-wave events that were previously identified and compare it to algorithms used in the published GWTC \cite{GWTC1, GWTC2, LIGO2022GWTC21, GWTC3}, OGC \cite{OGC-1,OGC-2,OGC-3,OGC-4}, and IAS \cite{IAS_O3a,IAS_O3b,IAS_higher_harmonics} catalogs. It is essential to note that AresGW utilizes two channels; it is trained on data from the two LIGO detectors, when both were active (Virgo data are not used in the training)\footnote{If one or both LIGO detectors were inactive during a specific time frame, the network does not generate a trigger for that period.}. Therefore, our main analyses pertain to BBH events detected when both the Livingston and Hanford detectors were operational. Nevertheless, we show in Sec. \ref{Sec:Virgo} that AresGW generalizes well to events that were detected by one of the LIGO detectors, with Virgo as the second detector, even though it has not been trained on Virgo data.

The Gravitational-wave Transient Catalog (GWTC) \cite{GWTC1, GWTC2, LIGO2022GWTC21, GWTC3} represents a comprehensive compendium of gravitational wave events, published by the the LIGO/Virgo/KAGRA collaboration. Here, we will consider those GWTC events that have been characterized as confident detections.  These detections are deemed to have a probability greater than 0.5 of being a gravitational wave signal rather than instrumental noise, assuming a compact binary coalescence source, as determined by at least one of the associated search pipelines.  The algorithms contributing to these catalogs encompass {\tt cwb}, {\tt gstlal}, {\tt mbta}, {\tt pycbc\_bbh}, and {\tt pycbc\_broad}, which employ matched filtering using specific waveform models or wavelet techniques for unmodeled burst detection. Proceeding to the four OGC catalogs \cite{OGC-1,OGC-2,OGC-3,OGC-4}, they confirm known GWTC events and report additional candidate events, using a matched filtering technique as well. Meanwhile, in recent searches, the IAS algorithm \cite{IAS_O3a,IAS_O3b,IAS_higher_harmonics} specifically targeted signals that could originate from the lower or upper mass gap, but also reported new candidate events that fall within our effective training range, using different versions of their matched-filtering detection algorithm.

\subsection{Known events within our effective training range}

Considering all three event catalogs, GWTC, OGC, and IAS, along with their respective sublists, there is a total of 43 gravitational wave events within our designated training range. It is crucial to reiterate that although our training dataset comprises events with source masses falling within the range $(7 M_\odot, 50 M_\odot)$, we evaluate events within our effective training range, which has the additional restriction of ${\cal M}$ being in the range $(10 M_\odot, 40 M_\odot)$. In this range, there are 36 known GWTC events, another 4 that were first reported by OGC and 3 additional reported by IAS (a total of 43 different known events).

With AresGW, we confidently detect 34 of the 43 known detections in its effective training range, with astrophysical probability $p_{\rm astro} > 0.5$. Out of the 9 candidate events that AresGW reports with $p_{\rm astro} < 0.5$, 4 were previously reported with astrophysical probabilities only slightly greater than 0.5: 1 in the GWTC, 2 in the IAS, and 1 in the OGC catalogs, with $p_{\rm astro}$ values of 0.54, 0.63, 0.56, and 0.5, respectively.

\begin{table*}[ht]
    \centering
    \caption{AresGW Performance on all previously detected events in the dataset.}
    \label{AresGW_in}
    \renewcommand{\arraystretch}{1.2}

    \setlength{\extrarowheight}{0pt}
    \footnotesize
    \begin{tabular}{!{\vrule width 1pt}l|c|c|c|c|c|c|c|c|c|c|c!{\vrule width 1pt}}
        \specialrule{1pt}{0pt}{0pt}
        \# & Event Name & Catalog & $p_{\rm astro}$ & $p_{\rm astro}$\footnote{For GWTC, the $p_{\rm astro}$ given is the best one provided among all its algorithms.}  & FAR (1/yr) & FAR (1/yr)\footnote{ IAS FAR is computed within bank.} \footnote{ GTWC FAR is the one corresponding to the algorithm that provided the best $p_{\rm astro}$.} & $\langle{\mathcal R_s}\rangle$ &${\cal M}$ & $m_{1}$ & $m_{2}$ & Class \\
        \textbf{} & \textbf{} & \textbf{} & (AresGW) &  & (AresGW) &  & (AresGW) & ($M_\odot)$ & ($M_\odot)$ & ($M_\odot$)  &\textbf{}  \\
        \specialrule{1pt}{0pt}{0pt}
        \rowcolor{lightblue}
        1 & GW190408\_181802 & GWTC & 1.00 & 1.00 & $\leq$ 0.0020 & $\leq$ 0.00001 & $\geq$ 16.0 & 18.5 & 24.8 & 18.5 & Selective Passband  \\
        2 & GW190412\_053044 & GWTC & 1.00 & 1.00 & $\leq$ 0.0020 & $\leq$ 0.00001 & $\geq$ 16.0 & 13.3 & 27.7 & 9.0 & Selective Passband \\
        \rowcolor{lightblue}

        3 & GW190421\_213856 & GWTC & 1.00 & 1.00 & $\leq$ 0.0020 & 0.0028 & $\geq$ 16.0 & 31.4 & 42.0 & 32.0 & Selective Passband  \\

        4 & GW190513\_205428 & GWTC & 1.00 & 1.00 & $\leq$ 0.0020 & 0.000013 & $\geq$ 16.0 & 21.8 & 36.0 & 18.3 & Selective Passband  \\

        \rowcolor{lightblue}

        5 & GW190727\_060333 & GWTC & 1.00 & 1.00 & $\leq$ 0.0020 & $\leq$ 0.00001 & $\geq$ 16.0 & 29.4 & 38.9 & 30.2 & Selective Passband  \\

        6 & GW190803\_022701 & GWTC & 1.00 & 0.97 & $\leq$ 0.0020 & 0.39 & $\geq$ 16.0 & 27.6 & 37.7 & 27.6 & Selective Passband \\
       \rowcolor{lightblue}

        7 & GW190828\_063405 & GWTC & 1.00 & 1.00 & $\leq$ 0.0020 & $\leq$ 0.00001 & $\geq$ 16.0 & 24.6 & 31.9 & 25.8 & Selective Passband  \\
        
        8 & GW190915\_235702 & GWTC & 1.00 & 1.00 & $\leq$ 0.0020 & $\leq$ 0.00001 & $\geq$ 16.0 & 24.4 & 32.6 & 24.5 & Selective Passband  \\
    \rowcolor{lightblue}
        9 & GW191215\_223052 & GWTC & 1.00 & $\geq$ 0.99 & $\leq$ 0.0020 & $\leq$ 0.00001 & $\geq$ 16.0 & 18.4 & 24.9 & 18.1 & Selective Passband  \\
        10 & GW191222\_033537 & GWTC & 1.00 & $\geq$ 0.99 & $\leq$ 0.0020 & $\leq$ 0.00001 & $\geq$ 16.0 & 33.8 & 45.1 & 34.7 & Selective Passband\\

       \rowcolor{lightblue}
       11 & GW200128\_022011 & GWTC & 1.00 &  $\geq$ 0.99 & $\leq$ 0.0020 & 0.0043 & $\geq$ 16.0 & 32.0 & 42.2 & 32.6 & Selective Passband\\
        
        12 & GW200129\_065458 & GWTC & 1.00 &  $\geq$ 0.99 & $\leq$ 0.0020 & $\leq$ 0.00001 & $\geq$ 16.0 & 27.2 & 34.5 & 29.0 & Selective Passband\\
        \rowcolor{lightblue}
        13 & GW200208\_130117 & GWTC & 1.00 &  $\geq$ 0.99 & $\leq$ 0.0020 & 0.00031 & $\geq$ 16.0 & 27.7 & 37.7 & 27.4 & Selective Passband\\
        
        14 & GW200209\_085452 & GWTC & 1.00 & 0.97 & $\leq$ 0.0020 & 12 & $\geq$ 16.0 & 26.7 & 35.6 & 27.1 & Selective Passband\\
        \rowcolor{lightblue}
        15 & GW200219\_094415 & GWTC & 1.00 &  $\geq$ 0.99 & $\leq$ 0.0020 & 0.00099 & $\geq$ 16.0 & 27.6 & 37.5 & 27.9 & Selective Passband\\
        
        16 & GW200224\_222234 & GWTC & 1.00 &  $\geq$ 0.99 & $\leq$ 0.0020 & $\leq$ 0.00001 & $\geq$ 16.0 & 31.1 & 40.0 & 32.7 & Selective Passband\\
        \rowcolor{lightblue}

        17 & GW200311\_115853 & GWTC & 1.00 &  $\geq$ 0.99 & $\leq$ 0.0020 & $\leq$ 0.00001 & $\geq$ 16.0 & 26.6 & 34.2 & 27.7 & Selective Passband\\

        18 & GW190916\_200658 & OGC & 1.00 & 0.90 & 0.0022 & 4.55 & 15.9 & 27.9 & 43.9 & 24.8 & Selective Passband \\
        
       \rowcolor{lightblue}
        19 & GW190719\_215514 & GWTC & 1.00 & 0.92 & 0.0023 & 0.63 & 15.8 & 22.8 & 36.6 & 19.9 & Selective Passband  \\

        20 & GW190512\_180714 & GWTC & 1.00 & 1.00 & 0.0029 & $\leq$ 0.00001 & 15.5 & 14.6 & 23.2 & 12.5 & Selective Passband \\
       \rowcolor{lightblue}
        21 & GW190731\_140936 & GWTC & 1.00 & 0.83 & 0.0037 & 1.9 & 15.2 & 29.7 & 41.8 & 29.0 & Selective Passband  \\

        22 & GW190527\_092055 & GWTC & 1.00 & 0.85 & 0.0073 & 0.23 & 14.3 & 23.9 & 35.6 & 22.2 & Selective Passband  \\
        \rowcolor{lightblue}

        23 & GW190521\_074359 & GWTC & 1.00 & 1.00 & 0.01 &  $\leq$ 0.00001 & 13.7 & 32.8 & 43.4 & 33.4 & Selective Passband  \\
  
        24 & GW190413\_052954 & GWTC & 1.00 & 0.93 & 0.033 & 0.82 & 12.3 & 24.5 & 33.7 & 24.2 & Selective Passband  \\
        \rowcolor{lightblue}

        25 & GW190517\_055101 & GWTC & 1.00 & 1.00 & 0.033 & 0.00035 & 12.3 & 26.5 & 39.2 & 24.0 & Selective Passband  \\
        
        26 & GW190503\_185404 & GWTC & 1.00 & 1.00 & 0.089 &  $\leq$ 0.00001 & 11.0 & 29.3 & 41.3 & 28.3 & Selective Passband  \\ 
        \rowcolor{lightblue}

        27 & GW200305\_084739 & OGC & 1.00 & 0.59 & 0.1 & 50 & 10.8 & 24.0 & 33.8 & 27.5 & Selective Passband\\

        28 & GW191230\_180458 & GWTC & 1.00 & 0.96 & 0.16 & 0.42 & 10.2 & 36.5 & 49.4 & 37.0 & Selective Passband\\
       \rowcolor{lightblue}
        29 & GW191204\_110529 & GWTC & 1.00 & 0.74 & 1.5 & 3.3 & 7.3 & 19.8 & 27.3 & 19.2  & Selective Passband \\   
        
        30 & GW200225\_060421 & GWTC & 1.00 &  $\geq$ 0.99 & 4.3& $\leq$ 0.000011 & 9.8 & 14.2 & 19.3 & 14.0 & Selective Noise Rejection \\
        \rowcolor{lightblue}

        \rowcolor{lightblue} 
        31 & GW190906\_054335 & IAS  & 0.99 & 0.61 & 0.22 & 1.4 & 15.9 & 21.90 & 37.00 & 24.00 & Selective Noise Rejection \\
        
        32 & GW200220\_124850 & GWTC& 0.99 & 0.83 & 4.1 & 1800 & 9.9 & 28.2 & 38.9 & 27.9 &  Selective Noise Rejection \\
        \rowcolor{lightblue}

        33 & GW200106\_134123 & OGC & 0.95 & 0.69 & 7.9 & 17 & 5.1 & 29.7 & 44.0 & 27.4 & Selective Passband\\
    
        34 & GW200322\_091133 & GWTC & 0.55 & 0.62 & 18.2 & 450 & 4.0 & 15.0 & 38.0 & 11.3 & Selective Passband\\
        \rowcolor{lightblue}
        
        35 & GW190926\_050336 & OGC & 0.42 & 0.92 & 31 & 3.7 & 5.7 & 26.2 & 40.0 & 23.8 & Selective Noise Rejection\\

        36 & GW190828\_065509 & GWTC & 0.22 & 1.00 & 91 & 0.000035 & 4.6 & 13.4 & 23.7 & 10.4 & Default Low-Pass \\
       \rowcolor{lightblue}
        
        37 & GW190514\_065416 & GWTC & 0.16 & 0.76 & 64 & 2.8 & 4.2 & 29.1 & 40.9 & 28.4 & Selective Noise Rejection\\
        
        38 & GW190805\_211137 & GWTC & 0.15 & 0.95 & 71 & 0.63 & 4.0 & 31.9 & 46.2 & 30.6 & Selective Noise Rejection\\
        \rowcolor{lightblue}
        39 & GW200306\_093714 & GWTC & 0.14 & 0.81 & 78 & 410 & 3.8 & 17.5 & 28.3 & 14.8 & Selective Noise Rejection\\
        40 & GW200318\_191337  & OGC & 0.0 &   0.97 & 582 & 2.0 & 3.1 &  33.5
        & 49.1 & 31.6 & Default Low-Pass \\
         \rowcolor{lightblue}
        41 & GW190530\_030659 & IAS  & 0.0 & 0.63 & 1815 & 0.63 & 2.6 & 21.90 & 36.0 & 18.0 & Default Low-Pass\\
        42 & GW190404\_142514 & OGC & 0.0 & 0.50 & 4309 & 50 & 2.3 & 13.8 & 21.6 & 12.1 & Default Low-Pass\\
         \rowcolor{lightblue}
        43 & GW200301\_211019 & IAS & 0.0 & 0.56 & 80380 & 0.4 & 2.1 & 14.74 & 22.0 & 13.2 & Default Low-Pass\\
        \specialrule{1pt}{0pt}{0pt}
    \end{tabular}
\end{table*}

Of the 34 known events that we confirm with AresGW, we obtain 30 events with $p_{\rm astro} = 1.0$. Furthermore, we find 28 events with FAR $< 1/{\rm y}$.  In Table \ref{AresGW_in} we list the published $p_{\rm astro}$, FAR and source parameters $m_1, m_2, {\cal M}$  for the 43 known events. In addition, we report the $p_{\rm astro}$, FAR, and ranking statistic $\langle{\mathcal R_s}\rangle$, as obtained with AresGW. We find 33 out of 34 events that we confirm, to have $p_{\rm astro}$ higher than or equal to their published value.

In Section \ref{Comparison}, we present a more detailed comparison between AresGW and the individual pipelines involved in the GWTC/OGC/IAS catalogs.

\begin{table*}[ht]
    \centering
    \caption{AresGW Performance on events out of the dataset.}
    \label{AresGW_out}
    \renewcommand{\arraystretch}{1.2}
    \setlength{\extrarowheight}{0pt}
    \footnotesize
    \begin{tabular}{!{\vrule width 1pt}l|c|c|c|c|c|c|c|c|c|c|c!{\vrule width 1pt}}
        \specialrule{1pt}{0pt}{0pt}
        \# & Event Name & Catalog & $p_{\rm astro}$ & $p_{\rm astro}$  & FAR (1/yr) & FAR (1/yr)& $\langle{\mathcal R_s}\rangle$ &${\cal M}$ & $m_{1}$ & $m_{2}$ & Class \\
        \textbf{} & \textbf{} & \textbf{} & (AresGW) &  & (AresGW) &  & (AresGW) & ($M_\odot)$ & ($M_\odot)$ & ($M_\odot$) & \textbf{}\\
        \specialrule{1pt}{0pt}{0pt}
        \rowcolor{lightblue}

        1 & GW191127\_050227 & GWTC & 1.00 & 0.74 & $\leq$ 0.0020 & 4.1 & $\geq$ 16.0 & 29.90 & 53.0 & 24.0 & Selective Passband\\
        2 & GW190413\_134308 & GWTC & 1.00 & 0.99 &  0.0027 & 0.18 & 15.6 & 33.30 & 51.3 & 30.4 & Selective Passband\\
        
        \rowcolor{lightblue}

        3 & GW191204\_171526 & GWTC & 1.00 & $\geq$ 0.99 & 0.0029 &  $\leq$ 0.00001 & 15.5 & 8.56 & 11.7 & 8.4 & Selective Passband\\

        4 & GW190706\_222641 & GWTC & 1.00 & 1.00 & 0.083 & 0.00005 & 11.1 & 45.60 & 74.0 & 39.4 & Selective Passband\\
        \rowcolor{lightblue}
        
        5 & GW190519\_153544 & GWTC & 1.00 & 1.00 &  0.15 &  $\leq$ 0.00001 & 10.3 & 44.30 & 65.1 & 40.8 & Selective Passband\\    
    
        6 & GW191109\_010717 & GWTC & 1.00 & $\geq$ 0.99 & 0.24 & 0.00018 & 9.7 & 47.50 & 65.0 & 47.0 & Selective Passband\\
        
        \rowcolor{lightblue}
        7 & GW190701\_203306 & GWTC & 1.00 & 1.00 & 0.32 & 0.56 & 15.2 & 40.20 & 54.1 & 40.5 & Selective Noise Rejection\\
        8 & GW190707\_093326 & GWTC & 1.00 & 1.00 & 3.2 &  $\leq$ 0.00001 & 6.3 & 8.40 & 12.1 & 7.9 & Selective Passband \\
        \rowcolor{lightblue}
        9 & GW190728\_064510 & GWTC & 1.00 & 1.00 & 3.9 &  $\leq$ 0.00001 & 10.0 & 8.6 & 12.5 & 8.0 & Selective Noise Rejection\\
        10 & GW190711\_030756 & IAS &  0.99& 1.00 & 5.5 & 0.02584 & 9.3 & 29.8 & 71.7 & 17.9 & Selective Noise Rejection\\

        \specialrule{1pt}{0pt}{0pt}
    \end{tabular}
\end{table*}

\subsection{Known events outside our effective training range} 

The GWTC/OGC/IAS catalogs contain 55 events that fall outside of our effective training range (the majority of these were reported in the IAS catalog). Even though AresGW was trained mainly on data that fall within our effecive training range, so that it has not seen samples for most of these 55 events, we find that it was still able to detect 10 of them with $p_{\rm astro} \geq 0.99$ (see 
Table \ref{AresGW_out}).

The observation that our algorithm recognizes one-fifth of all events outside its effective training range underscores the robustness of our deep learning detection code in identifying gravitational wave signals. It also implies that  with additional training in other mass regions, AresGW should be able to detect events with the same efficiency as in the current $(10 M_\odot < {\cal M} < 40 M_\odot)$ range.

To further emphasize the network's effectiveness beyond its familiar waveform region, we highlight its detection of a signal like GW191109\_010717. This event features source masses of 65.0 $M_\odot$ and 47.0$M_\odot$, and ${\cal M}$ of 47.50$M_\odot$, parameters considerably larger than those in the effective training dataset. AresGW also detected several other signals with at least one parameter significantly different from those in the effective training dataset. Notable examples include GW190519\_153544, GW190706\_222641, GW191109\_010717 and GW190711\_030756. In fact, the first three of these events were detected with the maximum $\langle{\mathcal R_s}\rangle$ value allowed by double machine precision ($\langle{\mathcal R_s}\rangle=16$).

\subsection{Comparison of sensitivity between AresGW and other search pipelines} \label{Comparison}

Commencing with the individual algorithms that contribute to the GWTC catalog, among the 43 published events within our effective training range, the {\tt mbta} algorithm detects 27, {\tt pycbc\_bbh} 31, {\tt pycbc\_broad} 20, {\tt gstlal} 27, and {\tt cwb} 7 events, respectively. Note that several events are reported by multiple pipelines. The number of different events reported by all GWTC pipelines within our effective training range is 36. On the other hand, our enhanced version of AresGW detects 34 events in the same range.   

The algorithm used in the OGC catalog detected 37 events in this range. For events reported in the IAS publications, we distinguish between those reported in  \cite{IAS_O3a,IAS_O3b}, which we designate as IAS$_a$ and those in \cite{IAS_higher_harmonics}, which we  designate as IAS$_b$. 
IAS$_a$ detected 34 events in this range, whereas IAS$_b$ detected 36 events in this range.

In Section \ref{AresGW_Events} we present 8 new candidate events, detected by AresGW. This brings the total number of events detected by AresGW within its effective training range to 42, exceeding the total number of events detected by other pipelines.

Table \ref{table:AresGW_vs_All_in} summarizes the above comparisons of the number of distinct BBH events within our effective training range that are reported with $p_{\rm astro} > 0.5$ by the different detection pipelines.

Furthermore, considering different events in this range, it is interesting to compare the detections that have been reported with $p_{\rm astro} = 1.0$ by the different pipelines. With this requirement, the GWTC algorithms detected 21 events and AresGW detected 30 (without including new events). The algorithm used in the OGC catalog detected 24 events, and the IAS$_a$ and IAS$_b$ algorithms detected 20 and 18, respectively. In \cite{2024arXiv240310439K}, 21 events with $p_{\rm astro} = 1.0$ were reported with the pycbc\_KDE algorithm. Table \ref{table:AresGW_p1} summarizes the above comparison.

In short, AresGW shows exceptional sensitivity in the detection of gravitational waves, within its effective training range. It detects a larger number of events compared to other algorithms that are based on matched filtering or unmodeled search. A key factor for this is the fact that AresGW is simultaneously trained on injections in real O3a noise for both detectors, reporting a single ranking statistic for a pair of detectors. In contrast, other pipelines may reject candidate events if they don't pass an SNR threshold in individual detectors.

\begin{table*}[ht]
    \centering
        \caption{Performance of AresGW in comparison to the GWTC, OGC, IAS and pycbc\_KDE \cite{2024arXiv240310439K} algorithms on all 51 candidate events (43 previously published plus 8 new events detected by AresGW) that are within the effective training range.}
    \label{table:AresGW_vs_All_in}
    \renewcommand{\arraystretch}{1.2}
    \setlength{\extrarowheight}{0pt}
    \footnotesize
    \rowcolors{2}{white}{lightblue}
    \begin{tabular}{|*{10}{>{\centering\arraybackslash}p{1.6cm}|}}
        \specialrule{1pt}{0pt}{0pt}
        cwb & pycbc\_broad & mbta & gstlal & pycbc\_KDE & pycbc\_bbh & IAS\_a & IAS\_b & OGC  & AresGW  \\
        \specialrule{1pt}{0pt}{0pt}
        7 & 20 & 27 & 27 & 29 & 31 & 34 & 36 & 37 & 42 \\
        \specialrule{1pt}{0pt}{0pt}
    \end{tabular}
\end{table*}

\begin{table*}[ht]
    \centering
        \caption{Performance of AresGW in comparison to the GWTC, OGC, IAS and pycbc\_KDE \cite{2024arXiv240310439K} algorithms on all {\it previously published} events that were reported with $p_{\rm astro}=1$ and are within the effective training range of AresGW (this does not include the new AresGW candidate events). }
    \label{table:AresGW_p1}
    \renewcommand{\arraystretch}{1.2}
    \setlength{\extrarowheight}{0pt}
    \footnotesize
    \rowcolors{2}{white}{lightblue}
    \begin{tabular}{|*{6}{>{\centering\arraybackslash}p{1.6cm}|}}
        \specialrule{1pt}{0pt}{0pt}
        GWTC & pycbc\_KDE & IAS\_a & IAS\_b & OGC  & AresGW  \\
        \specialrule{1pt}{0pt}{0pt}
        21 & 21 & 20 & 18 & 24 & 30\\
        \specialrule{1pt}{0pt}{0pt}
    \end{tabular}
\end{table*}

\begin{table*}[htb]
    \centering
    \caption{AresGW Performance on BBH events detected while only one of the LIGO detectors and the Virgo detector were active (this list does not include detections, when all three detectors were observing). }
    \label{table:LV-HV}
    \renewcommand{\arraystretch}{1.2}
    \setlength{\extrarowheight}{0pt}
    \footnotesize
    \begin{tabular}{!{\vrule width 1pt}l|c|c|c|c|c|c|c!{\vrule width 1pt}}
        \specialrule{1pt}{0pt}{0pt}
        \# & Event Name & Catalog & Detectors & $\langle{\mathcal R_s}\rangle$ & ${\cal M}$ & $m_{1}$ & $m_{2}$  \\
        \textbf{} & \textbf{} & \textbf{} &\textbf{} & \textbf{} &($M_\odot)$ & ($M_\odot)$ & ($M_\odot$) \\
        \specialrule{1pt}{0pt}{0pt}
        \rowcolor{lightblue}
        1 & GW191216\_213338 & GWTC & HV & 11.5 & 8.33 & 12.1 & 7.7  \\
    
        2 & GW190630\_185205 & GWTC & LV & 8.2 & 25.1 & 35.1 & 24.0 \\
        \rowcolor{lightblue}
        3 & GW200112\_155838 & GWTC & LV & 7.2 & 27.4 & 35.6 &  28.3 \\
        
        4 & GW190708\_232457 & GWTC & LV & 5.6 & 13.1 & 19.8 & 11.6  \\
        \rowcolor{lightblue}
        5 & GW190620\_030421 & GWTC & LV & 3.2 & 38.1 & 58.0 & 35.0  \\
        6 & GW200302\_015811 & GWTC & HV & 2.5 & 23.4 & 37.8 & 20.0 \\
       \rowcolor{lightblue}
        7 & GW190925\_232845 & GWTC & HV & 2.3 & 15.6 & 20.8 & 15.5  \\

        8 & GW190910\_112807 & GWTC & LV & 1.4 & 33.5 & 43.8 & 34.2 \\
        \specialrule{1pt}{0pt}{0pt}
    \end{tabular}
\end{table*}

\subsection{Generalization to Virgo data}
\label{Sec:Virgo}

In Sections \ref{sec:Training_Data} and \ref{sec:Detection_of_known_GW_Events} we discussed AresGW's performance on O3 data when considering events that were detected by the two LIGO detectors (without taking into account Virgo data, when they were available). 
 To further assess the generalizability of our detection code, we extended our analysis to data from the Virgo detector, when combined with one of the two LIGO detectors. In essence, we evaluated published BBH merger events that were detected by either the LIGO Hanford and Virgo network or the LIGO Livingston and Virgo network of detectors (when the third detector was not operating). In the GWTC catalogs there are 8 such events during the O3 period, see Table \ref{table:LV-HV} (this number does not include detections, when all three detectors were observing). Although we did not estimate a false alarm rate for these events, it is encouraging that half of them were detected with a $\langle{\mathcal R_s}\rangle \geq 5.6$.

 This highlights the impressive ability of AresGW to generalize. Its performance on previously unseen, noisier compared to the LIGO detectors, data from the Virgo detector, suggests the potential for deployment in future observing runs without necessitating retraining. Furthermore, there is promise for uncovering new gravitational wave events by adapting the code using specifically Virgo detector data alongside that of another detector through direct analysis of noise from another observing period. Moreover, in the future, the use of a three-channel AresGW pipeline seems to hold great potential in achieving higher detection sensitivities, as the code will be trained on simultaneous injections for three different detectors and will evaluate the three-detector data as a single, concatenated data segment, producing a single ranking statistic.

\subsection{Generalization to O1 and O2 data}
\label{Sec:O2}

Upon observing AresGW's ability to adjust to Virgo's noise, we aimed to further assess the generalization capacity of our code by evaluating it on the data of the O1 and O2 observation periods \cite
{KAGRA:2023pio}. As shown in Table \ref{table:O2}, our machine learning code validated 5 of the 6 events that were previously detected during O2 with a maximum ranking statistic value $\langle{\mathcal R_s}\rangle$ of 16.0, while it, also, detected the sixth event with ${\mathcal R_s} = 15.4$. Moreover, it identified the first gravitational wave event ever detected in the O1 period (GW150914) with the maximum ranking statistic value of $\langle{\mathcal R_s}\rangle = 16$. Therefore, AresGW detected all events (but one) within its training range in the O1 and O2  periods, with a $\langle{\mathcal R_s}\rangle\geq 15.4$. The only exception was the O1 event GW151012, which was detected with a lower ranking statistic\footnote{The other O1 event, GW151226, was assigned a very low ranking statistic by AresGW, but it's outside its nominal effective training range.}.

The above findings, alongside our neural network's capacity to generalize effectively to Virgo's data, highlight AresGW's impressive capability for generalization. This aspect inspires confidence in the possibility of detecting, with further analysis, new gravitational waves using O1 and O2 data too, all without necessarily retraining the network. Furthermore, it holds the same promise for the O4 observing run, in which the noise levels for the two LIGO detectors have decreased significantly.

\begin{table}[htb]
    \centering
    \caption{AresGW Performance on BBH events detected on O1 and O2 from both LIGO detectors.}
    \label{table:O2}
    \renewcommand{\arraystretch}{1.2}
    \setlength{\extrarowheight}{0pt}
    \footnotesize
    \begin{tabular}{!{\vrule width 1pt}l|c|c|c|c|c|c!{\vrule width 1pt}}
        \specialrule{1pt}{0pt}{0pt}
        \# & Event Name & Catalog & $\langle{\mathcal R_s}\rangle$& ${\cal M}$ & $m_{1}$ & $m_{2}$  \\
        \textbf{} & \textbf{} &\textbf{} & \textbf{} &($M_\odot)$ & ($M_\odot)$ & ($M_\odot$) \\
        \specialrule{1pt}{0pt}{0pt}
        \rowcolor{lightblue}
        1 & GW170104 & GWTC & $\geq$ 16.0 & 21.1 & 28.7 & 20.8 \\ 
        2 & GW170729 & GWTC & $\geq$ 16.0 & 34.6 & 54.7 &  30.2 \\
        \rowcolor{lightblue}
        3 & GW170809 & GWTC & $\geq$ 16.0 & 24.8 & 	34.1 & 24.2 \\
        4 & GW170814 & GWTC & $\geq$ 16.0 & 24.1 & 30.9 &  24.9 \\
        \rowcolor{lightblue}
        5 & GW170823 & GWTC &  $\geq$ 16.0 & 28.6 & 38.3 & 29.0 \\
        6 & GW150914  & GWTC &  $\geq$ 16.0 & 27.9 & 34.6 & 30.0 \\
       \rowcolor{lightblue}
        7 & GW170818 & GWTC & 15.4 & 26.8 & 
        34.8 & 27.6 \\
        8 & GW151012 & GWTC &  2.5 & 15.6 & 24.8 & 13.6 \\
        \rowcolor{lightblue}
        9 & GW151226 & GWTC & 1.3 & 8.9  & 14.2 & 7.5 \\
        \specialrule{1pt}{0pt}{0pt}
    \end{tabular}
\end{table}

\section{New AresGW candidate events} \label{AresGW_Events}

In addition to confirming previously published events, we have obtained 8 new triggers with $p_{\rm astro}\geq 0.5$. We stress that only one of these new triggers, GW190607\_093827, is included in the Deep Extended Catalog of GWTC-2.1, which comprises 1201 subthreshold triggers with a FAR less than 2 per day \cite{LIGO2022GWTC21}. Here, we present several tests to establish the astrophysical relevance of these new candidate events. We also perform parameter estimation and discuss their population properties.

\begin{table*}[t!]
    \centering
    \caption{New candidate events identified by AresGW.}
    \label{table:AresGW_New_Detections}
    \renewcommand{\arraystretch}{1.2}
    \setlength{\extrarowheight}{0pt}
    \footnotesize
    \begin{tabular}{!{\vrule width 1pt}c|l|c|c|c|c|c|c|c|c!{\vrule width 1pt}}
        \specialrule{1pt}{0pt}{0pt}
        \# & Event Name & GPS Time & $p_{\rm astro}$ & FAR & $\langle{\cal R}_s\rangle$ & Time delay & $\chi_L^2$ & $\chi_H^2$ & Class \\
         \textbf{} &  \textbf{} & (s) &  \textbf{} & (1/yr) &  \textbf{} & (s) &  \textbf{} &  \textbf{} &  \textbf{} \\
        \specialrule{1pt}{0pt}{0pt}
        \rowcolor{lightblue}
        1 & GW190511\_125545 & 1241614563.77 & 1.00 & 0.27 & 9.54 & 0.0027 & 1.16 & 1.46 & Selective Passband   \\
        2 & GW190614\_134749 & 1244555287.93 & 0.99 & 4.6 & 5.80 & 0.0012 & 0.65 & 0.80 & Selective Passband  \\
        \rowcolor{lightblue}
        3 & GW190607\_083827 & 1243931925.99 & 0.99 & 6.5 & 8.95 & 0.0056 & 0.90 & 0.48 & Selective Noise Rejection \\
        4 & GW190904\_104631 & 1251629209.01 & 0.72 & 14 & 4.35 & 0.0002 & 0.38 & 0.71 & Selective Passband  \\
        \rowcolor{lightblue}
        5 & GW190523\_085933 & 1242637191.44 & 0.68 & 20 & 6.60 & 0.0054 & 0.75 & 1.39 & Selective Noise Rejection \\
        6 & GW200208\_211609 & 1265231787.68 & 0.55 & 18 & 4.0 & 0.0063 & 0.69 & 0.98 & Selective Passband  \\
        \rowcolor{lightblue}
        7 & GW190705\_164632 & 1246380410.88 & 0.51 & 49 & 5.82 & 0.0103 & 1.05 & 0.98 & Default Low-Pass$^{*}$ \\
        8 & GW190426\_082124 & 1240302101.93 & 0.50 & 20 & 3.91 & 0.0007 & 1.48 & 0.53 & Selective Passband \\
        \specialrule{1pt}{0pt}{0pt}
    \end{tabular}
\\    $^{*}$ This event also classified as Selective Noise Rejection, but it has the best $p_{\rm astro}$ as Default Low-Pass.
\end{table*}

\subsection{False alarm rate and consistency tests}
Among all triggers with a ranking statistic $\langle{\mathcal R_s}\rangle$ exceeding 3.5 (or 2.0 for the Selective Passband class) we focused only on those exhibiting visible features consistent with the expected chirp produced by BBH mergers in time-frequency spectrograms. These triggers were further investigated, including time and consistency assessments. From this analysis, we present the triggers that have successfully passed all relevant tests, thus identifying them as new candidate events. 

 In Table \ref{table:AresGW_New_Detections} we present several properties of each new candidate signal, such as their GPS timestamp, ranking statistic $\langle{\mathcal R_s}\rangle$, false alarm rate FAR and astrophysical probability $p_{\rm astro}$, as well as the trigger class, according to the hierarchical classification we introduced in Section \ref{sec:New_filters}. Furthermore, after performing the parameter estimation for each candidate signal (see Section \ref{sec:pe}) we obtained the time delay of each signal between the two LIGO detectors and the $\chi^2$ for the signal at the two detectors, $\chi_L^2$ and $\chi_H^2$, which we also list in Table \ref{table:AresGW_New_Detections}. We note that all 8 of our new candidate events successfully pass those necessary consistency tests that were also used in previously published catalogs of GW events.

\begin{figure}[t]
  \centering
  \includegraphics[width=0.95 \linewidth]{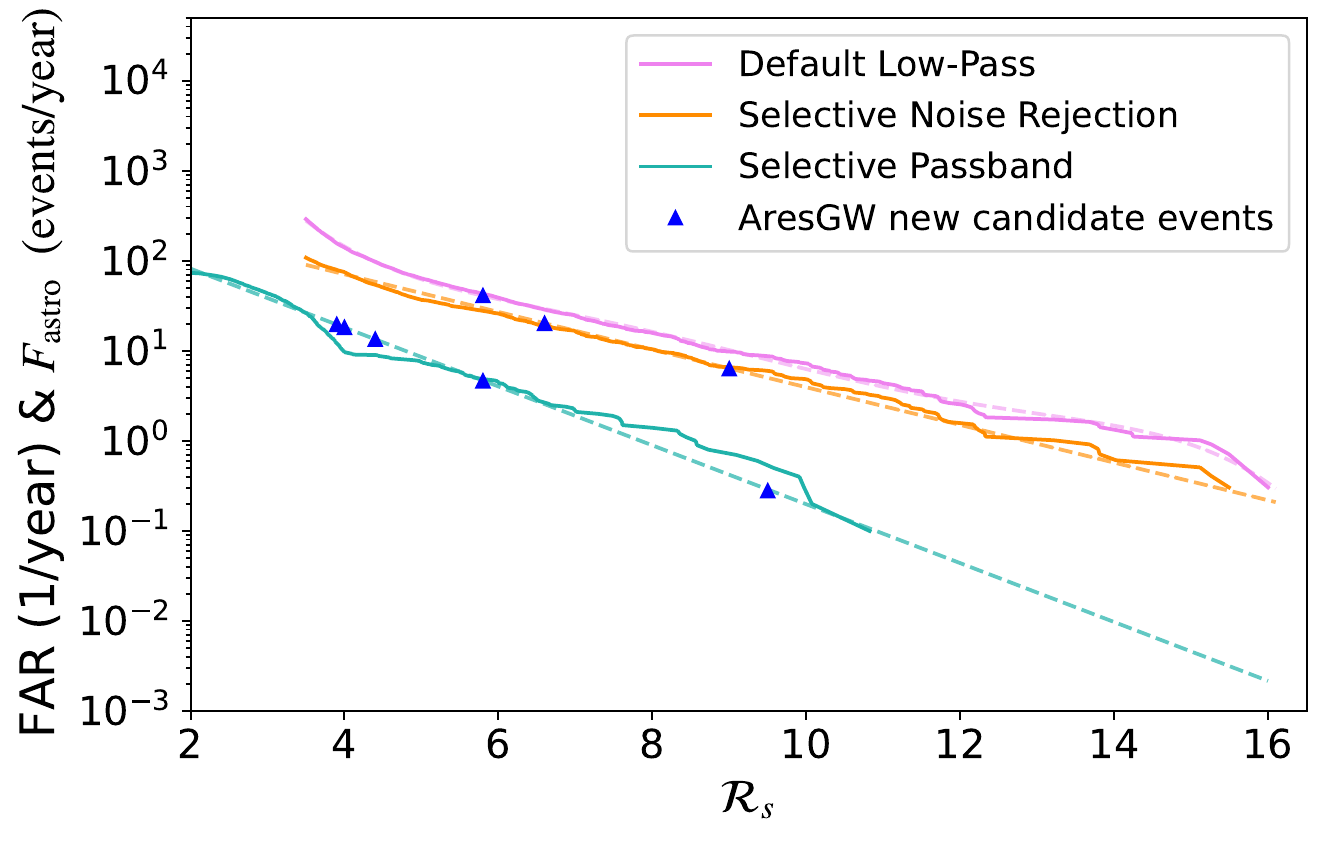}
  \caption{False alarm rate based on 10-year time shifts and corresponding analytic fits (dashed lines) for the three classes of triggers. The new events identified by AresGW are also shown.}
 
  \label{fig:far}
\end{figure}

In Fig.\ref{fig:far} we present our estimate of the cumulative ${\rm FAR}({\cal >R}_s)$ as a function of the ranking statistic for the new candidate events. The estimates are based on the analytical fits for each trigger class, see Eqs.(\ref{eq:Far_Initial}), 
(\ref{eq:Far_Selective Noise Rejection}) and (\ref{eq:Far_Selective Passband }) in Section \ref{sec:far} . Evidently, the Selective Noise Rejection triggers exhibit a lower FAR for equivalent ranking statistic values compared to the Default Low-Pass triggers. Furthermore, the Selective Passband  triggers demonstrate a significant reduction in the FAR across several orders of magnitude, compared to both the Selective Noise Rejection and the Default Low-Pass triggers. This observation underscores the superiority of the Selective Passband  trigger class, in terms of a lower FAR estimate.

Had we kept all triggers above $\langle{\mathcal R_s}\rangle$ as Default Low-Pass and without removing known glitches, this would amount to $O(150)$ triggers for further examination. Through the application of the classification system alongside the glitch removal process though, we observed 22 noise triggers in the Selective Passband class and 39 in the Selective Noise Rejection class\footnote{Following glitch removal, the Default Low-Pass class exhibited 76 noise triggers.}, therefore, we constructed wavelet-based Qp-transform spectrograms \cite{2024PhRvD.109j2010V, Virtuoso2024Wavelet} 
for those triggers. The Qp-transform spectrograms were then visually expected, and a few obvious glitches, which were not caught by Gravity Spy, were also eliminated. 

Specifically for the Selective Passband  triggers, we followed the exact same procedure described above, but starting from a ranking statistic threshold of $\langle{\mathcal R_s}\rangle$. Between the ranking statistic values of 2.0 and 3.5, approximately 40 additional triggers were included for inspection.
Subsequently, we identified a number of $O(10)$ promising gravitational wave candidate events for further inspection.

\begin{figure}[H]
  \centering
  \includegraphics[width=1.0 \linewidth]{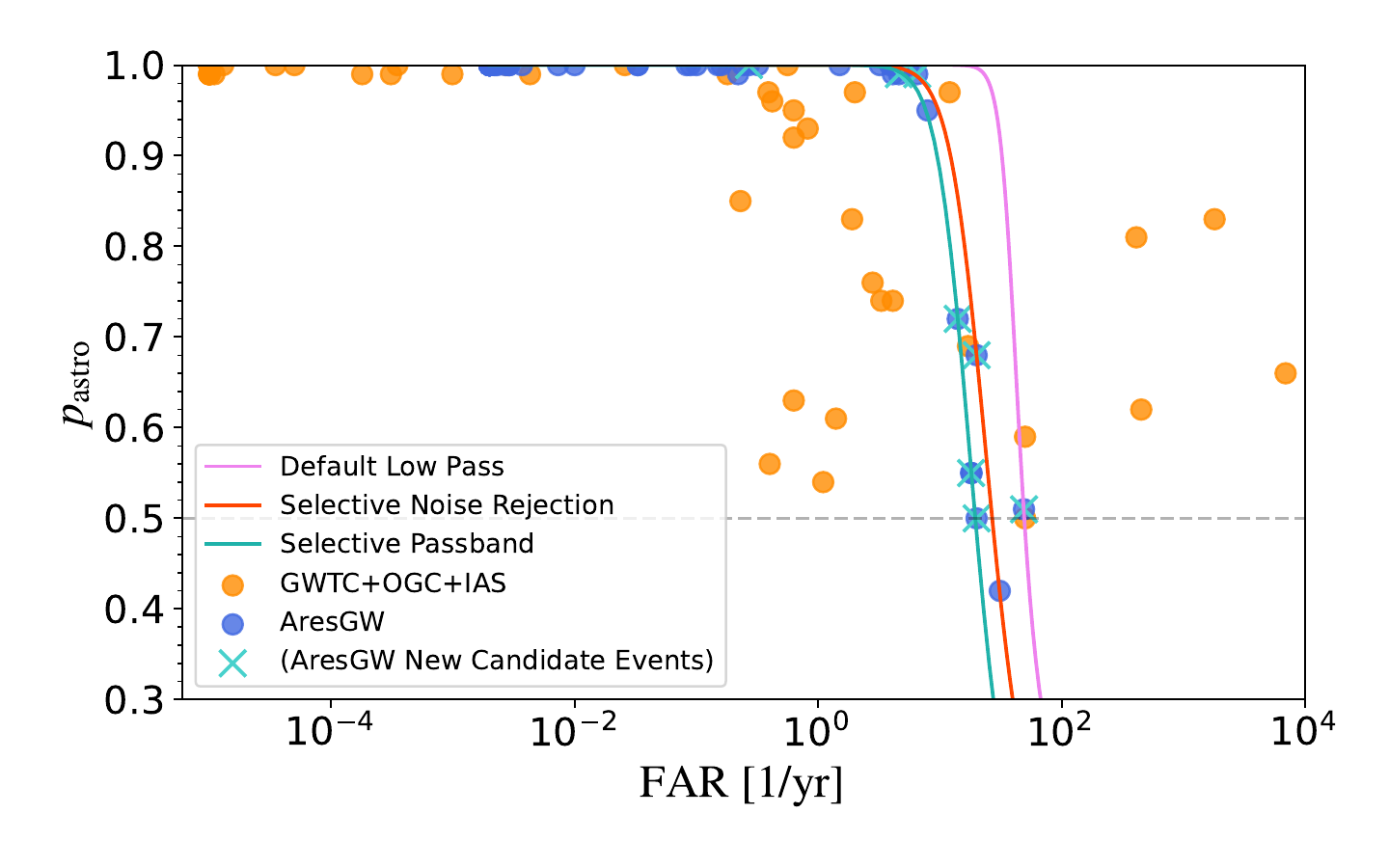}
  \caption{Astrophysical probability vs. false alarm rate for candidate events identified with AresGW. Both previously published and new events arer shown. For comparison, the $p_{\rm astro}$ vs. FAR values of previously published events in the GWTC/OGC/IAS catalogs are shown.}
 
  \label{fig:pastrofar}
\end{figure}

Fig. \ref{fig:pastrofar} shows the 
astrophysical probability vs. 
the false alarm rate, as obtained with AresGW,  for both previously published and new candidate events identified. 
The $p_{\rm astro}$ and FAR values obtained with AresGW are within the total spread of values of previously published events in the GWTC/OGC/IAS catalogs.

\begin{figure*}[t!]
    \centering
    \begin{tabular}{cc}
        \includegraphics[width=0.49\textwidth]{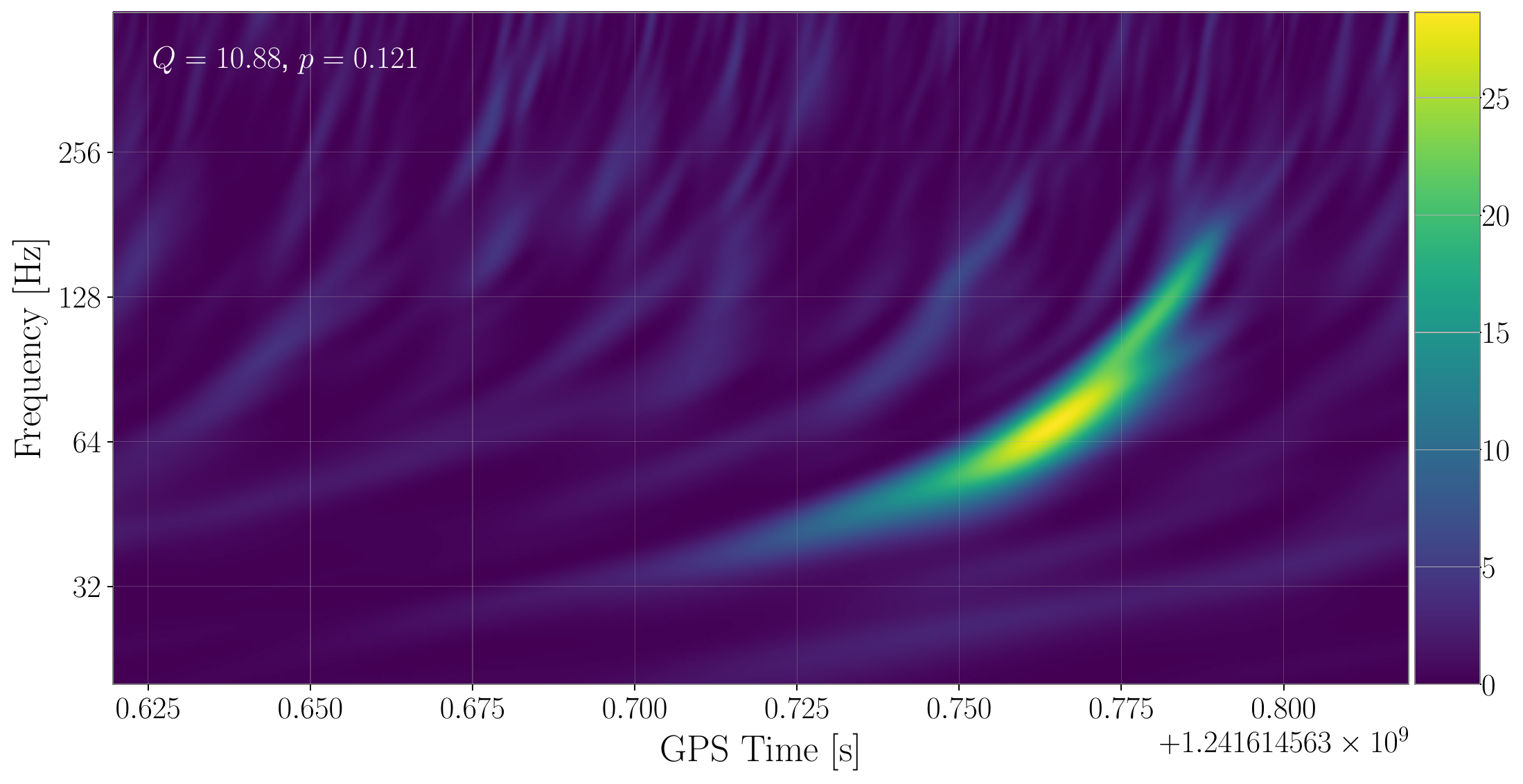} &
        \includegraphics[width=0.49\textwidth]{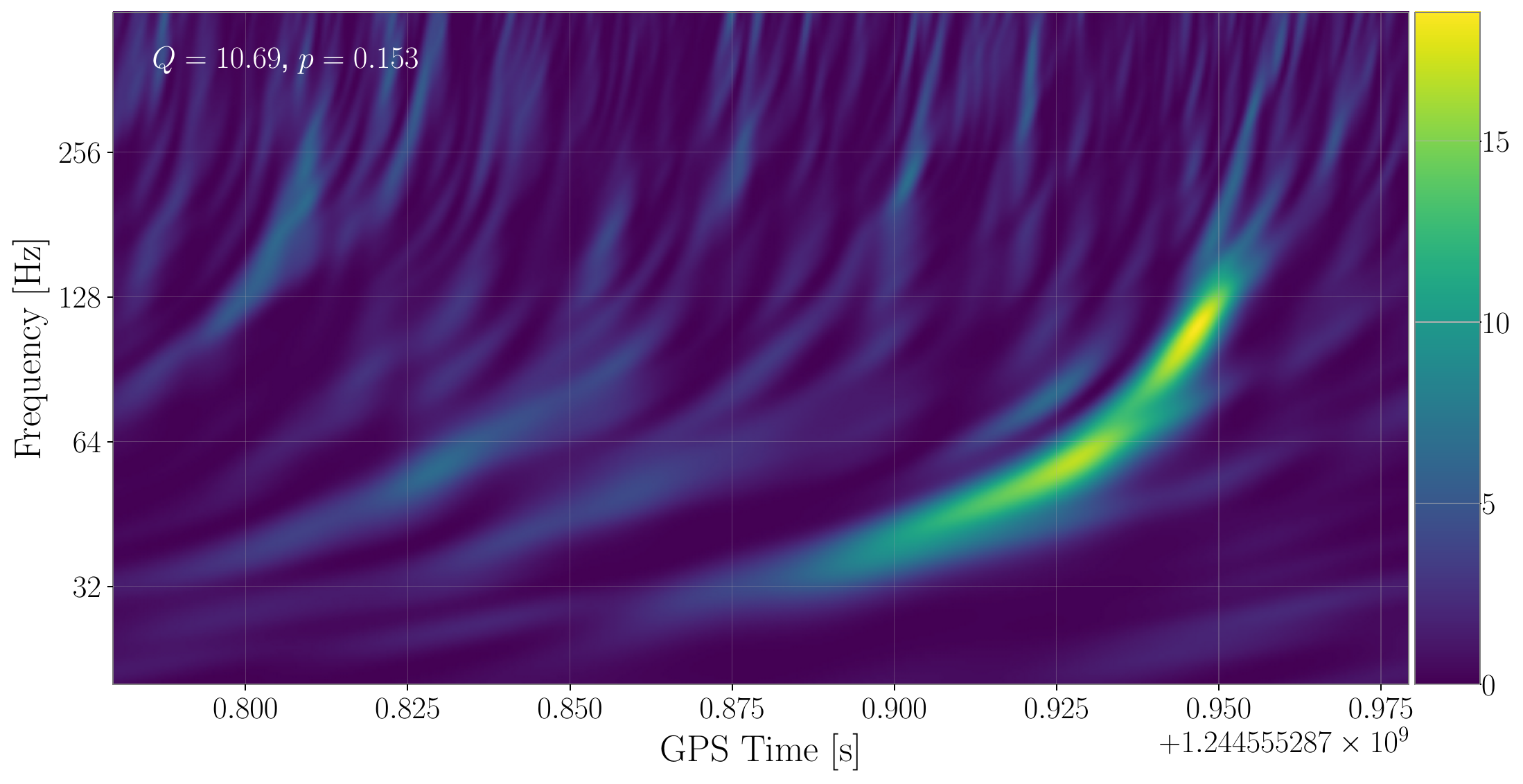} \\
        \includegraphics[width=0.49\textwidth]{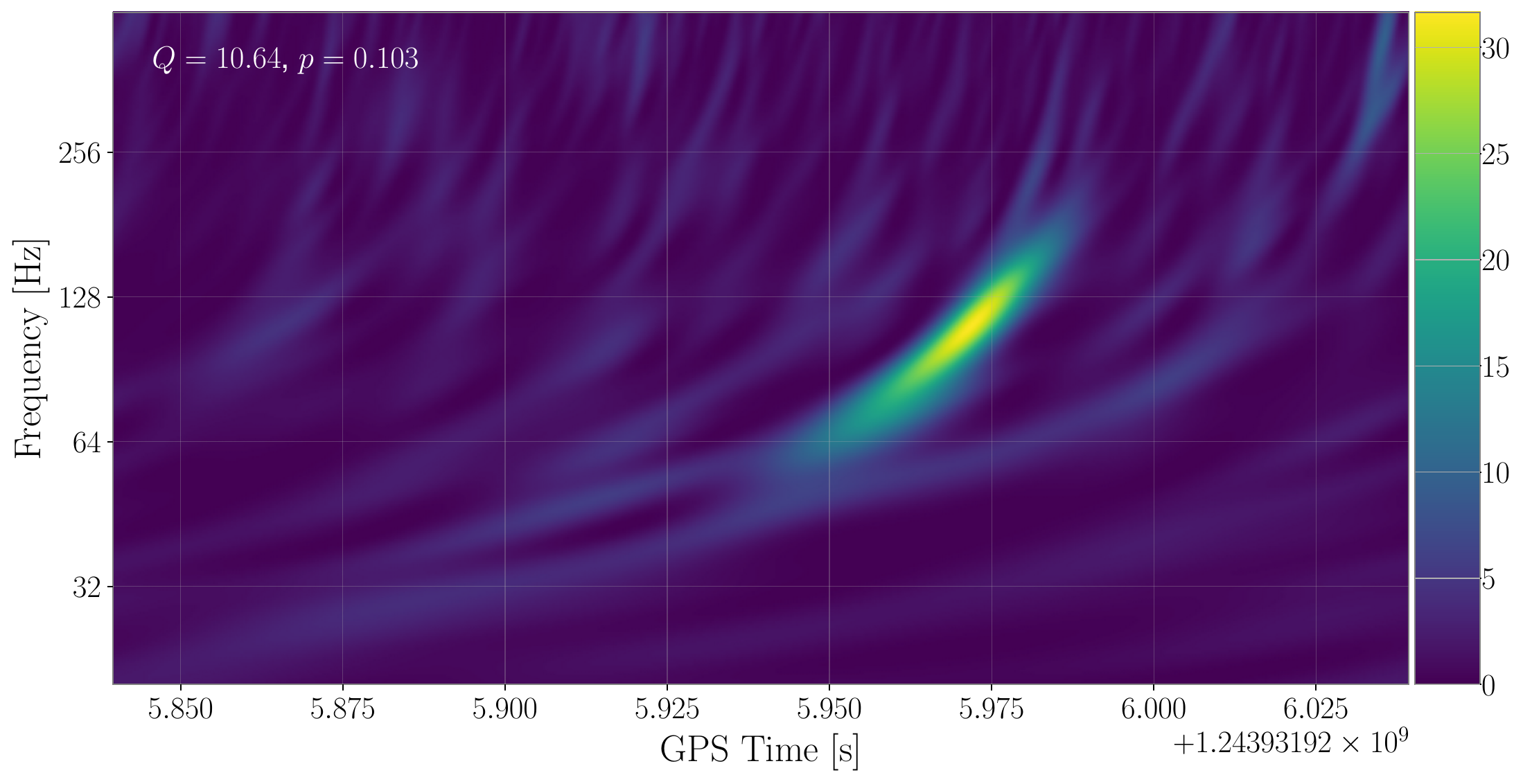} &
        \includegraphics[width=0.49\textwidth]{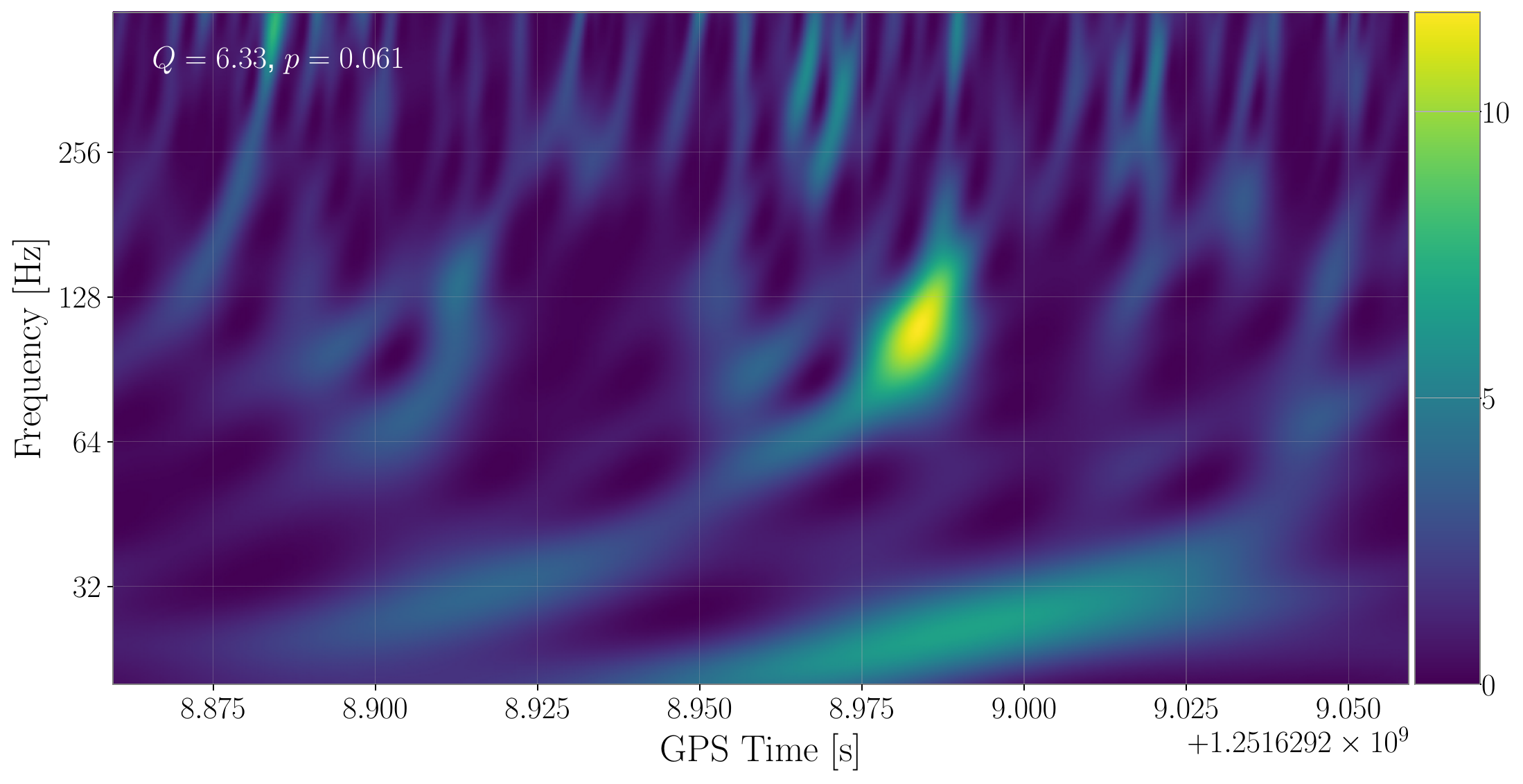} \\
        \includegraphics[width=0.49\textwidth]{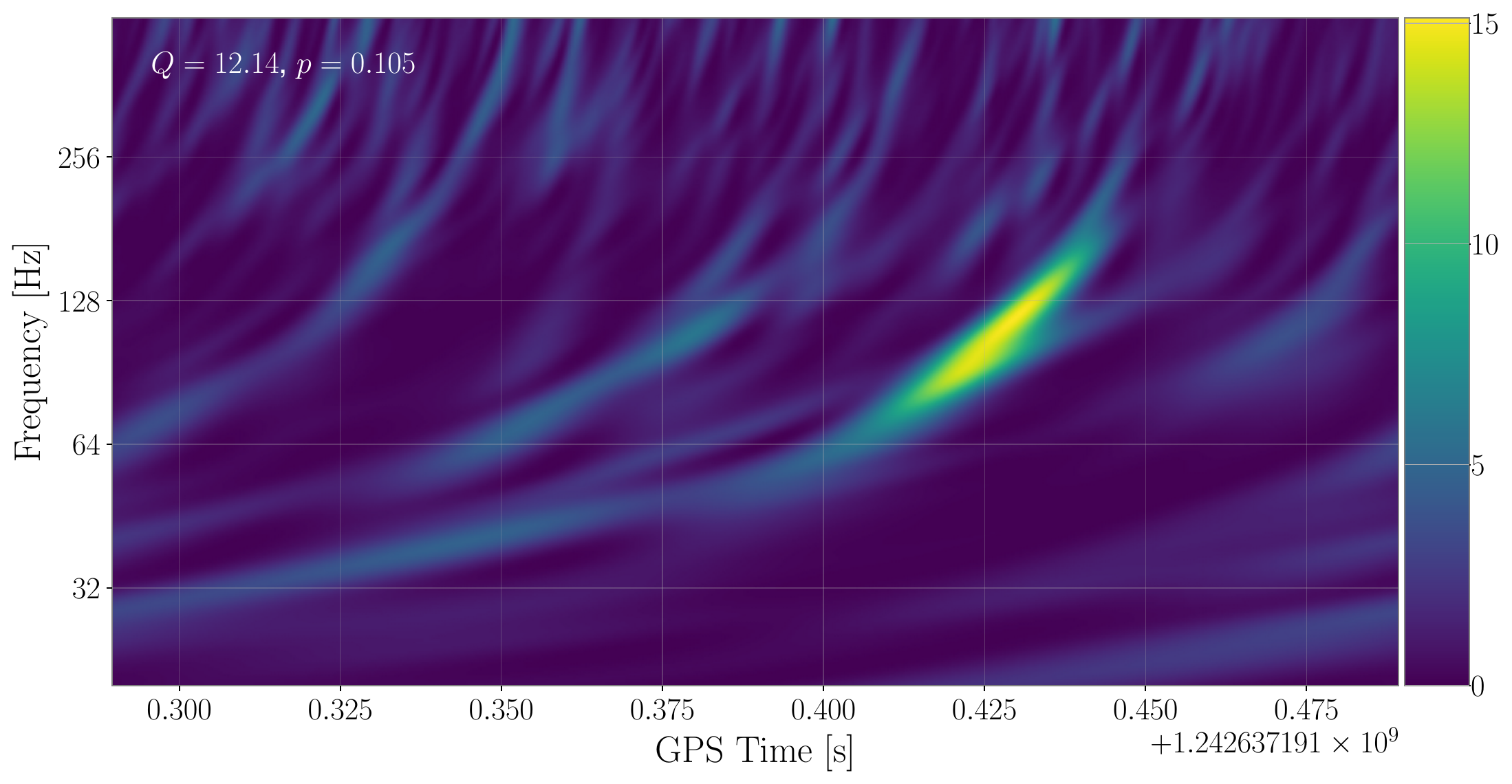} &
        \includegraphics[width=0.49\textwidth]{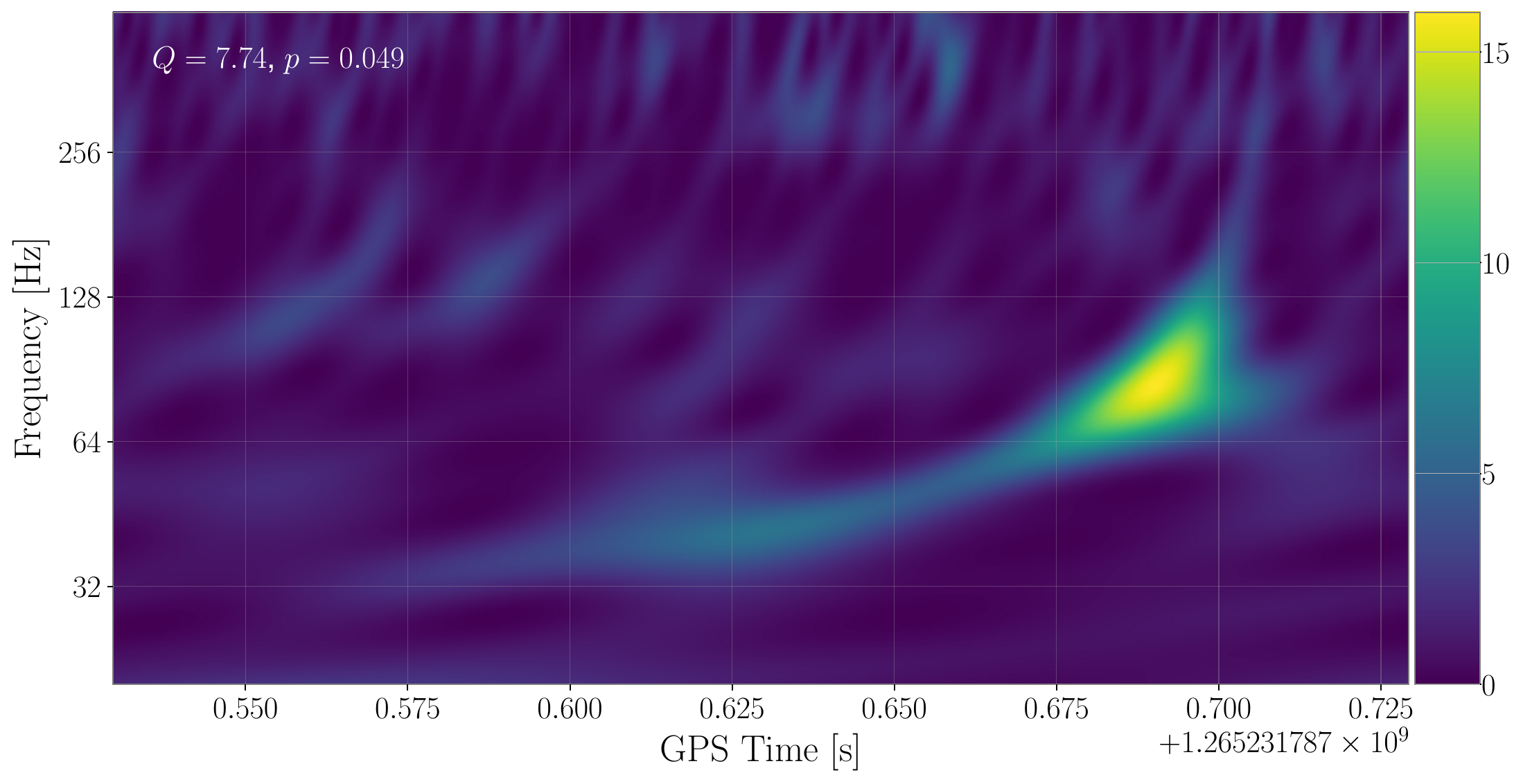} \\
        \includegraphics[width=0.49\textwidth]{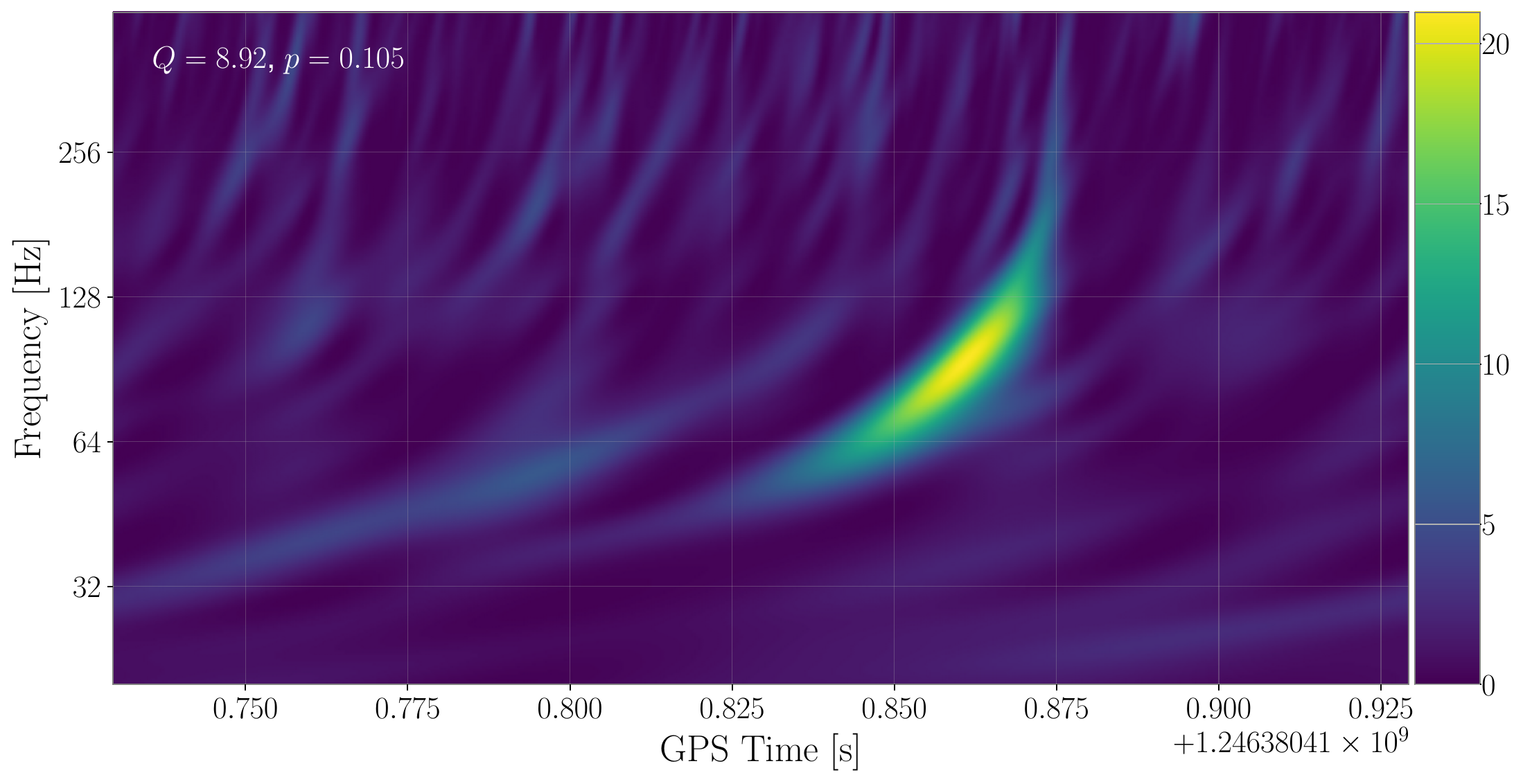} &
        \includegraphics[width=0.49\textwidth]{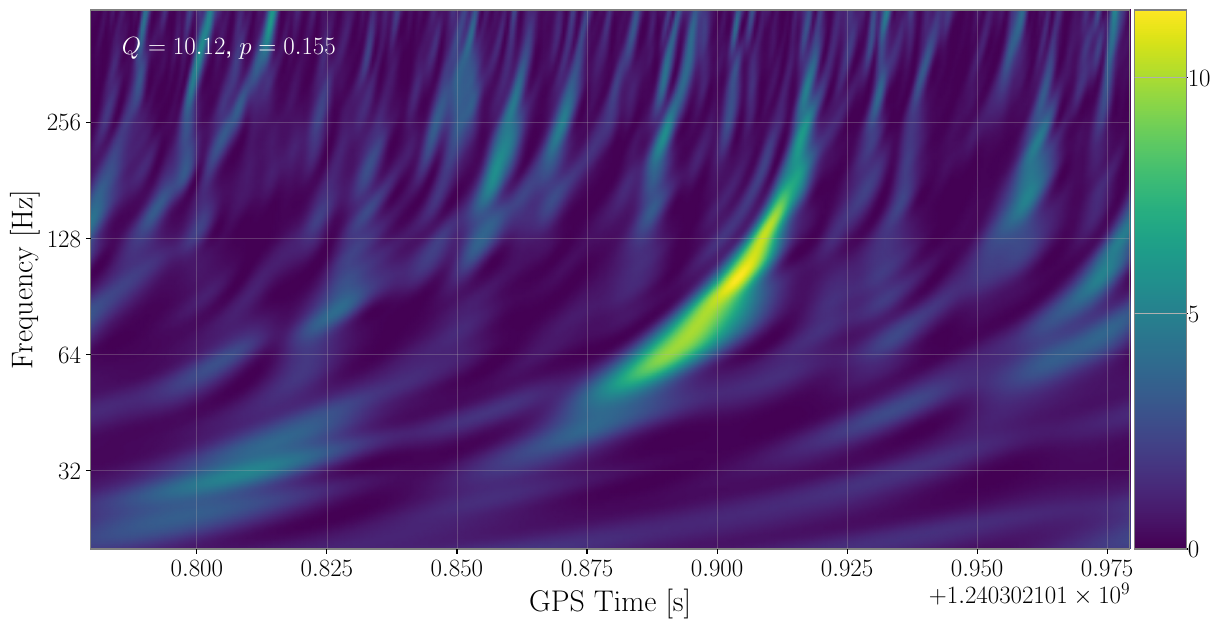} \\
    \end{tabular}
    \caption{Qp-transform spectrogram of the eight new candidate events identified by AresGW in Table \ref{table:AresGW_New_Detections}. From left to right and top to bottom: 1st row: GW190511\_125545 \& GW190614\_134749. 2nd row: GW190607\_083827 \& GW190904\_104631. 3rd row: GW190523\_085933 \& GW200208\_211609. 4th row: GW190705\_164632 \&  GW190426\_082124. All plots are for the LIGO Livingston detector, except for the last one, which is for the LIGO Hanford detector.}
    \label{fig:grid}
\end{figure*}

Fig. \ref{fig:grid} shows the Qp-transform spectrograms for the 8 new GW candidate events identified by AresGW. The familiar shape of a chirping signal is prominent in these spectrograms.

\begin{table*}[ht]
    \centering
    \caption{Parameter estimation for the new AresGW candidate events.}
    \label{table:Parameter_Estimation}
    \renewcommand{\arraystretch}{1.2}
    \setlength{\extrarowheight}{0pt}
    \footnotesize
    \begin{tabular}{!{\vrule width 1pt}l|c|c|c|c|c|c|c|c|c|c!{\vrule width 1pt}}
        \specialrule{1pt}{0pt}{0pt}
        \# & Event Name & ${\cal M}$ & $q$  & $m_{1}$ & $m_{2}$ & $D_{\rm L}$ & $\chi_{eff}$ & ${\rm SNR}$  & ${\rm SNR}$ & ${\rm SNR}$ $\hat{\rho}$\\
        \textbf{} & \textbf{} & ($M_\odot)$ & \textbf{} & ($M_\odot)$ & ($M_\odot)$ & (Mpc) & \textbf{} & ${(\rm H1})$ &  ${(\rm L1})$ & \textbf{}(${\rm network}$) \\
        \specialrule{1pt}{0pt}{0pt}
        \rowcolor{lightblue}
        1 & GW190511\_125545 & $28.95^{+9.45}_{-6.86}$ & $0.72^{+0.25}_{-0.36}$ & $40.7^{+16.2}_{-10.5}$ & $28.2^{+11.6}_{-11.2}$ & $3707^{+3471}_{-2173}$ & $0.23^{+0.25}_{-0.29}$ &
        2.29 & 7.34 & 7.29 \\
        2 & GW190614\_134749 & $25.97^{+16.59}_{-6.20}$ & $0.70^{+0.27}_{-0.36}$ & $37.0^{+31.8}_{-10.7}$ & $25.2^{+15.2}_{-9.7}$ & $6551^{+9562}_{-3558}$ & $0.05^{+0.34}_{-0.34}$ &
        3.51 & 6.08 & 7.02 \\
        \rowcolor{lightblue}
        3 & GW190607\_083827 & $30.48^{+7.21}_{-4.68}$ & $0.78^{+0.19}_{-0.29}$ & $40.5^{+12.0}_{-7.6}$ & $31.0^{+9.1}_{-8.2}$ & $4928^{+2725}_{-2435}$ & $0.01^{+0.26}_{-0.30}$ &
        4.04 & 7.29 & 8.33 \\
        4 & GW190904\_104631  & $21.24^{+5.76}_{-4.40}$ & $0.64^{+0.31}_{-0.33}$ & $31.3^{+14.5}_{-8.5}$ & $19.7^{+7.1}_{-7.2}$ & $5614^{+4441}_{-2864}$ & $0.05^{+0.30}_{-0.37}$ &
        4.50 & 4.88 & 6.64 \\
        \rowcolor{lightblue}
        5 & GW190523\_085933 & $23.82^{+10.24}_{-7.95}$ & $0.49^{+0.45}_{-0.32}$ & $41.7^{+19.3}_{-15.5}$ & $19.4^{+14.6}_{-10.5}$ & $6091^{+6613}_{-3702}$ & $0.42^{+0.31}_{-0.45}$ &
        3.48 & 5.14 & 6.02 \\
        6 & GW200208\_211609 & $18.83^{+4.68}_{-3.18}$ & $0.69^{+0.28}_{-0.40}$ & $26.9^{+14.6}_{-6.3}$ & $18.0^{+6.4}_{-6.9}$ & $3669^{+3413}_{-1985}$ & $0.01^{+0.37}_{-0.37}$ &
        4.75 & 6.22 & 7.83 \\
        \rowcolor{lightblue}
        7 & GW190705\_164632 & $27.21^{+7.34}_{-5.24}$ & $0.52^{+0.41}_{-0.32}$ & $44.7^{+24.8}_{-12.8}$ & $23.0^{+11.7}_{-9.8}$ & $5692^{+4030}_{-2863}$ & $0.29^{+0.26}_{-0.34}$ &
        4.42 & 6.88 & 8.11 \\
        8 & GW190426\_082124 & $17.93^{+4.12}_{-3.42}$ & $0.45^{+0.45}_{-0.28}$ & $31.5^{+22.5}_{-11.3}$ & $13.8^{+6.9}_{-5.2}$ & $3213^{+4555}_{-1573}$ & $-0.01^{+0.39}_{-0.50}$ &
        5.15 & 4.46 & 6.41 \\
        \specialrule{1pt}{0pt}{0pt}
    \end{tabular}
\end{table*}

\subsection{Parameter estimation}
\label{sec:pe}

 We performed parameter estimation for all of our new candidate events, using Bilby \cite{bilby} (see Section \ref{sec:PE}) and assuming the IMRPhenomXPHM waveform model, which has the component black hole masses $m_{1,2}$, their respective dimensionless spin magnitutes $s_{1,2}$, the tilt angle between their spins and the orbital angular momentum $\theta_{1,2}$, and the two spin vectors describing the azimuthal angle separating the spin vectors $\delta\phi$ and the cone of precession about the system's angular momentum 
 as the intrinsic parameters. Together with the extrinsic parameters which are the luminosity distance $d_{L}$, the right ascension $\alpha$ and declination $\delta$, the inclination $\iota$ of the system with respect to the observer's line of sight, the polarization angle $\psi$, the coalescence phase $\phi_c$ and time $t_{c}$, this amounts to a total of 15 parameters that were inferred. In Table \ref{table:Parameter_Estimation}, we display only the main parameters inferred for each new candidate event: the source-frame chirp mass ($\cal M$), the component masses ($m_{1}$ and $m_{2}$), the mass ratio ($q$), the luminosity distance ($D_{L}$), the effective spin ($\chi_{eff}$) and the reweighted network SNR ($\hat \rho$). 

Based on the median values of the individual masses and the chirp mass, we find that all our new candidate events fall within the effective training data range of AresGW and are detected with an reweighted network SNR in the range $6.22 \leq \hat{\rho} \leq 8.33$. This range is not surprising. Other algorithms have already detected events with higher network SNR, which we have also confirmed in Section \ref{sec:Detection_of_known_GW_Events}. Moreover, the other algorithms evaluate the data of each detector independently and impose a threshold on the individual SNR they consider acceptable. For some of our new events, with low SNR values at the Hanford detector, this could explain why these algorithms did not detect them. Note that most of the injections on which AresGW was trained had a single-detector SNR $< 4$ (see \cite{AresGW_model}), which explains the good sensitivity of AresGW to low-SNR signals. In Section \ref{sec:Population} we analyze further the distribution of the network SNR for our new candidate events, compared to the events detected by other algorithms.

In Appendix \ref{AppendixA}, we present corner plots of the 1-D and 2-D posterior distributions of the main parameters for all 8 new candidate events. The values shown for each 1-D distribution correspond to the median, along with the marginalized values for the 90\% confidence interval. Since Bilby assumes a Gaussian likelihood, it is expected that the resulting posterior distributions will be Gaussian-like, in the absence of loud non-Gaussian noise features. Hence, a prominent Gaussian morphology of the posterior distributions favors the robustness of the parameter estimation. For all of our new candidate events, the Bayes factor was larger than unity.

\begin{figure*}[t!]
  \centering
  \includegraphics[width=0.99\linewidth]{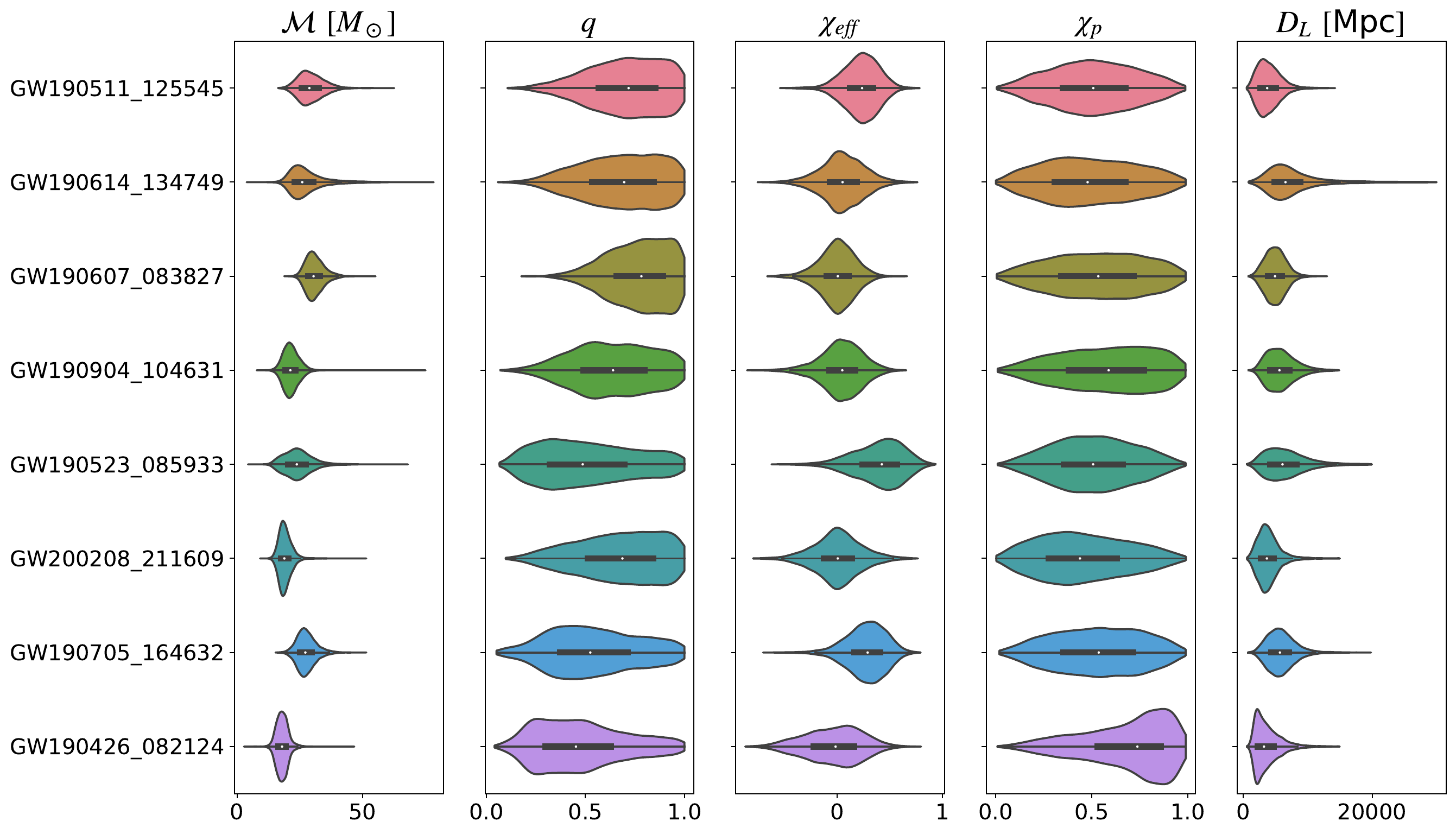}
  \caption{Marginal probability distributions for the source chirp mass ${\cal M}$, mass ratio $q$, effective inspiral spin $\chi_{\rm eff}$, effective precession spin $\chi_{p}$ and luminosity distance $D_{\rm L}$ for the new candidate events identified by AresGW.}  \label{fig:violin_plot}
\end{figure*}

In Fig. \ref{fig:violin_plot} we present the marginal posterior probability distributions corresponding to all 8 new candidate events in the form of violin plots. 
The morphology and choice of the showcased parameters aligns with the respective version included in GWTC-3.

\subsection{Population properties of new GW candidate events} \label{sec:Population}

In the following section, we will examine the distribution of various properties of our newly identified candidate events and compare them to the properties of previously published BBH  merger events. We stress that we are only comparing {\it median} values, and one should take into account the uncertainties in the corresponding posterior distributions to arrive at quantitative statements.  

The chirp mass distribution of BBH mergers, as inferred in \cite{Mchirp_distribution_ref}, using observations in the GWTC-2 catalog \cite{GWTC2},  shows multiple peaks, starting around 8 $M_\odot$. Beyond this primary peak, secondary peaks were inferred in \cite{Mchirp_distribution_ref} at approximately 14 $M_\odot$ and 26 $M_\odot$, while an outlier appeared at $\sim 45$ $M_\odot$. For further astrophysical considerations regarding the expected chirp mass distribution, see, e.g. \cite{2023ApJ...950L...9S}.  

\begin{figure}[t]
  \centering
  \includegraphics[width=0.99\linewidth]{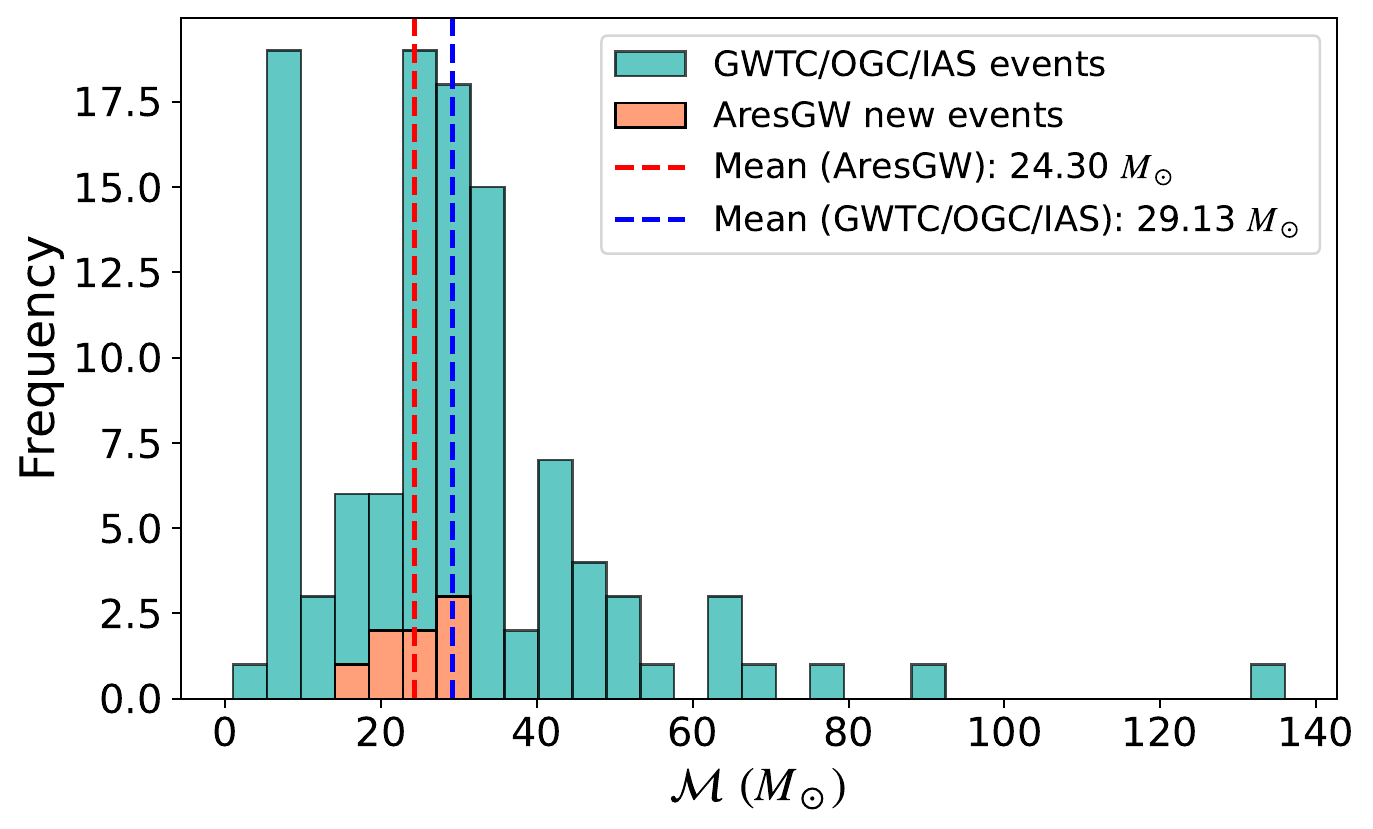}
    \caption{
    Distribution of chirp masses (${\cal M}$) for the new candidate events indentified by AresGW (orange), compared to the corrsponding distribution of previously published events (green).}
    \label{fig:Mchirp_Comparison_Histogram}
\end{figure}

\begin{figure}[t]
  \centering
  \includegraphics[width=0.99\linewidth]{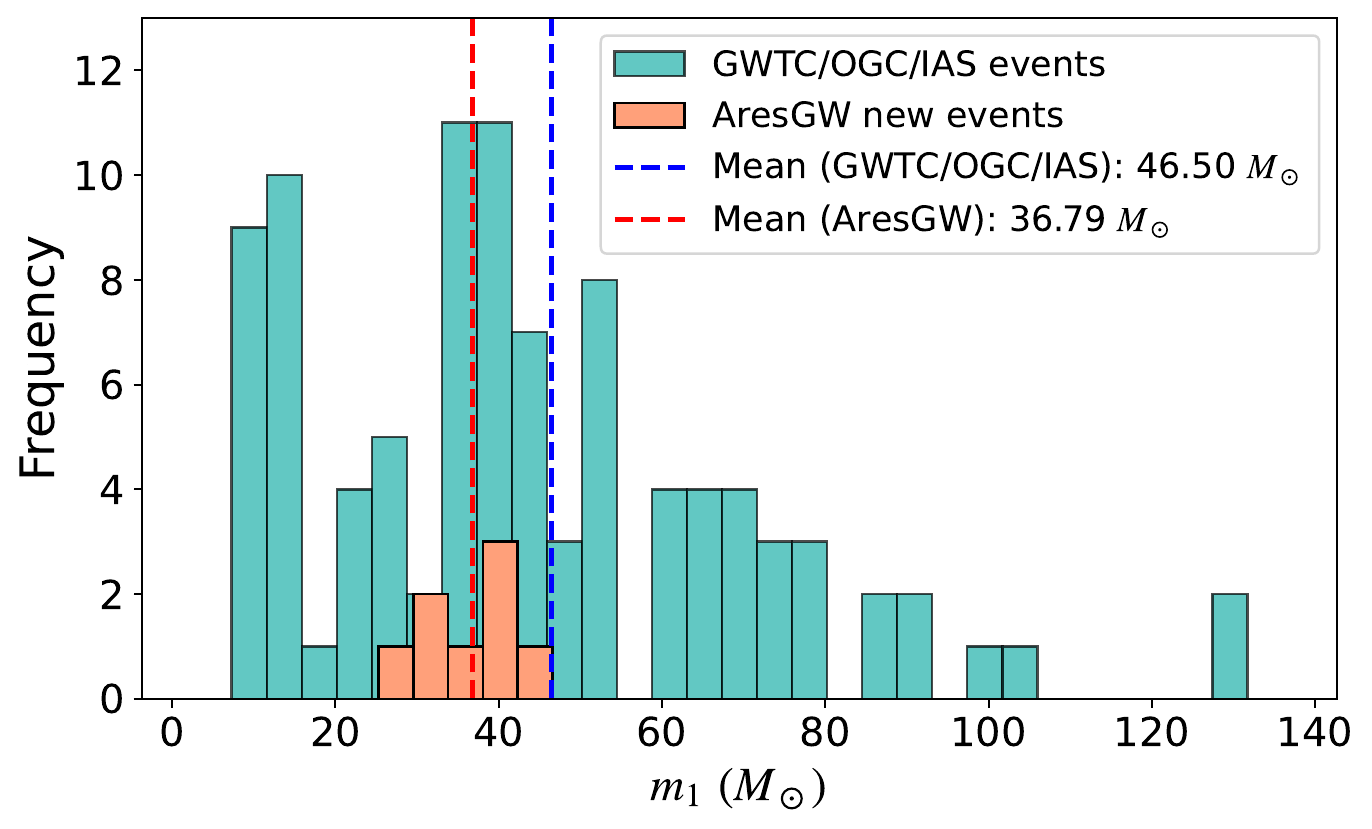}
    \caption{Same as Fig. \ref{fig:Mchirp_Comparison_Histogram}, but for $m_1$.}
    \label{fig:m1_Comparison_Histogram}
\end{figure}

By incorporating known events detected by both LIGO detectors, as reported in all previous catalogs (GWTC, OGC, and IAS), we obtain the histogram of chirp masses shown in Fig.~\ref{fig:Mchirp_Comparison_Histogram}. This distribution shows a mean of 29.13 $M_\odot$, and the majority of events previously detected are distributed around this value. Additionally, there is a secondary peak for masses smaller than 10 $M_\odot$, while the frequency of samples continuously decreases beyond the highest peak.

In Fig.~\ref{fig:Mchirp_Comparison_Histogram} we also display the distribution of our new candidate events, which is Gaussian-like with a mean of 24.30 $M_\odot$. This specific value can be attributed to two main factors. Firstly, the astrophysical distribution of the sources, as implied by the previously published events, appears to have a peak near this value. Additionally, AresGW likely became more sensitive around the mean of the training data, which closely aligns, by design, with this value. The fact that \emph{our distribution aligns with that from the other catalogs} serves as a positive indicator for the astrophysical origin  of our new candidate events.  Similar conclusions can be drawn when one examines the distribution of the primary mass, $m_1$, see Fig. \ref{fig:m1_Comparison_Histogram}.

\begin{figure}[t]
  \centering
  \includegraphics[width=0.99\linewidth]{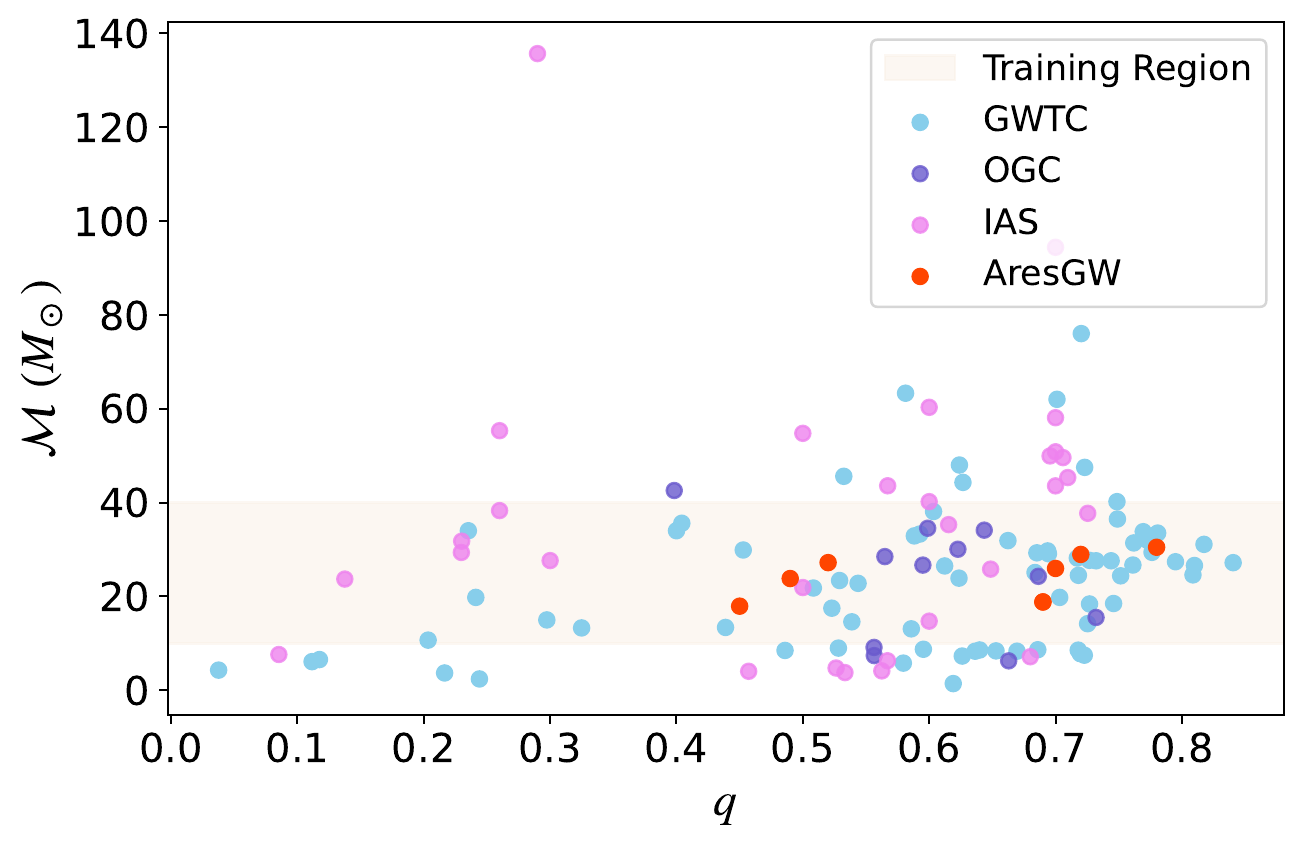}
    \caption{Chirp mass ${\cal M}$ vs. mass ratio $q$ for the new candidate events identified by AresGW (red dots) in comparison to published events in the GWTC, OGC and IAS catalogs. The effective traning region of AresGW is also shown.}
    \label{fig:M-q}
\end{figure} 

\begin{figure}[t]
  \centering
  \includegraphics[width=0.99\linewidth]{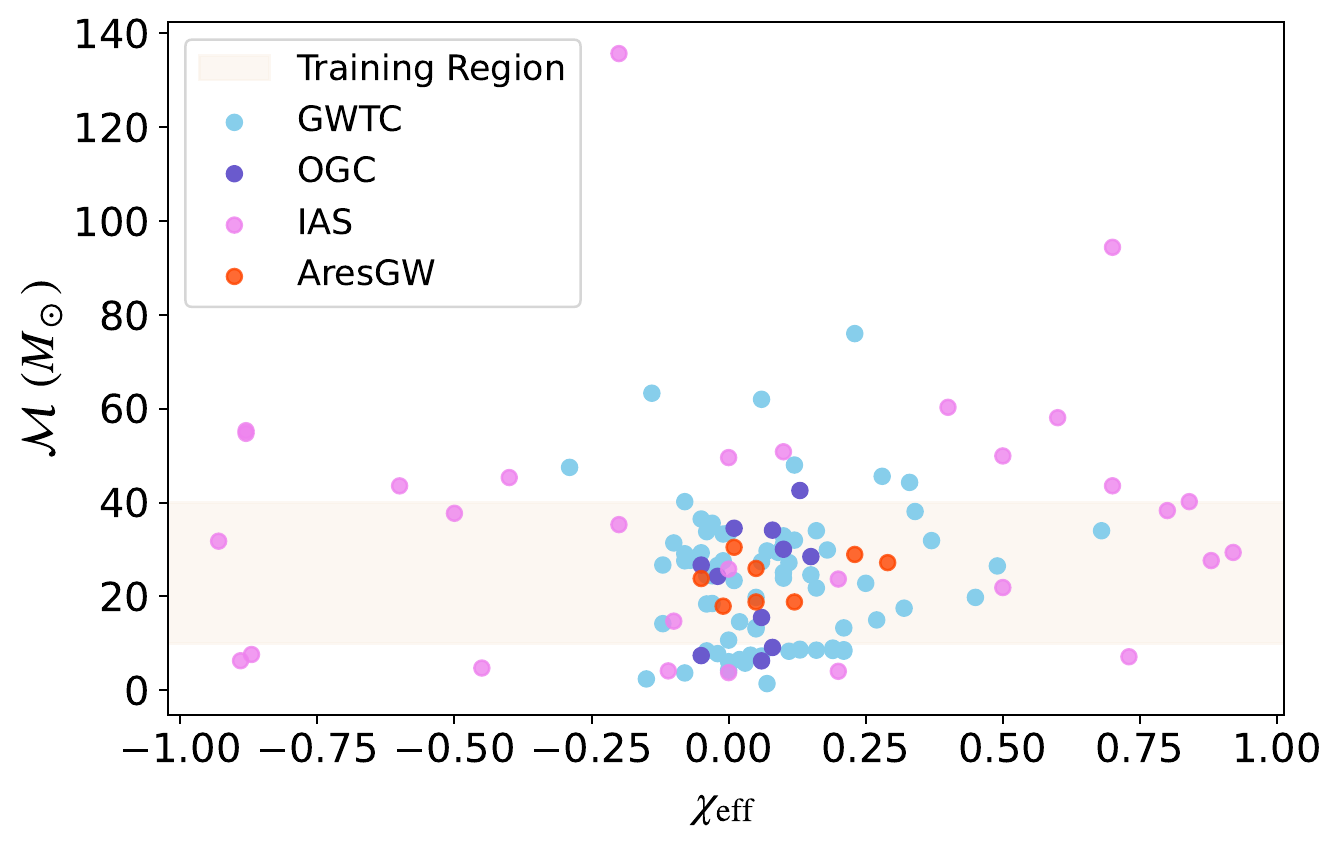}
    \caption{Same as Fig. \ref{fig:M-q}, but for the chirp mass ${\cal M}$ vs. effective inspiral spin $\chi_{\rm eff}$.}
    \label{fig:M-Xeff}
\end{figure} 

\begin{figure}[t]
  \centering
  \includegraphics[width=0.99\linewidth]{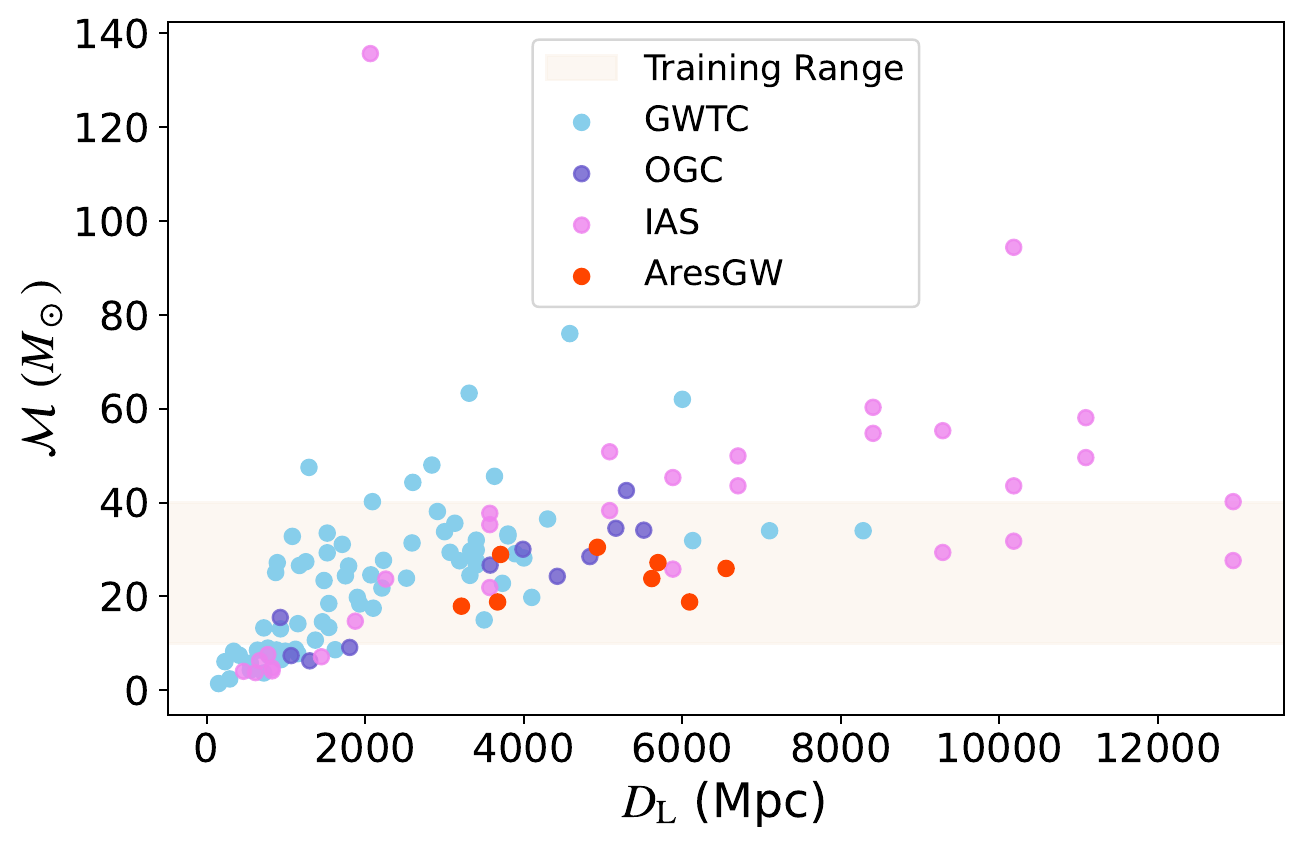}
    \caption{Same as Fig. \ref{fig:M-q}, but for the chirp mass ${\cal M}$ vs. luminosity distance $D_{\rm L}$ .}
    \label{fig:M-D}
\end{figure} 

Actually, the broad agreement between the distribution of the new AresGW candidate events and those previously detected by other algorithms is evident through various comparisons. For example, in Fig. \ref{fig:M-q}, the distribution of the chirp mass $\cal M$ vs the mass ratio $q$  for our new candidate events is in agreement with the corresponding distribution for most previously published events. 
Similarly, the chirp mass $\cal M$ vs. effective spin $\chi_{\rm eff}$ distribution of our new candidate events is comparable to the corresponding distribution of the events listed in the catalogs of the GWTC and OGC catalogs. The distribution of the IAS is wider, as depicted in Fig. \ref{fig:M-Xeff}.

\begin{figure}[t]
  \centering
  \includegraphics[width=0.99\linewidth]{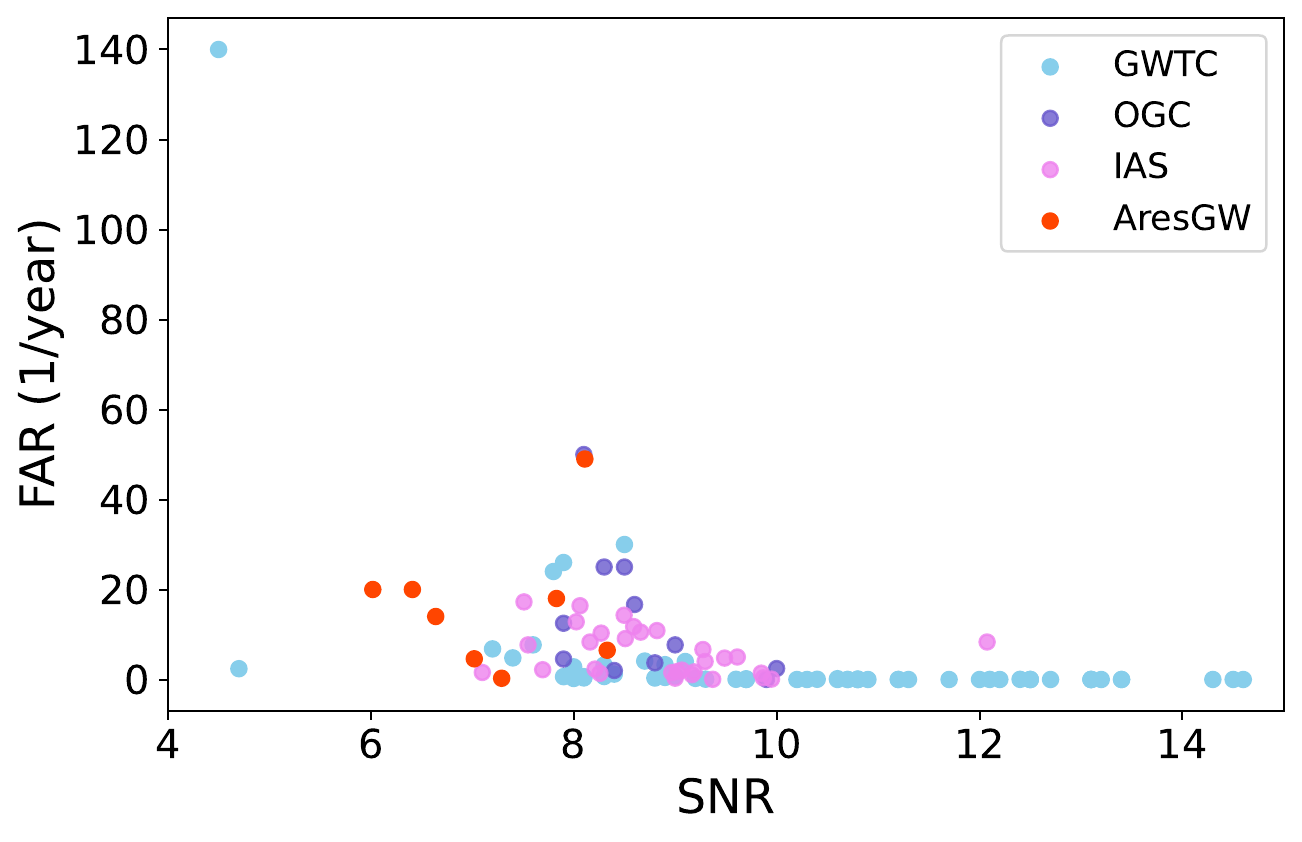}
    \caption{False alarm rate  vs. signal-to-noise ratio for the new candidate events identified by AresGW (red dots) in comparison to published events in the GWTC, OGC and IAS catalogs.}
    \label{fig:FAR-SNR}
\end{figure} 

However, our new candidate events exhibit some distinctive characteristics. For example, as illustrated in Fig. \ref{fig:M-D}, our candidate events tend to exhibit higher luminosity distances $D_{\rm L}$ compared to the average of previously confirmed events. This observation underscores the high sensitive distance of AresGW.

The successful detection of distant, high-redshift black hole binaries can provide valuable insights into the evolution of their merger rate over different redshifts \cite{redshift}. This can further support the potential correlation with other properties \cite{spinshift}.
 All of our new candidate events exceed the median value of $D_{\rm L} \sim 3000$ Mpc for published GW events. Interestingly, our new candidate event GW190614\_144749 constitutes the third most distant BBH merger ever detected in AresGW's effective training range.

Moreover, our new candidate events tend to have a lower reweighted network SNR $\hat \rho$ compared to those found in the GWTC, OGC, and IAS catalogs. As observed in Fig. \ref{fig:FAR-SNR}, there is a general tendency for detected events to exhibit a higher FAR at lower SNR. The majority of our new candidate events follows this trend. However, our 
new candidate events can reach lower SNR compared to other algorithms, while maintaining a comparable FAR.

This observation suggests that, although our new candidate events have a relatively low signal-to-noise ratio, our ML-based detection pipeline has effectively detected them. This increased sensitivity and efficacy in event detection distinguish AresGW from traditional detection algorithms.

\section{Conclusions and Discussion} 
The AresGW neural network has undergone enhancements, including the utilization of a larger dataset for training, double precision for results, and a 350 Hz low-pass filter for both training and detection. Furthermore, the development of new filters for data analysis, coupled with the implementation of glitch removal techniques, utilizing the capabilities of Gravity Spy, has resulted in the categorization of triggers into three distinct classes: Default Low-pass, Selective Noise Rejection, and Selective Passband  triggers. The second and third classes employ different filters, such as noise filters of high-frequency cut-offs at 400 Hz or 500 Hz, to reduce false alarms. As a result, false alarms were reduced by at least 70\% in the first class and 90\% in the second class, when compared to false alarms in the first class alone.  

AresGW utilizes two input channels and is trained on BBH merger signal injections in O3a noise from the two LIGO detectors. In the effective training range, we considered events detected when both the Livingston and Hanford detectors were operational during the O3 period, with source masses in the range $m_{1,2} \in (7 M_\odot, 50 M_\odot)$ and chirp mass in the range ${\cal M} \in (10 M_\odot, 40 M_\odot)$.  With the aforementioned improvements, AresGW successfully confirmed 34 of the 43 signals in its effective training range, that were previously detected by the GWTC, OGC, or IAS algorithms. The four out of the nine undetected signals had a rather low published astrophysical probability of $p_{\text{astro}} \leq 0.63$.

Beyond the confines of the training dataset, a total of 55 detections were recorded when both LIGO detectors were active. Our algorithm identified 10 of these detections with $p_{\rm astro} \geq 0.99$, showcasing the resilience of our neural network in detecting gravitational wave signals even beyond its training dataset. This underscores the potential for AresGW to maintain high sensitivity in various mass ranges with appropriate training.

In addition to identifying published events in the GWTC, OGC, and IAS catalogs, our machine algorithm detected 8 new gravitational wave candidate events within its effective training range. With the addition of these new candidate events,  AresGW reached a total of 42 detections in its effective training range, making AresGW the most efficient detection pipeline in this range. Parameter estimation for all these signals was performed using Bilby, while the appropriate time and consistency tests were also successfully passed. 

Furthermore, we compared our new candidate events with previously detected signals and found that their distribution aligns with those from other catalogs, a positive indicator that our new candidate events are likely of astrophysical origin. A distinctive characteristic of these new events is that they exhibit luminosity distances higher than the average value for previously confirmed events, further highlighting the sensitivity of AresGW. Furthermore, our new candidate events tend to have a lower signal-to-noise ratio than previously detected signals. Here again, our pipeline distinguishes itself by reaching lower signal-to-noise ratio levels at a false alarm rate comparable to other algorithms. 

AresGW also confirms several events with high astrophysical probabilities: GW190916\_200658, originally detected by mbta with \( p_{\text{astro}} = 0.66 \) and by pycbc\_bbh with \( p_{\text{astro}} = 0.65 \), now confirmed with \( p_{\text{astro}} = 1.00 \); GW200305\_084739, found by OGC with \( p_{\text{astro}} = 0.59 \), is now confirmed with \( p_{\text{astro}} = 1.00 \); GW190906\_054335, detected by IAS with \( p_{\text{astro}} = 0.61 \), is now confirmed with \( p_{\text{astro}} = 0.99 \); and GW200106\_134123, found by OGC with \( p_{\text{astro}} = 0.69 \), is now confirmed with \( p_{\text{astro}} = 0.95 \).

Moreover, to evaluate the adaptability of AresGW, we also applied it on data in which the Virgo detector was operational at the same time of one of the other two LIGO detectors, without any retraining. Our findings show that AresGW can adapt to Virgo's noise characteristics and, when coupled with an additional detector, it effectively identifies signals burried in noise. In fact, AresGW successfully detected five of the eight events in the data with a ranking statistic $\langle{\mathcal R_s}\rangle$ exceeding 3.5 (the threshold used in the analysis presented above). This underscores once again the significant generalizability of our network.

The analysis of AresGW's adaptability to Virgo's noise led us to investigate its generalization capacity by testing it on data from the O1 and O2 observing periods, too. Our results revealed that our machine learning code successfully validated all, but one, events previously detected during O2, achieving a maximum value $\langle{\mathcal R_s}\rangle$ of 16.0. It also identified the inaugural gravitational wave event, GW150914, from the O1 period with the same ranking statistic. This indicates that AresGW effectively detected almost all events within its training range in both O1 and O2 periods. Notably, the neural network's ability to generalize effectively also on data from those observing periods underscores its generalization capability, suggesting the potential for detecting new gravitational waves in O1 and O2 data without necessarily  retraining AresGW, as well as its potential applicability to various phenomena characterized by time series with non-Gaussian noise. Furthermore, this bodes well for the O4 observing run, anticipated to benefit from significantly reduced noise levels in the LIGO detectors. 

Looking ahead, there are several avenues to further advance the capabilities of the AresGW neural network. For example, retraining AresGW on the noise characteristics of each observing run could enhance its adaptability and detection efficiency.
Another avenue for exploration lies in training the algorithm on different combinations of detectors, including single, double, and triple detector setups. In particular, the implementation of a three-channel AresGW pipeline holds significant promise, as it would leverage data from multiple detectors to improve detection accuracy and reduce false alarms. Furthermore, directly integrating glitch identification in  AresGW, could help mitigate false positives by excluding glitches from trigger selections.

Additionally, the utilization of a four-channel network, incorporating both time series and spectrogram data, presents an intriguing opportunity to enhance the algorithm's sensitivity and precision. By training the network to dynamically adjust low-pass filters based on spectrogram analysis, we could potentially optimize detection thresholds for various signal segments, further refining AresGW's performance in capturing gravitational wave events.

Moreover, one more promising avenue is to extend the application of AresGW to the detection of gravitational waves emitted from the merger of compact binary systems involving at least one neutron star, to supernovae, and other astrophysical events. Neutron star mergers, in particular, are of significant interest, due to their potential to produce observable electromagnetic counterparts, known as kilonovae, enabling multimessenger studies.

\section{Acknowledgements}
We are grateful to Thomas Dent, Melissa Lopez and Viola Sordini for useful discussions. We are indebted to Derek Davis, Thomas Dent and Agata Trogato for detailed comments that improved the manuscript. We are also grateful to Andrea Virtuoso and Edoardo Milotti, for making their Qp-transform code publicly available and to the organizers of the MLGWSC-1 challenge, for providing routines for creating different datasets.
We acknowledge the support
provided by the IT Center of the Aristotle University of
Thessaloniki (AUTh) as our results have been produced
using the AUTh High-Performance Computing Infrastructure and Resources. We also acknowledge support
from COST Action (European Cooperation in Science
and Technology) CA17137 (G2Net).
This research has made use of data or software obtained from the Gravitational Wave Open Science Center (gw-openscience.org), a service of LIGO Laboratory,
the LIGO Scientific Collaboration, the Virgo Collaboration, and KAGRA. LIGO Laboratory and Advanced
LIGO are funded by the United States National Science Foundation (NSF) as well as the Science and Technology Facilities Council (STFC) of the United Kingdom, the Max-Planck-Society (MPS), and the State of
Niedersachsen/Germany for support of the construction
of Advanced LIGO and construction and operation of
the GEO600 detector. Additional support for Advanced
LIGO was provided by the Australian Research Council. Virgo is funded, through the European Gravitational Observatory (EGO), by the French Centre National de Recherche Scientifique (CNRS), the Italian Istituto Nazionale di Fisica Nucleare (INFN) and the Dutch
Nikhef, with contributions by institutions from Belgium,
Germany, Greece, Hungary, Ireland, Japan, Monaco,
Poland, Portugal, Spain. The construction and operation of KAGRA are funded by Ministry of Education,
Culture, Sports, Science and Technology (MEXT), and
Japan Society for the Promotion of Science (JSPS), National Research Foundation (NRF) and Ministry of Science and ICT (MSIT) in Korea, Academia Sinica (AS)
and the Ministry of Science and Technology (MoST) in
Taiwan.

\newpage

\appendix

\section{Corner Plots of AresGW New Detections}
\label{AppendixA}

\begin{figure}[t!]
  \centering
  \includegraphics[width=0.99\linewidth]{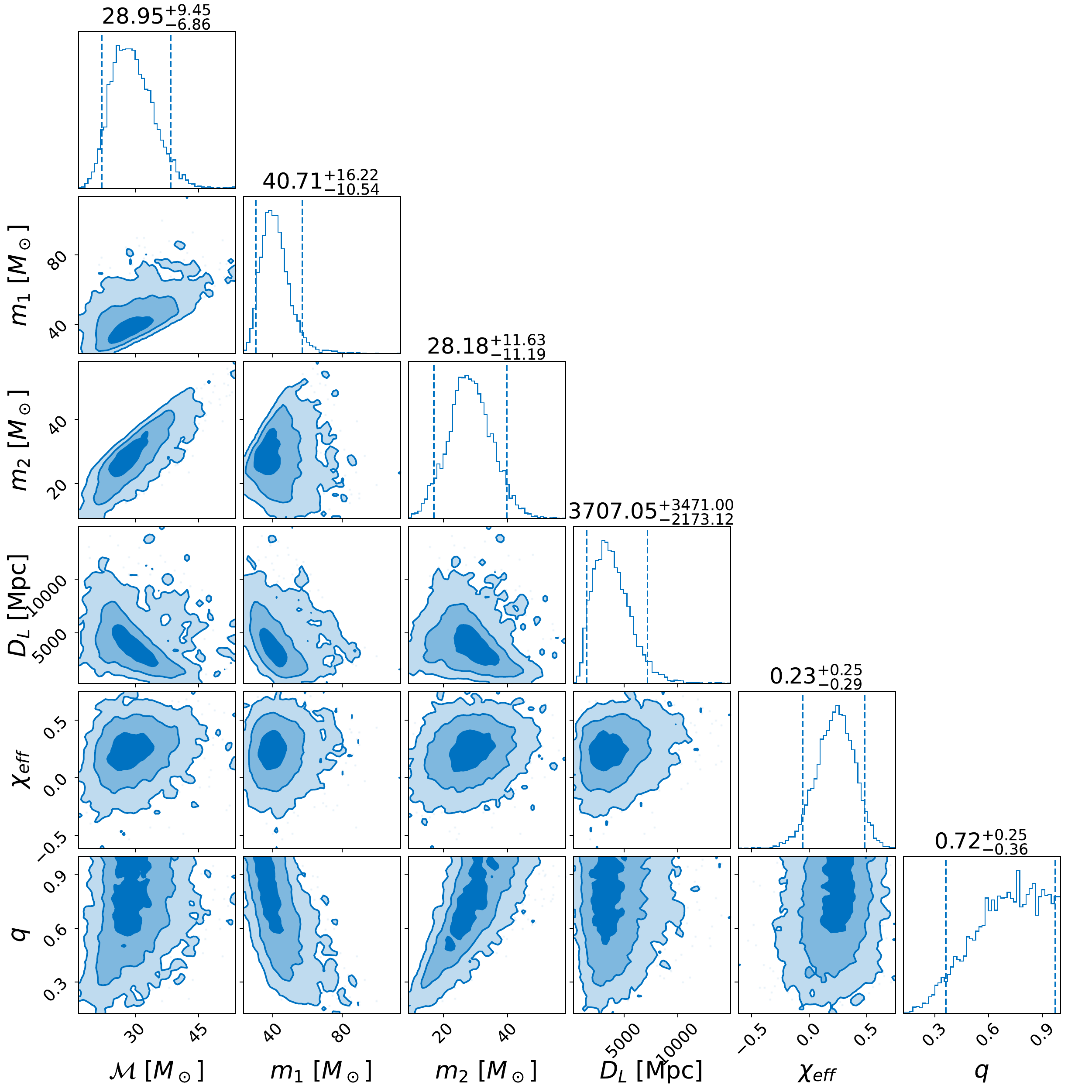}
    \caption{1-D and 2-D Marginal Posterior Distributions corresponding to the inferred parameters of the new event  GW190511\_125545, identified by AresGW.}
    \label{fig:gw190511post}
\end{figure}

\begin{figure}[t!]
  \centering
  \includegraphics[width=0.99\linewidth]{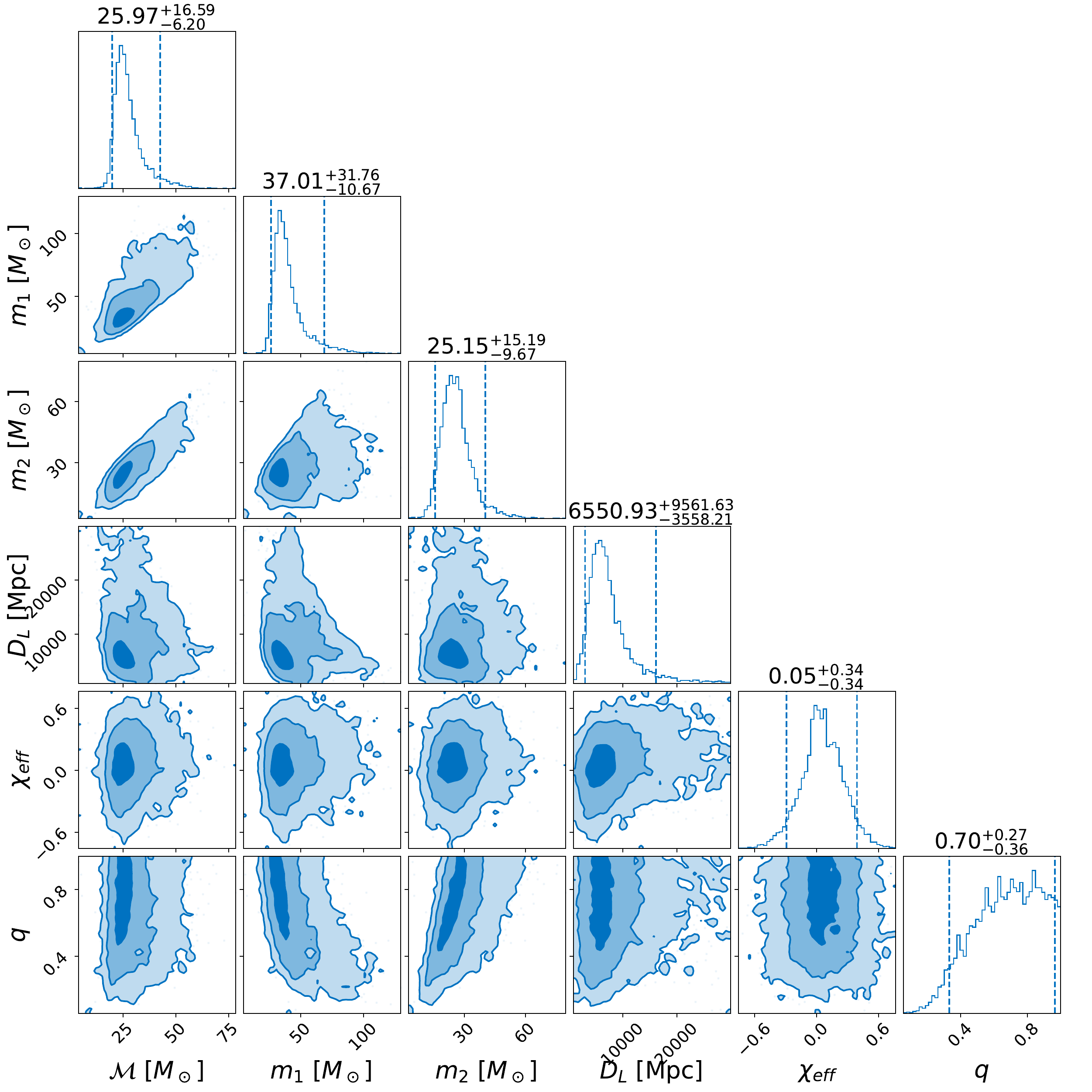}
    \caption{Same as Fig. \ref{fig:gw190511post}, but for the new event GW190614\_134749.}
    \label{fig:gw190614post}
\end{figure}

\begin{figure}[t!]
  \centering
  \includegraphics[width=0.99\linewidth]{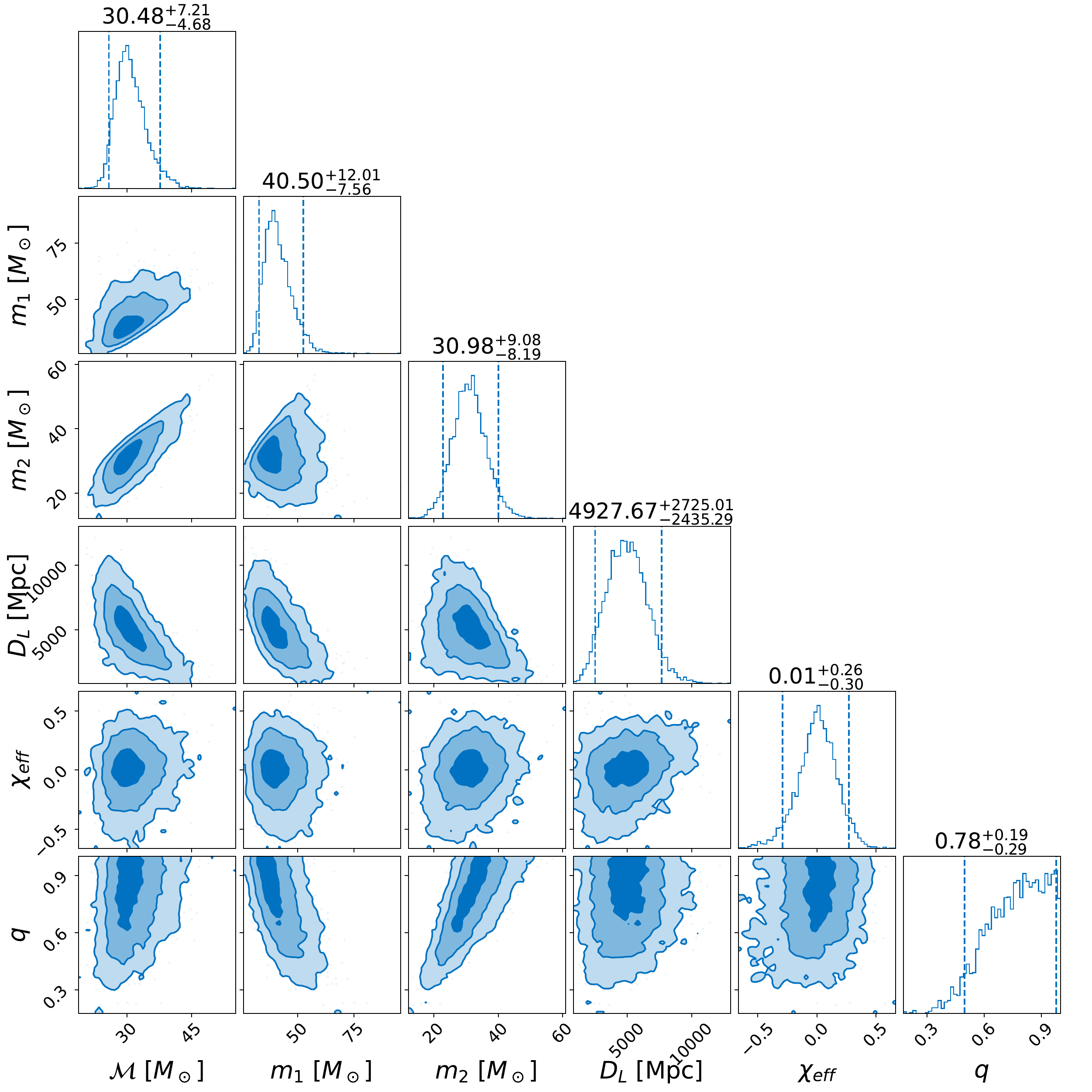}
    \caption{Same as Fig. \ref{fig:gw190511post}, but for the new event GW190607\_083827.}
    \label{fig:gw190607post}
\end{figure}

\begin{figure}[t!]
  \centering
  \includegraphics[width=0.99\linewidth]{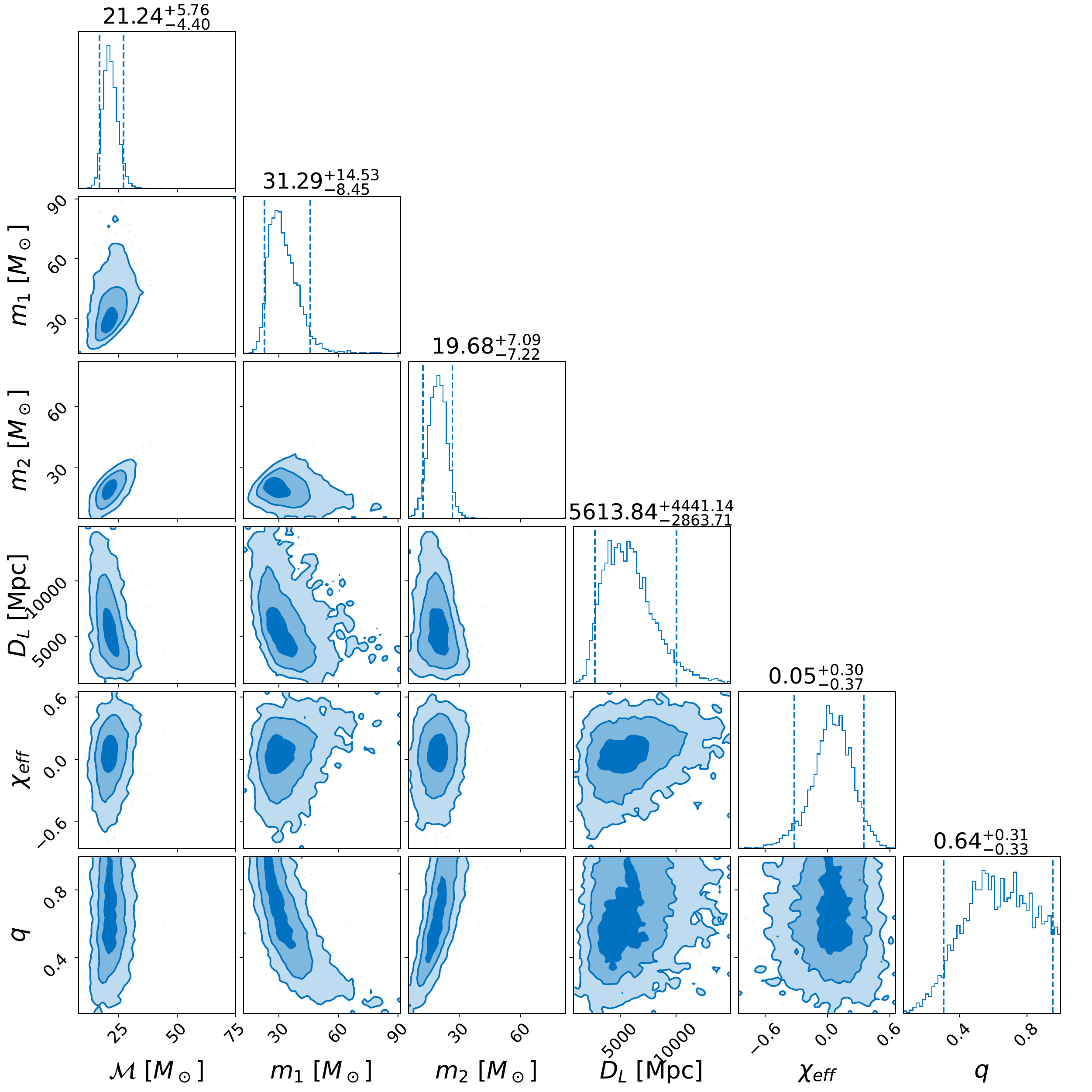}
    \caption{Same as Fig. \ref{fig:gw190511post}, but for the new event GW190904\_104631.}
    \label{fig:gw190904post}
\end{figure}

\begin{figure}[t!]
  \centering
  \includegraphics[width=0.99\linewidth]{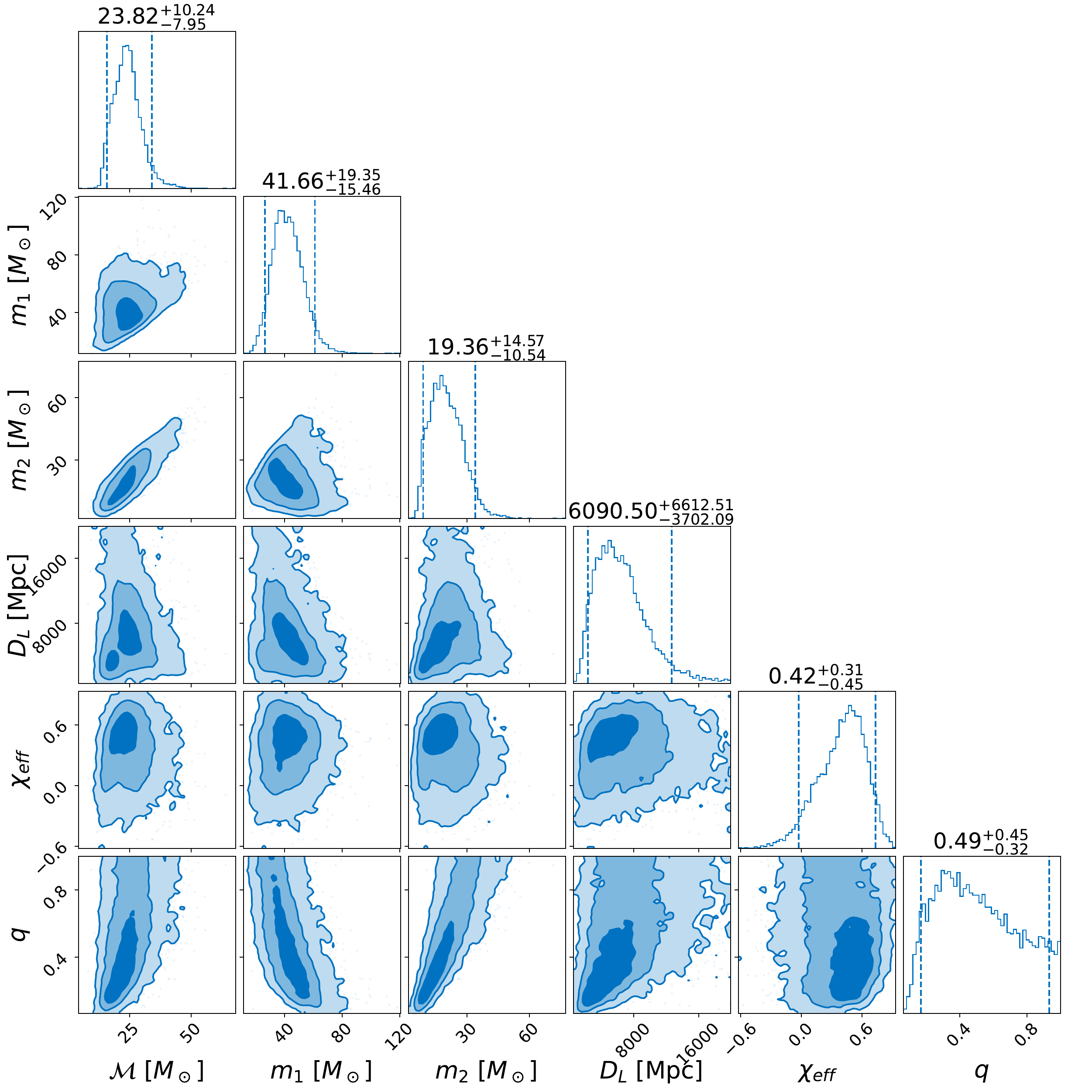}
    \caption{Same as Fig. \ref{fig:gw190511post}, but for the new event GW190523\_085933.}
    \label{fig:gw190523post}
\end{figure}

\begin{figure}[t!]
  \centering
  \includegraphics[width=0.99\linewidth]{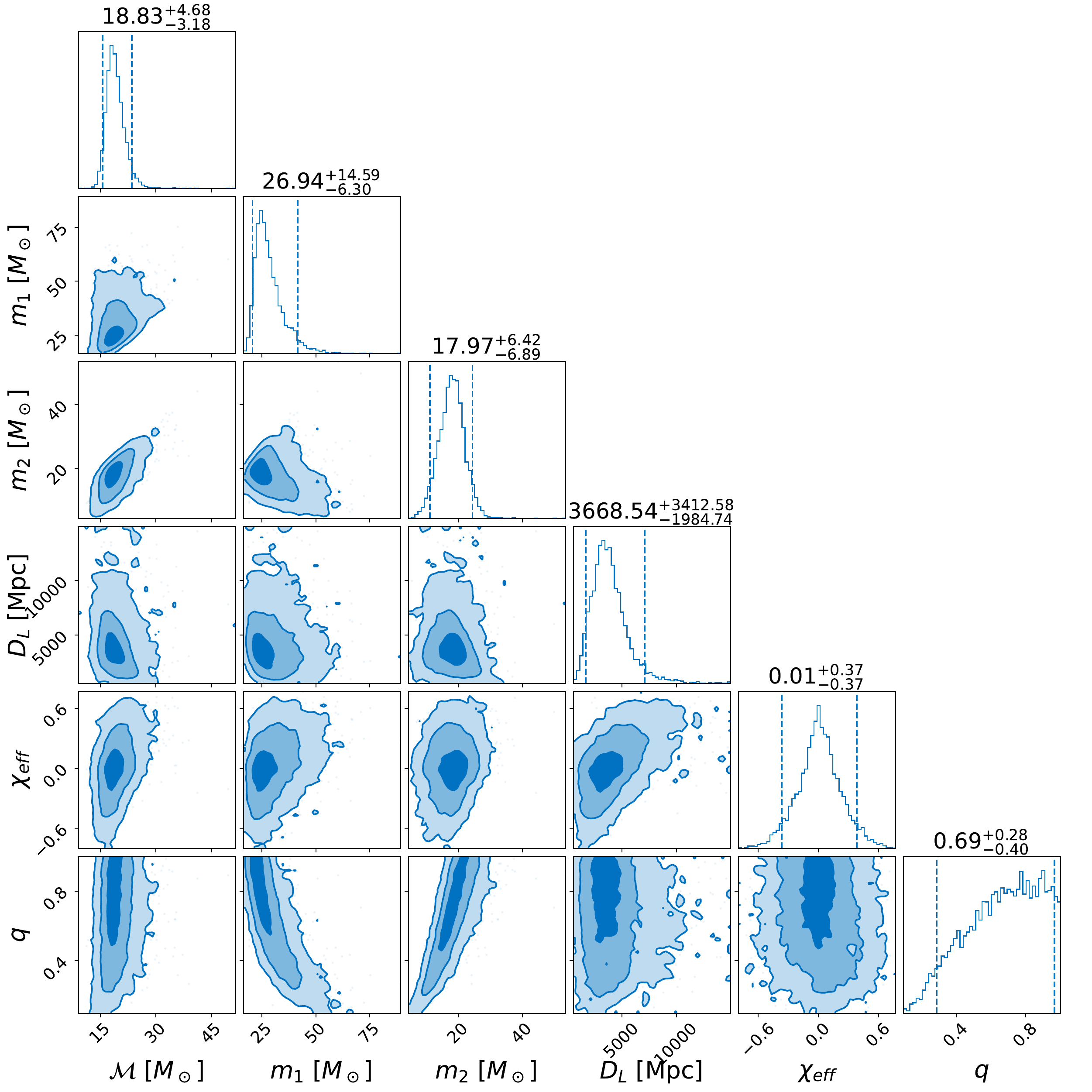}
    \caption{Same as Fig. \ref{fig:gw190511post}, but for the new event GW200208\_211609.}
    \label{fig:gw200208post}
\end{figure}

\begin{figure}[t!]
  \centering
  \includegraphics[width=0.99\linewidth]{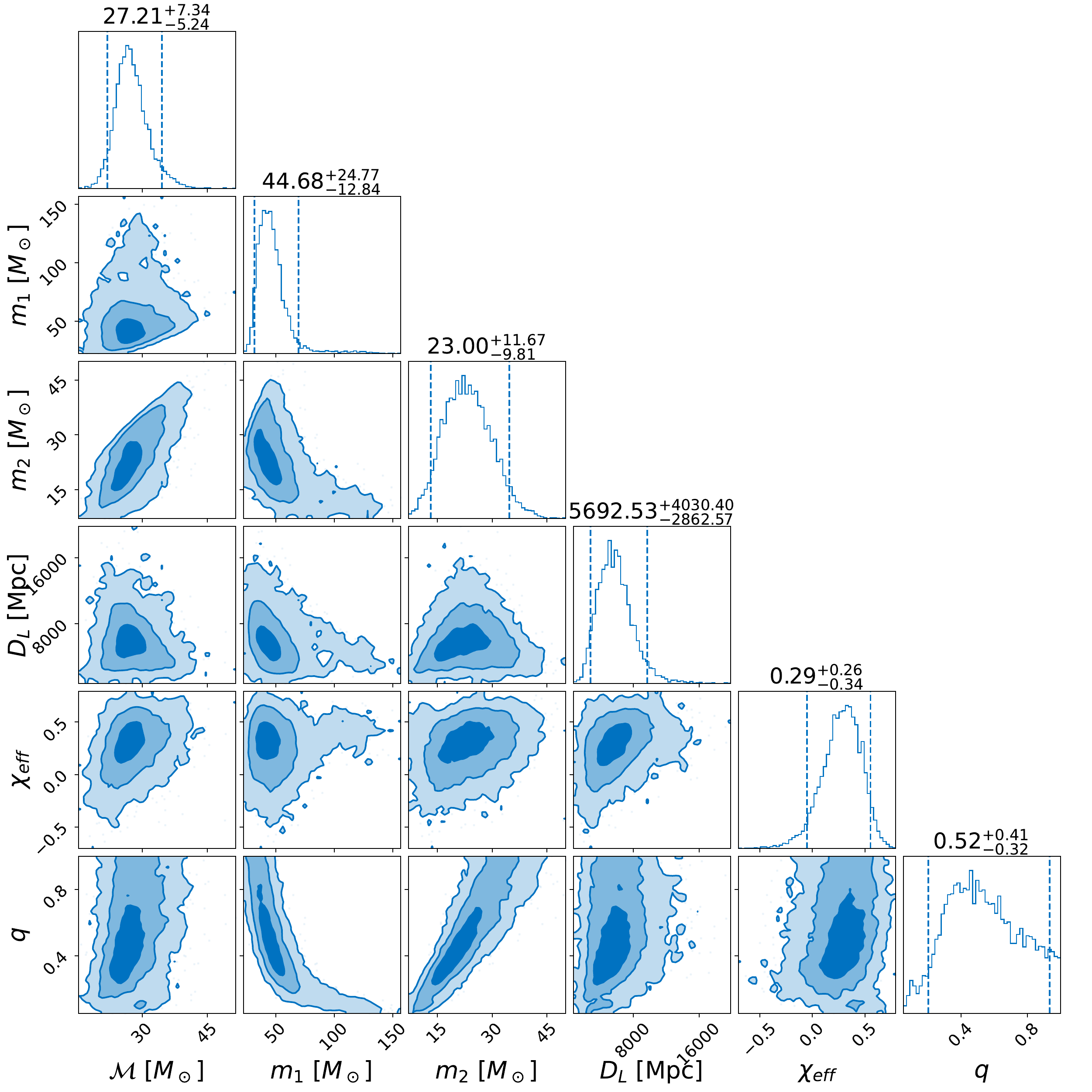}
    \caption{Same as Fig. \ref{fig:gw190511post}, but for the new event GW190705\_164632.}
    \label{fig:gw190705post}
\end{figure}

\begin{figure}[t!]
  \centering
  \includegraphics[width=0.99\linewidth]{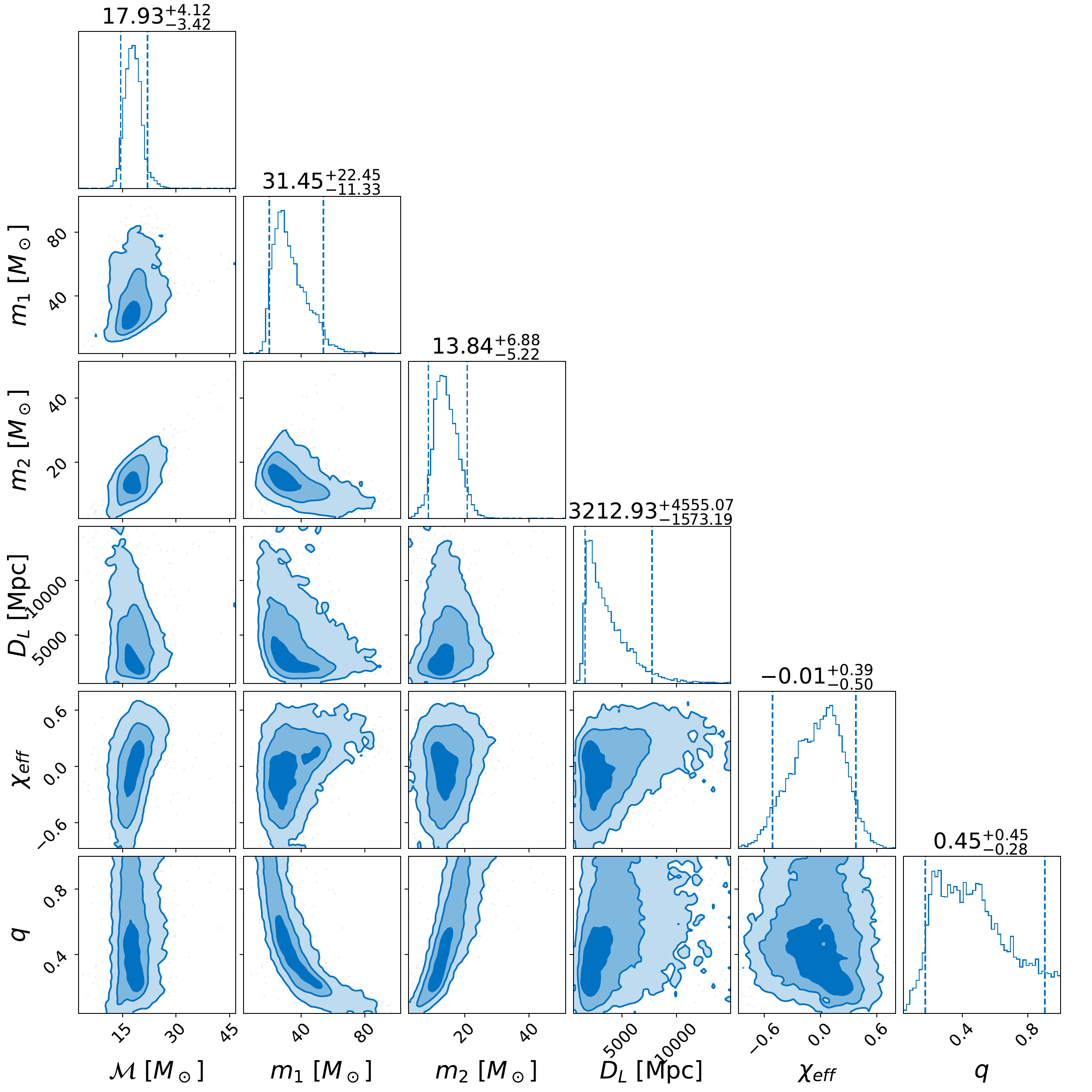}
    \caption{Same as Fig. \ref{fig:gw190511post}, but for the new event GW190426\_082124.}
    \label{fig:gw190426post}
\end{figure}

Figs. \ref{fig:gw190511post} -- \ref{fig:gw190426post} display corner plots of the most important parameters for the 8 new GW candidate events, identified by AresGW.

\newpage

\section{Time-Domain Reconstruction of AresGW New Detections}
\label{AppendixB}

\begin{figure}[t!]
  \centering
  \includegraphics[width=0.99\linewidth]{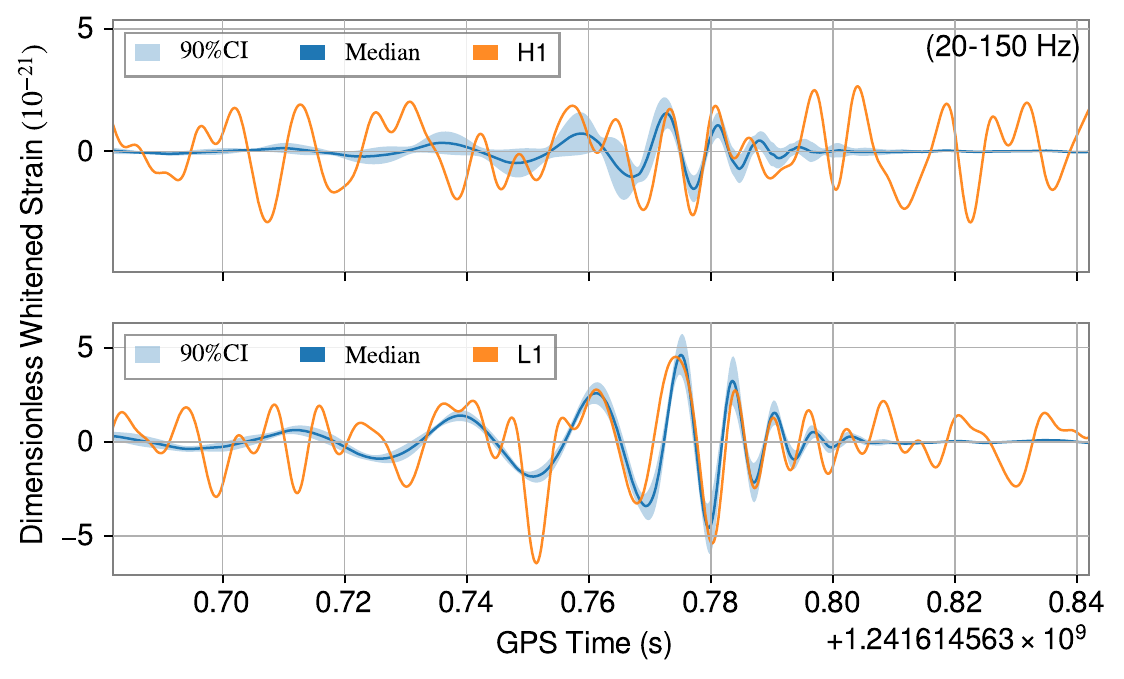}
    \caption{Whitened, bandpassed strain data and reconstructed waveform for the new event
     GW190511\_125545 identified by AresGW.}
    \label{fig:gw190511rec}
\end{figure}

\begin{figure}[t!]
  \centering
  \includegraphics[width=0.99\linewidth]{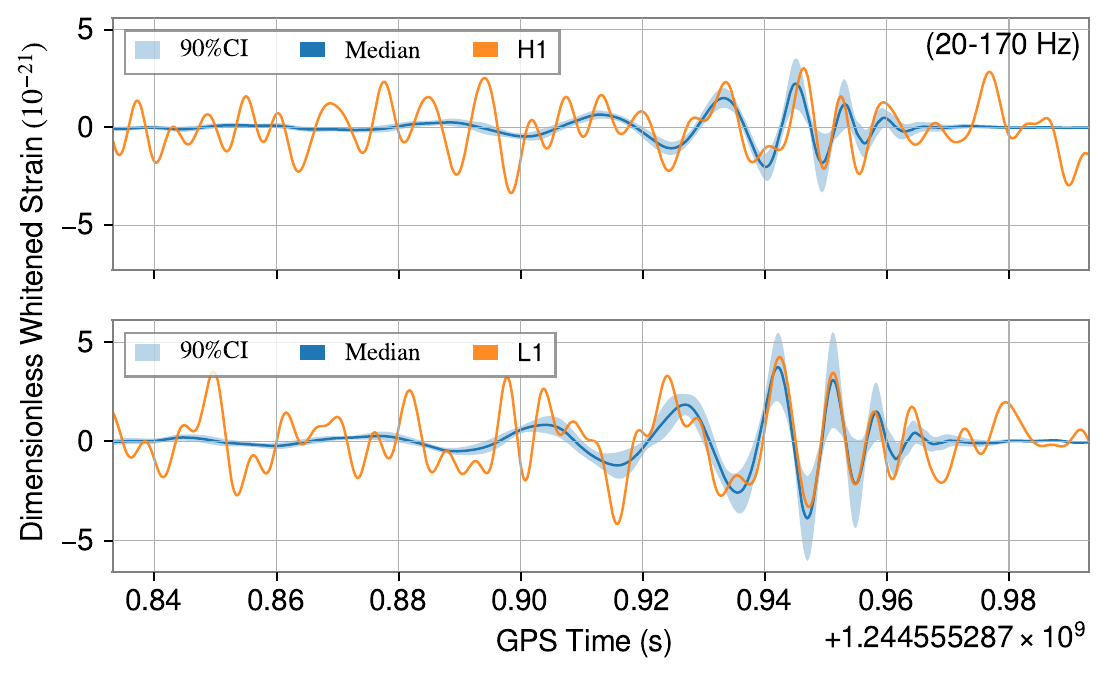}
    \caption{Same as Fig. \ref{fig:gw190511rec}, but for the new event GW190614\_134749.}
    \label{fig:gw190614rec}
\end{figure}

\begin{figure}[t!]
  \centering
  \includegraphics[width=0.99\linewidth]{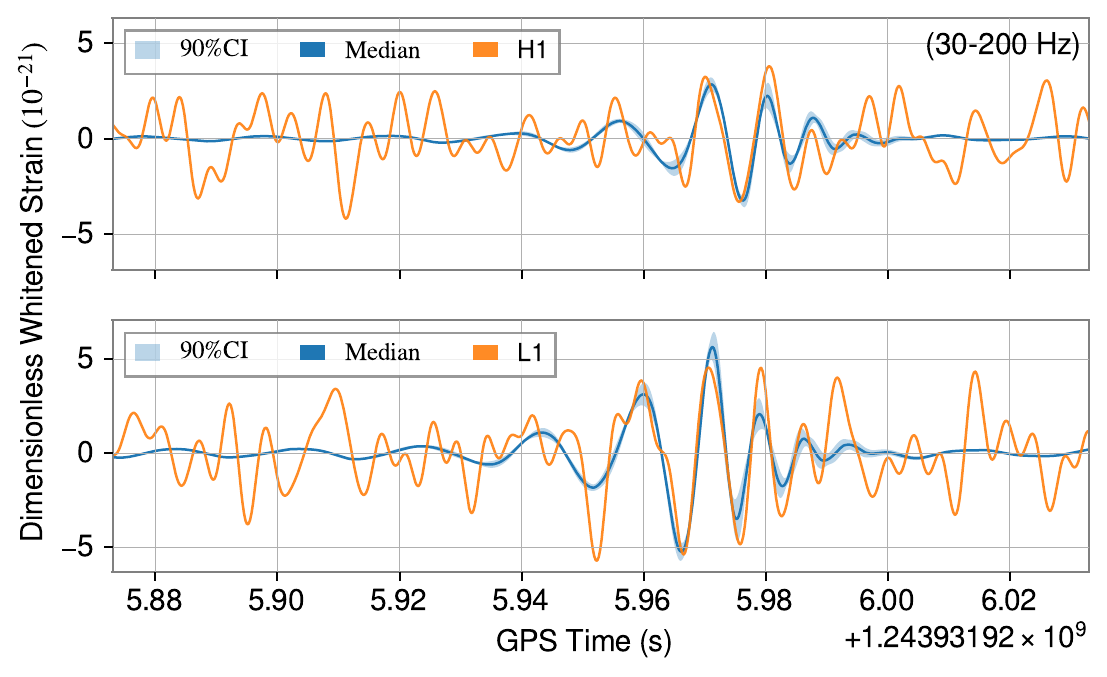}
    \caption{Same as Fig. \ref{fig:gw190511rec}, but for the new event GW190607\_083827.}
    \label{fig:gw190607rec}
\end{figure}

\begin{figure}[t!]
  \centering
  \includegraphics[width=0.99\linewidth]{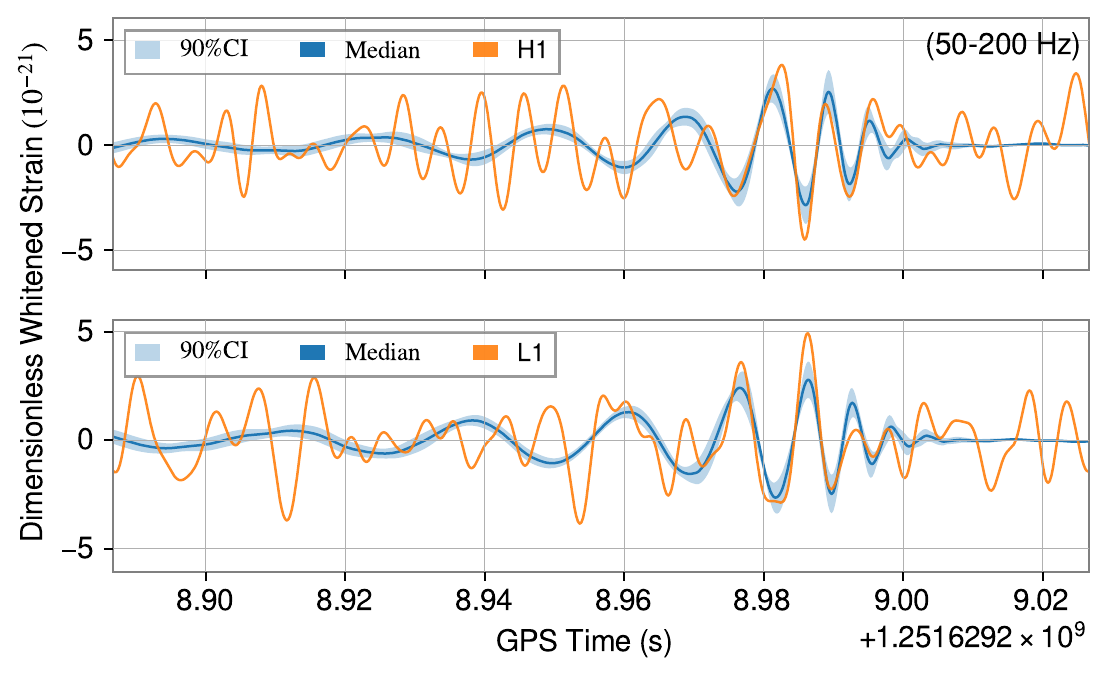}
    \caption{Same as Fig. \ref{fig:gw190511rec}, but for the new event GW190904\_104631.}
    \label{fig:gw190904rec}
\end{figure}

\begin{figure}[t!]
  \centering
  \includegraphics[width=0.99\linewidth]{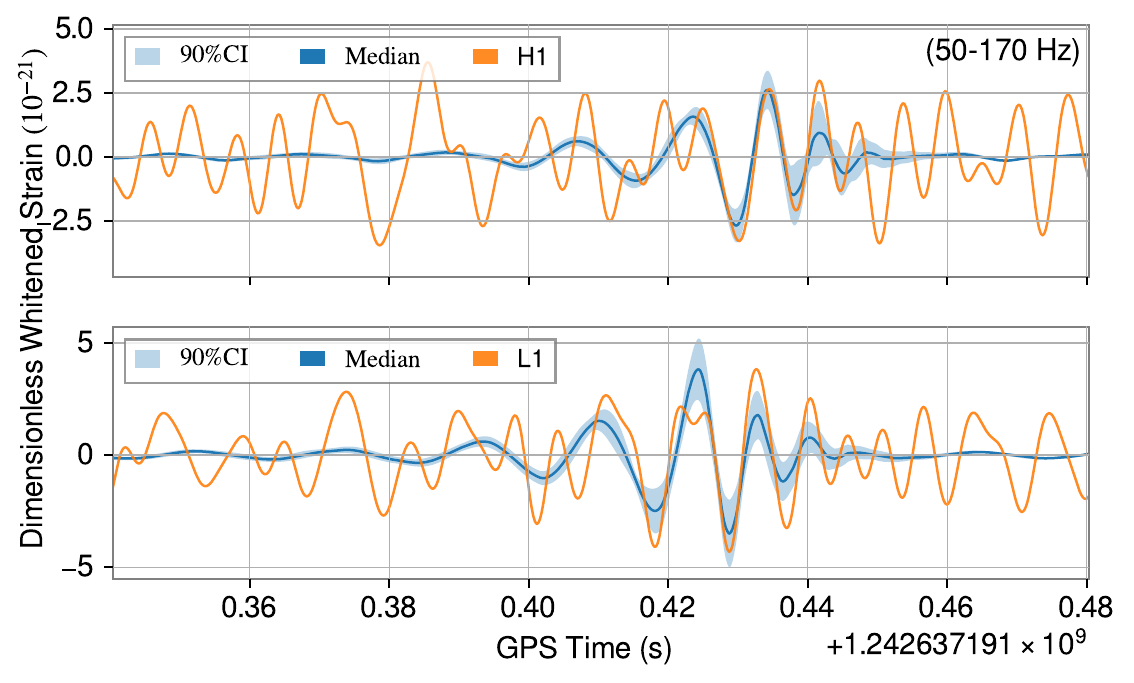}
    \caption{Same as Fig. \ref{fig:gw190511rec}, but for the new event GW190523\_085933.}
    \label{fig:gw190523rec}
\end{figure}

\begin{figure}[t!]
  \centering
  \includegraphics[width=0.99\linewidth]{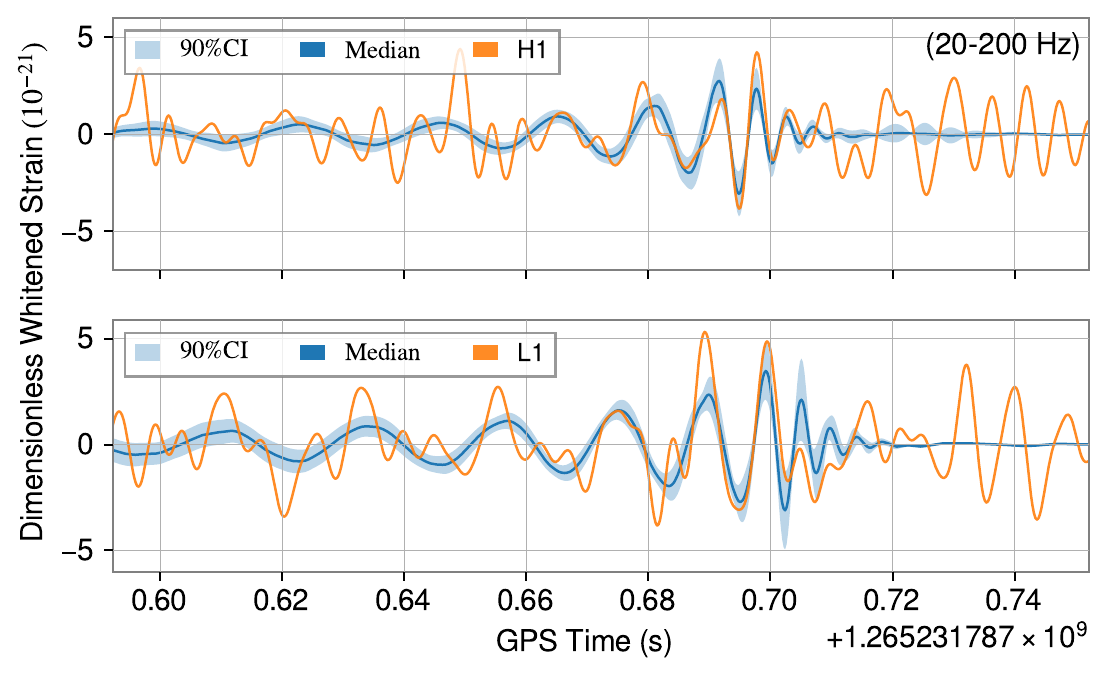}
    \caption{Same as Fig. \ref{fig:gw190511rec}, but for the new event GW200208\_211609.}
    \label{fig:gw200208rec}
\end{figure}

\begin{figure}[t!]
  \centering
  \includegraphics[width=0.99\linewidth]{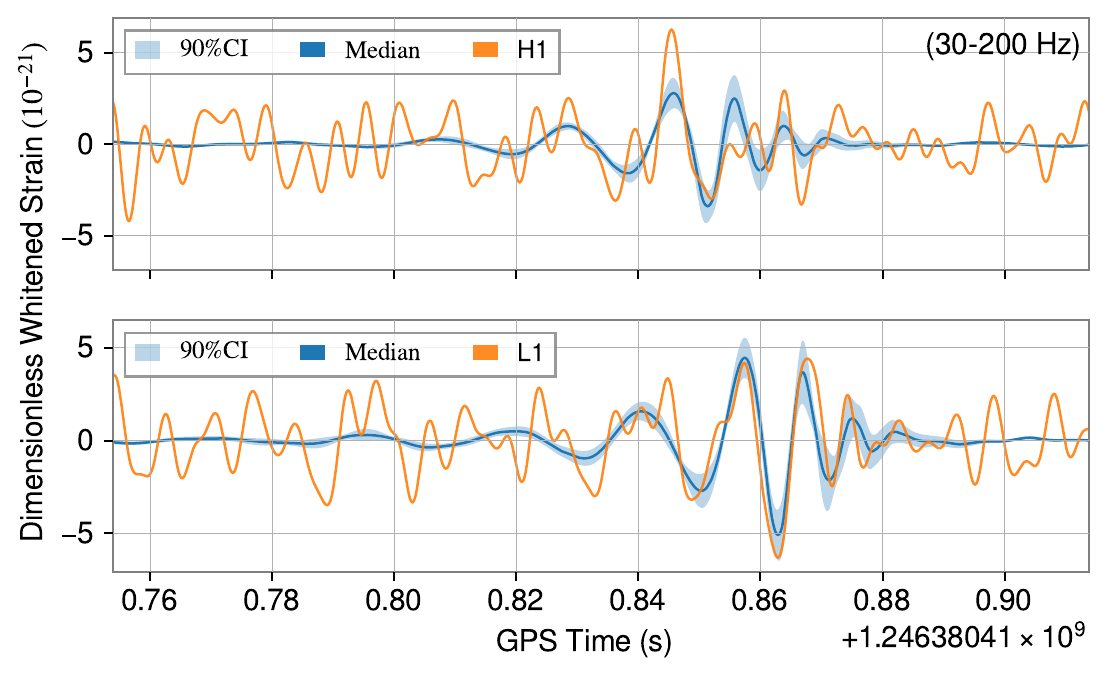}
    \caption{Same as Fig. \ref{fig:gw190511rec}, but for the new event GW190705\_164632.}
    \label{fig:gw190705rec}
\end{figure}

\begin{figure}[t!]
  \centering
  \includegraphics[width=0.99\linewidth]{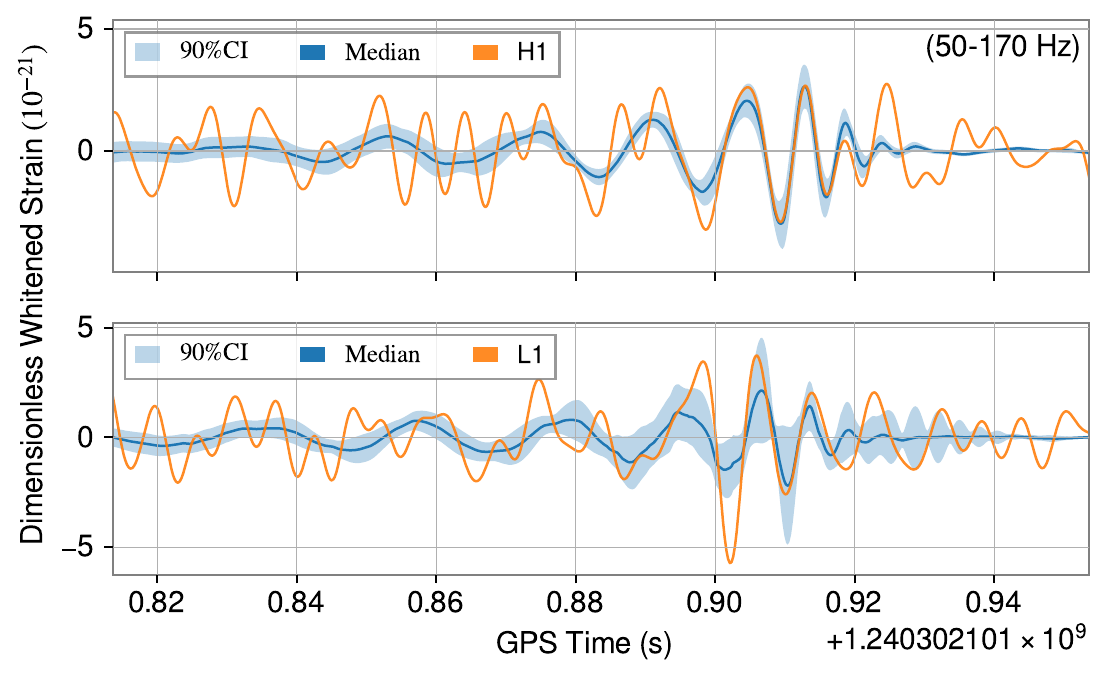}
    \caption{Same as Fig. \ref{fig:gw190511rec}, but for the new event GW190426\_082124.}
    \label{fig:gw190426rec}
\end{figure}

To provide further visual context, we present the timeseries data from both LIGO detectors for the 16 new detections made by AresGW in Figs. \ref{fig:gw190511rec}-\ref{fig:gw190426rec}. In these figures, the whitened data are depicted in orange, the reconstructed whitened-bandpassed median waveforms in blue, and their corresponding 90$\%$ confidence intervals in light blue. Additionally, the frequency band used is indicated in the upper right corner of each figure. 

\bibliography{AresGW-O3-final}%

\end{document}